\documentclass[%
 aip,
 amsmath,amssymb,
 reprint,%
]{revtex4-1}

\usepackage{graphicx}% Include figure files
\usepackage{dcolumn}% Align table columns on decimal point
\usepackage{bm}% bold math
\usepackage[utf8]{inputenc}
\usepackage[T1]{fontenc}
\usepackage{mathptmx}
\usepackage{etoolbox}
\usepackage{dsfont}
\usepackage{xcolor}

\newcommand{\be}{\begin{equation}}
\newcommand{\ee}{\end{equation}}
\newcommand{\bea}{\begin{eqnarray}}
\newcommand{\eea}{\end{eqnarray}}
\newcommand{\df}{{\operatorname d}}

\makeatletter
\def\@email#1#2{%
 \endgroup
 \patchcmd{\titleblock@produce}
  {\frontmatter@RRAPformat}
  {\frontmatter@RRAPformat{\produce@RRAP{*#1\href{mailto:#2}{#2}}}\frontmatter@RRAPformat}
  {}{}
}%
\makeatother
\begin{document}

\preprint{AIP/123-QED}

\title[Sensing with submarine optical cables]{Sensing with submarine optical cables}
% Force line breaks with \\
\author{Antonio Mecozzi}
\email{antonio.mecozzi@univaq.it}
\affiliation{ 
Department of Physical and Chemical Sciences, University of L'Aquila, L'Aquila, Italy%\\This line break forced with \textbackslash\textbackslash
}%
%\altaffiliation[Also at ]{Physics Department, XYZ University.}%Lines break automatically or can be forced with \\
%\author{B. Author}%
% \email{Second.Author@institution.edu.}
% \author{C. Author}
%\homepage{http://www.Second.institution.edu/~Charlie.Author.}
%\affiliation{%
%Second institution and/or address%\\This line break forced% with \\
%}%

\date{\today}% It is always \today, today,
             %  but any date may be explicitly specified

\begin{abstract}
{ In this paper, we establish the theoretical framework for understanding the sensing capabilities of megameter-long submarine optical cables.} We show the distinct advantage of polarization over phase in detecting subhertz environmental processes. Subsequently, we propose a scheme capable of extracting the spectrum of perturbations affecting a specific section at any position along an optical fiber by detecting the state of polarization of the backreflected light. We discuss two examples of earthquake detection and the detection of sea swells and ocean tides through the analysis of the state of polarization of an optical signal reconstructed by the receiver of a transoceanic cable, obtained from an online database.\cite{zhongwenzhan_2020} Finally, we provide the analytical expression for the cross-correlation of the polarization perturbations of two wavelength division multiplexed channels, and show that the analysis of the polarization correlations between adjacent channels can provide valuable insights into the localization of earthquakes. 
\end{abstract}

\maketitle

\section{Introduction}

Fiber sensing technologies have emerged as powerful tools for environmental monitoring, enabling precise and real-time data collection over large geographical areas. Taking advantage of the inherent properties of optical fibers, such as low loss transmission and sensitivity to external perturbations, researchers have developed innovative techniques to detect various environmental parameters, including seismic activity, ocean dynamics, and submarine fault detection. Marra et al. introduced ultrastable laser interferometry for earthquake detection using terrestrial and submarine cables,\cite{Marra:18} paving the way towards exciting progress towards disaster management and early warning systems. In Ref. \onlinecite{Zhan:21,Mecozzi:21} the authors demonstrated the use of optical coherent detection for environmental sensing, showing the potential of polarization sensing in applications such as earthquake detection and tsunami warning. In addition to seismic monitoring, fiber sensing technologies have been applied to understand ocean dynamics and detect underwater phenomena. Lindsey et al.\cite{Lindsey:19} demonstrated the use of dark fiber distributed acoustic sensing for mapping seafloor faults and monitoring ocean dynamics, while Landr{\o} et al.\cite{Landrø2022} utilized arctic fiber optic cables for sensing whales, storms, ships, and earthquakes, proving the versatility of fiber sensing in harsh environments. Skarvang et al.\cite{Skarvang:23} presented observations of local small magnitude earthquakes using state-of-polarization monitoring in a passive arctic submarine communication cable, highlighting the potential of polarization-based sensing techniques in seismic monitoring. These developments underscore the importance of fiber sensing in enhancing our understanding of environmental processes and facilitating early detection of natural hazards. Using the capabilities of optical fibers, researchers continue to push the boundaries of environmental sensing, enabling more efficient and reliable monitoring of the planet.

{ Although optical sensing is a well-established field \cite{Lu:10,Pastor-Graells:16,Chen:17,Guerrier:20,Westbrook:20} (see the excellent review provided by Ref. \onlinecite{Ping_Lu:19} and references therein), in this paper we will mainly concentrate on techniques capable of providing sensing capabilities to submarine optical cables of megameter lengths. The paper is organized as follows. In Section \ref{Basic}, we set up the basic equations describing the propagation of a polarized optical field in an unperturbed optical fiber. We introduce here the two quantities that play a central role for sensing, specifically the polarization-averaged phase and the polarization rotation vector, and show how these quantities are related to the fiber polarization-averaged wavevector and birefringence vector. In Sec. \ref{Pert}, we consider the effect that time-dependent external perturbations produce on the polarization-averaged wavevector and on the birefringence vector of a fiber with random polarization coupling, like those universally used in optical communication systems. In Secs. \ref{Phase} and \ref{Pbs}, we set the theoretical basis for understanding how these perturbations can be used for sensing. In Sec. \ref{Phase}, we show the limitations imposed by the laser phase noise on the performance of phase-based sensing when applied to very low-frequency processes. We also show here the benefit of low-pass filtering the frequency deviations (and not the phase deviations) over the smallest possible bandwidth compatible with the signal bandwidth. In Sec. \ref{Comp}, we study how the signal accumulates along the link in sensing approaches based on transmission configurations, and show that while polarization is almost immune to the orientation of a seismic perturbation because the signal accumulation is incoherent, the phase signal is potentially sensitive to the orientation of a seismic wave, with higher sensitivity to waves orthogonal to the cable orientation in the perturbed area. This result may explain a trend observed in the outcomes of the experimental campaign reported in Ref. \onlinecite{Donadello:24}. Section \ref{Rot} is propedeutical to the analysis in Sec. \ref{Loc}, and discusses the importance of using a rotating frame in the analysis of polarization perturbation. The use of this frame is equivalent to the use of the interaction representation in quantum mechanics.\cite{sakurai:17} In Sec. \ref{Loc}, we propose a scheme that can extract the spectrum of the perturbation acting at a given position along the cable by detecting the state of polarization of the backreflected light, paving the way to the possibility of localizing the perturbation. Finally, in Sec. \ref{exp}, we delve into the vast database in Ref. \onlinecite{zhongwenzhan_2020} to show a few examples of earthquake detection using the reconstructed state of polarization of the receiver of the Curie cable system connecting Los Angeles, California, to Valparaiso, Chile. In the same section, we also present an analytical expression of the cross-correlation of the state of polarization between two wavelength-division multiplexed channels and highlight how studying the correlations between the polarization of nearby channels may yield valuable insights into the localization of earthquakes.}

%We will first set up the theoretical basis for the understanding of the sensing capability of optical fibers. We will show the pros and cons of the use of the polarization-averaged optical phase and of the polarization as a sensing tool. We will propose a scheme that can provide the spectrum of the perturbation acting upon the cable by detecting the state of polarization of the backreflected light. We will then delve into the vast database in Ref. \onlinecite{zhongwenzhan_2020} to show a few examples of earthquake detection using the reconstructed state of polarization of the receiver of the Curie cable system connecting Los Angeles, in California, to Valparaiso in Chile. Finally, we will present an analytical expression of the cross-correlation of the state of polarization between two wavelength division multiplexed channels and highlight how studying the correlations between the polarization of nearby channels may yield valuable insights into localization of earthquakes.

\section{Basic equations} \label{Basic}
Let the transverse electric field propagating in a single mode fiber be represented by the column vector
\be \begin{bmatrix}
\mathcal {E}_x(z) \\
\mathcal{E}_y(z) 
\end{bmatrix} = \begin{bmatrix}
E_x(z) \\
E_y(z) 
\end{bmatrix} \exp\left(-i \omega_0 t\right), \ee
where $\omega_0$ is the center optical frequency. Let us define
\be \begin{bmatrix}
E_x(z) \\
E_y(z) 
\end{bmatrix} = A(z) |\vec s(z) \rangle, \ee
where $A(z) = \left[|E_x(z)|^2 + |E_y(z)|^2 \right]^{1/2}$ and 
\be |\vec s(z) \rangle = \begin{bmatrix}
s_x(z) \\
s_y(z) 
\end{bmatrix}, \ee
is a Jones vector, normalized such that 
\be \langle \vec s(z) |\vec s(z) \rangle = 1. \ee
Here, following Dirac's bra-ket notation, we defined $\langle \vec s(z) |=(|\vec s(z) \rangle)^\dagger$, with the dagger standing for hermitian conjugation (the transpose conjugate of the vector). Assuming negligible polarization dependent loss and gain, the evolution of the amplitude of the field $A(z)$ is only determined by the gain and loss profile of the fiber, and decouples to that of the polarization. In a right-handed reference frame with $x$ and $y$ in the transverse plane and $z$ in the propagation direction, the propagation of the Jones vector can then be described by the equation
\be \frac{\df }{\df z} |\vec s(z) \rangle = i \, \mathbf{V}(z) \, \mathbf{B}(z) \, \mathbf{V}^\dagger(z) \, |\vec s(z) \rangle, \label{dsdz} \ee
where $\mathbf{V}(z)$ is the unitary matrix
\be \mathbf{V}(z) = \begin{bmatrix}
\cos \theta(z) & -\sin \theta(z) \\
\sin \theta(z) & \cos \theta(z) 
\end{bmatrix} \ee
representing an anticlockwise rotation by the angle $\theta(z)$ of in the $x-y$ plane with respect to the $z$ axis, so that $\mathbf{V}(z) \, \mathbf{B}(z) \, \mathbf{V}^\dagger(z)$ produces a rotation by the same angle of the birefringence vector, and
\be \mathbf{B}(z) = \begin{bmatrix}
\beta_1(z) & 0 \\
0 & \beta_2(z) 
\end{bmatrix}, \ee
with $\beta_1(z)$ and $\beta_2(z)$ the magnitudes of the (local) eigenvalues of the transmission matrix. 
Defining the Pauli matrices as
\be \sigma_1 = \begin{pmatrix}
1 & 0 \\
0 & -1 
\end{pmatrix}, \quad \sigma_2 = \begin{pmatrix}
0 & 1 \\
1 & 0 
\end{pmatrix}, \quad \sigma_3 = \begin{pmatrix}
0 & i \\
-i & 0 
\end{pmatrix}, \ee
and the Pauli spin vector as the Stokes vector $\vec \sigma = \sigma_1 \hat e_1 + \sigma_2 \hat e_2 + \sigma_3 \hat e_3$, where $\hat e_i$ are the canonical basis vectors of Stokes space, Eq. (\ref{dsdz}) becomes
\be \frac{\df }{\df z} |\vec s(z) \rangle = i \left[\beta_0(z) \mathds 1 + \frac{\vec \beta(z)}{2} \cdot \vec \sigma \right] \, |\vec s(z) \rangle, \ee
where $\mathds 1$ is the two-by-two unit matrix, and
\bea \beta_0(z) &=& \frac{1}{2} \left[\beta_1(z) + \beta_2(z)\right], \\
\vec \beta(z) &=& \beta(z) \cos [2 \theta (z)] \hat e_1 +  \beta(z) \sin[2 \theta(z)] \hat e_2, \eea
with
\be \beta(z) = \beta_1(z) - \beta_2(z). \ee
Solution of Eq. (\ref{dsdz}) is

\be |\vec s(z) \rangle = \mathbf{U}_0(z)|\vec s(0) \rangle, \label{s0} \ee
where $\mathbf{U}_0(z)$ satisfy the equation
\be \frac{\df \mathbf{U}_0(z)}{\df z} = i \left[\beta_0(z) \mathds 1 + \frac{\vec \beta(z)}{2} \cdot \vec \sigma \right] \, \mathbf{U}_0(z), \label{dU0dz}\ee
with initial condition $\mathbf{U}_0(0) = \mathds 1$. Using in Eq. (\ref{dU0dz}) the expansion
\be \mathbf{U}_0(z) = \exp[i \varphi_0(z)] \mathbf{U}(z), \label{Udef} \ee
we obtain the equation
\be i \frac{\df \varphi_0(z)}{\df z} \mathbf{U}(z) + \frac{\df \mathbf{U}(z)}{\df z} 
= i \left[\beta_0(z) \mathds 1 + \frac{\vec \beta(z)}{2} \cdot \vec \sigma \right] \, \mathbf{U}(z), \ee
which is verified if
\be \frac{\df \varphi_0(z)}{\df z} = \beta_0(z), \label{phi0} \ee
and
\be \frac{\df \mathbf{U}(z)}{\df z} = \frac{i}{2} \left[ \vec \beta(z) \cdot \vec \sigma \right] \mathbf{U}(z),\label{dUdz0}\ee
with initial conditions $\varphi_0(0) = 0$ and $\mathbf{U}(0) = \mathds 1$. The solution of Eq. (\ref{dUdz0}) can be formally written as 
\be \mathbf{U}(z) = \prod_{z'=0}^z\exp \left[\frac{i}{2} \, \vec \beta(z') \cdot \vec \sigma \, \, \df z'\right], \ee
with the individual members of the product belonging to $\mathrm{SU}(2)$ (the special unitary group of degree 2, which is the group of unitary two-by-two matrices of unit determinant). Being $\mathrm{SU}(2)$ closed with respect to matrix multiplication,\footnote{This property is an immediate consequence of the isomorphism between $\mathrm{SU}(2)$ and $\mathrm{SO}(3)$ (the group of three dimensional rotations around the origin of Stokes space) and of the fact that the concatenation of rotations is still a rotation.} $\mathbf{U}(z)$ itself is a member of $\mathrm{SU}(2)$.\cite{Gordon:00,sakurai:17} { Performing the trace of $1/(2 i) \log \mathbf{U}_0$ averages out the polarization dependent term of the argument of the exponential in $\mathbf{U}_0$ and is therefore equivalent to the average over all the polarization states.\cite{Gordon:00} Consequently, this operation returns the phase of the optical field averaged over all polarization states,
\be \overline \varphi_0 = \frac 1 {2 i} \, \mathrm{trace} \log  \mathbf{U}_0 = \frac 1 {2i} \log \det \mathbf{U}_0. \ee
It is easy to show, using Eq. (\ref{Udef}) and the property that $\mathbf{U}(z)$ belongs to $\mathrm{SU}(2)$ hence its determinant is one, that
\be \overline \varphi_0 = \varphi_0 + \frac 1 {2 i} \, \log  \det \mathbf{U} = \varphi_0. \ee
Consequently, from now on we will refer to $\varphi_0$, defined as the solution of Eq. (\ref{phi0}),  as the polarization-averaged phase.} A member of the $\mathrm{SU}(2)$ group can always be expressed as\cite{Gordon:00,sakurai:17}
\be \mathbf{U}(z) = \exp\left[\frac{i} 2 \vec \varphi(z) \cdot \vec \sigma\right], \label{Uz}\ee
so that
\be \mathbf{U}_0(z) = \exp\left[i \varphi_0(z) \mathds 1 + \frac{i} 2 \vec \varphi(z) \cdot \vec \sigma\right]. \ee
{ Since the vector $\vec \varphi(z)$ contains all the information on the rotation of the field polarization from the input to the position $z$ along the fiber we will refer to this vector as the polarization rotation vector.} Being $\varphi_0(z)$ solution of Eq. (\ref{phi0}), it is not affected by the rotation of the birefringence axes. Consequently, the fluctuations of the average phase are insensitive to the rotation of the birefringence axes, or equivalently, they are uncoupled to the fluctuations that leave unchanged the modulus of the birefringence eigenvalues. Consequence of this property is that, as we will see, fiber twist does not affect the polarization-averaged phase. The evolution of the phase fluctuations can be expressed in terms of the polarization-averaged wavevector by simply integrating Eq. (\ref{phi0})
\be \varphi_0(z) = \int_0^z \beta_0(z') \df z'. \label{varphi0} \ee
Let us define the Stokes vector corresponding to $|\vec s(z)\rangle$ as\cite{Gordon:00}
\be \vec s(z) = \langle \vec s(z)| \vec \sigma | \vec s(z) \rangle. \ee
This vector describes in Stokes space the evolution of the field polarization. The equation describing the evolution of $\vec s(z)$ can be obtained using Eq. (\ref{s0}) in the definition of Stokes vector { and using that $\mathbf{U}_0(z) = \exp[i \varphi_0(z)] \mathbf{U}(z)$}
\be \vec s(z) = \langle \vec s(0)|\mathbf{U}^\dagger(z) \vec \sigma \mathbf{U}(z)| \vec s(0) \rangle,  \ee
then differentiating both terms and using Eq. (\ref{dUdz0}) and the equality $(\vec a \cdot \vec \sigma) \vec \sigma = \vec a \mathds 1 + i \vec a \times \vec \sigma$ and that $\vec \sigma  (\vec a \cdot \vec \sigma) = \vec a \mathds 1 - i \vec a \times \vec \sigma$. The final result is\cite{Gordon:00}
\be \frac{\df \vec s(z)}{\df z} = \vec \beta(z) \times \vec s(z). \label{dsdz1}\ee
{ Equation (\ref{varphi0}) shows that the value of the polarization-averaged phase is the effect of the accumulation of the infinitesimal contributions $\beta_0(z') \df z'$ and is therefore natural referring to $\beta_0(z')$ as the polarization-averaged wavevector. Equation (\ref{dsdz1}) in turn shows that the Stokes vector of the field evolves by accumulating infinitesimal rotations described by the rotation vectors $\vec \beta(z)$, and we will therefore refer to this vector as the birefringence vector.} Having derived the fundamental equations, the following section will be devoted to the analysis of the effect of external perturbations on the polarization-averaged wavevector $\beta_0$ and on the birefringence vector $\vec \beta$.

\section{Wavevector perturbations} \label{Pert}

Let us consider a perfectly cylindrical fiber and treat any effects of the deviations from cylindrical symmetry (including birefringence) as a small perturbation. Assume first a perfectly cylindrical fiber with no preexisting birefringence. If some strain is applied to the fiber, the polarization-averaged wavevector $\beta_0 = 2 \pi n_0/\lambda$ turns into
\be \beta_0 = \frac{2 \pi}{\lambda} n(z), \ee
where $n(z)$ is the refractive index given by
\be n(z) = n_0 + C \epsilon(z). \ee
Here, $n_0$ is the unperturbed glass refractive index, $C$ is the photoelastic factor and $\epsilon(z)$ the strain.  The strain also affects the propagation distance at every $z$, becoming $z(\epsilon) = [1+\epsilon(z)]z$. Thus, $\df z$ turns into $\df z'(\epsilon) = [1+\epsilon(z')] \df z'$ and { if we retain the symbol $\varphi_0$ for the polarization-averaged phase change in the absence of perturbation, and use the symbol $\varphi(z)$ for the value in the presence of perturbation, we get}
\be \varphi(z) = \int_0^{z} \frac{2 \pi}{\lambda} [n_0 + C \epsilon(z')] \, [1+\epsilon(z')] \df z'. \ee
Neglecting the term of the order of $\epsilon^2$ we obtain
\be \varphi(z) = \beta_0 \int_0^{z} \left[1 + \xi \epsilon(z')\right] \, \df z'. \ee
where $\xi = 1+ C/n_0$ is the photoelastic scaling factor for longitudinal strain (for isotropic glass, $\xi \simeq 0.78$), and 
\be \beta_0 = \frac{2 \pi}{\lambda} n_0 \ee
is the unperturbed wavevector. The effect of strain is to produce to the polarization-averaged propagation constant the relative change 
\be \frac{\Delta \beta}{\beta_0} = \xi \epsilon. \label{deltabeta0} \ee
%
%If the strain is caused by hydrostatic pressure via the Poisson effect, the symmetry of the perturbation implies that it does not affect the orientation of the birefringence vector, that is that $\Delta \vec \beta$ is parallel to $\vec \beta$. In addition, a perturbation with cylindrical symmetry cannot generate a birefringence in a cylindrical fiber, because it would break the symmetry. This condition is fulfilled if we assume that the perturbation affects the eigenvector on each eigenpolarization proportionally to the magnitude of the wavevector itself, consistently with the equation for the polarization-averaged wavevector
Assume now that the fiber has some (small) birefringence and treat the effects of the birefringence as a perturbation. If the strain is induced by hydrostatic pressure via the Poisson effect, the cylindrical symmetry of the perturbation implies that it does not alter the orientation of the birefringence vector, meaning that $\Delta \vec{\beta}$ is parallel to $\vec{\beta}$. Moreover, a perturbation with cylindrical symmetry cannot induce birefringence in a cylindrical fiber, since any nonzero birefringence would break the cylindrical symmetry. This requirement is satisfied if we assume that the perturbation alters the eigenvector on each eigenpolarization proportionally to the magnitude of the wavevector itself.  Enforcing this condition alongside  the additional criterion that Eq. (\ref{deltabeta0}) is recovered when birefringence is absent results in
\be \frac{\Delta \beta_h}{\beta_h} = \xi \epsilon \quad h =1,2. \ee
The use of this equation for $h = 1$ and $2$ gives
\be \frac{\Delta \beta_1- \Delta \beta_2}{\beta_1-\beta_2} = \xi \epsilon, \ee
that is, being $\Delta \beta_1(z)- \Delta \beta_2(z) = \Delta \beta$ and $\beta_1-\beta_2 = \beta$, the equation $\Delta \beta = \xi \epsilon \beta$. The additional condition of parallelism between $\Delta \vec \beta$ and $\vec \beta$ yields the expression
\be \Delta \vec \beta = \xi \epsilon \, \vec \beta. \label{deltavecbeta} \ee
We obtain the important result that in quasi-cylindrical symmetric fibers, in which birefringence is a small perturbation, the dependence of the birefringence on strain can be inferred from the dependence on strain of the polarization-averaged wavevector, which is easier to characterize experimentally.\cite{Mecozzi:21}

Another process affecting birefringence is the twist of the fiber. We have shown in the previous section that the fluctuations of the average phase are insensitive to rotations of the birefringence axes that leave unchanged the modulus of the birefringence eigenvalues, so that fiber twist, unlike strain, does not modify the polarization-averaged wavevector and phase.
When the fiber is twisted by an angle $\alpha$ around its axis, the first-order change in birefringence is given by
\be \Delta \vec \beta = 2 \, \alpha \, \hat e_3 \times \vec \beta. \label{Deltabetaalpha} \ee
Notice that $\hat e_3$ and $\vec \beta$ are orthogonal so that $\Delta \vec \beta$ belongs to the $(\hat e_1,\hat e_2)$ plane of Stokes space and $\Delta \beta = 2 \, |\alpha| \beta$.

In both cases of strain and twist, the magnitude of the birefringence perturbation is proportional to the pre-existing static birefringence. This characteristic enables the establishment of a correlation between a fiber's sensitivity to external perturbations and the fiber's polarization mode dispersion, a correlation we will elaborate on in Section \ref{Pbs}.

\section{Phase-based sensing} \label{Phase}

Phase sensing relies on measuring the perturbation to the phase of an optical field propagating along an optical fiber due to strain induced by coupling with the surrounding environment. In the previous section, it was demonstrated that strain causes a local alteration in the polarization-averaged wavevector, as described by Eq. (\ref{deltabeta0}). Consequently, this modification affects the polarization-averaged phase accumulated over a distance $z$ adding to Eq. (\ref{varphi0}) the term
\be \Delta \varphi(z,t) = \int_0^z \Delta \beta(z',t) \df z', \ee
which is equal to, using Eq. (\ref{deltabeta0})
\be \Delta \varphi(z,t) = \frac{2 \pi}{\lambda} n_0 \int_0^z \xi \epsilon(z',t) \df z'. \label{phase} \ee
The primary noise source that limits the sensitivity of phase measurements is the phase noise of the laser probe. In the following, we examine the limitations that phase noise imposes on the sensitivity of phase measurements. Consider a laser with quantum-limited white frequency noise. Accurate detection of the phase deviation occurring over the time $T$
\be \Delta \varphi_\mathrm{sig}(T) = \Delta \varphi(z,t+T)- \Delta \varphi(z,t), \label{cond} \ee
requires that $|\Delta \varphi_\mathrm{sig}(T)| \gg \Delta \varphi_\mathrm{rms}$, where $\Delta \varphi_\mathrm{rms}$ is the root mean square of the phase deviation due to the laser phase noise. The phase fluctuations of a quantum limited laser are described by a Wiener process with diffusion constant $D_\varphi = 2 \pi \nu$, where $\nu$ is the laser linewidth, so that 
\be \langle \Delta \varphi_\nu^2(T) \rangle = D_\varphi T. \ee
{ Let us assume that we detect the phase over an observation time $T_\mathrm{win} = n T$ with sampling time $T$, and consider the instantaneous frequency deviations, which we define as the phase deviations divided by the time interval,
\be \dot \varphi(T) = \frac {\Delta \varphi(z,t+T) - \Delta \varphi(z,t)}{T}. \label{dotvarphi} \ee
An estimate of the phase deviation over the time $T_\mathrm{win}$ can be obtained with the following procedure. We first take the average of the $n$ frequency estimates over the time interval. Being the signal and the additive phase noise of the laser independent, we may consider the effect of the averaging on the noise and on the signal separately. Let us consider the noise first. The root mean square of each term of the average is 
\be \dot \varphi_\mathrm{rms}(T) = \frac 1 T \sqrt{\langle \Delta \varphi_\nu^2(T) \rangle} = \sqrt{D_\varphi/T}, \label{dotvarphi1} \ee
and because of the independence of the increments of a Wiener process, the root mean square of the average over $n$ consecutive realizations is $(1/\sqrt n)$ times the root mean square of each term, that is
\be \dot \varphi_\mathrm{rms}(T,n) = \sqrt{D_\varphi/(n T)} = \sqrt{D_\varphi/T_\mathrm{win}}. \label{dotvarphi2} \ee
This procedure is equivalent to applying a low-pass filter to the sequence with a bandwidth of the order of $1/T_\mathrm{win}$. 

Let us now move to the analysis of the signal. If we assume that the rate of variation of the signal is constant over the observation time $T_\mathrm{win}$, the averaging does not affect the signal estimate and we have $\dot \varphi_\mathrm{sig}(T_\mathrm{win}) = \dot \varphi$. The condition $|\dot \varphi_\mathrm{sig}(T_\mathrm{win})| \gg \dot \varphi_\mathrm{rms}(T_\mathrm{win})$ becomes, after using Eq. (\ref{dotvarphi}) and the expression of $D_\varphi$ in terms of the linewidth,
\be |\dot{\varphi}|  \gg \sqrt{2 \pi \nu / T_\mathrm{win}}, \ee
that is
\be \nu \ll \frac {\dot{\varphi}^2 T_\mathrm{win}} {2 \pi}. \label{nuTwin} \ee
Equation (\ref{nuTwin}) shows the beneficial effect of a large observation time $T_\mathrm{win}$, which should be chosen to be as large as possible while remaining compatible with the bandwidth of the observed process, that is, with the maximum time over which $\dot \varphi$ can be assumed to be constant. 

The analysis of Eq. (\ref{dotvarphi}) reveals that the root mean square of the frequency noise $\dot \varphi_\mathrm{rms}(T_\mathrm{win})$ depends only on $T_\mathrm{win} = n T$ and is independent of $n$ and $T$ separately. In particular, its value coincides with that for $n=1$, when no averaging is performed and $T_\mathrm{win}= T$.\footnote{The property that the root mean square of the frequency noise is independent of the number of averaged samples $n$ is a direct consequence of the white noise character of the frequency noise of the laser.} One may therefore be tempted to use a large sampling time and avoid filtering, using for the sampling time the desired value of $T_\mathrm{win} = T$.  In this case, however, the $2 \pi$ periodicity of the phase imposes that $|\Delta \varphi_\mathrm{sig}| = |\dot{\varphi}| T \le 2 \pi$. Using Eq. (\ref{nuTwin}) with $T_\mathrm{win}= T$ and for $T$ its maximum value $2 \pi/|\dot{\varphi}|$, we obtain 
\be \nu \ll |\dot{\varphi}|, \label{rest} \ee
which is a more restrictive condition than that given by Eq. (\ref{nuTwin}). 

It is very important to notice that inverting the order of the two operations of averaging (filtering) and differentiating makes the filtering ineffective. For a given $T_\mathrm{win}$, filtering of the frequency deviations does not reduce the frequency noise compared to the case without filtering and $T = T_\mathrm{win}$. Similarly, filtering of the phase noise does not reduce the phase fluctuations. The key benefit of low-pass filtering the frequency fluctuations is that it allows the use of longer integration times $T_\mathrm{win}$ (or equivalently narrower low-pass filters). Notice that, while the condition $|\dot{\varphi}| T \le 2 \pi$ must always be valid, the condition that the noise is smaller than the signal over the sampling time, $|\dot{\varphi}| T \gg \sqrt{2 \pi \nu T}$, does not.

In principle, one could also avoid filtering following the phase over time intervals $T_\mathrm{win}$ that span many $2 \pi$ periods using a sampling time sufficiently short to ensure that the phase changes do not exceed $2\pi$ over the sampling period. One can then calculate the frequency deviations by subtracting the initial values to the final values and dividing by $T_\mathrm{win}$. The condition for the linewith would still be Eq. (\ref{nuTwin}). Besides this procedure being less reliable due to the need to unwrap the phase, it has the additional disadvantage of not filtering out other non-white noise components that might affect the detected frequency, originating either from the laser itself or from the detection process.

Our analysis shows the importance, in phase-based sensing that does not employ ultra-narrow linewidth lasers, of transforming the phase samples, obtained with a sampling time $T$ sufficiently small to avoid that the phase increments over $T$ exceed $2\pi$, into frequency samples, and the need for low-pass filtering the resulting frequency samples over a sufficiently narrow bandwidth. In practice, one should choose the narrowest possible bandwidth compatible with the bandwidth of the observed process.

Let us now check the condition (\ref{nuTwin}) over the earthquake detection reported in Fig. 2 c) of Ref. \onlinecite{Mazur:24}. There, frequency oscillations of about 10 Hz amplitude can be clearly observed after a 60 mHz filtering of the phase traces. In Eq. (\ref{nuTwin}), $1/T_\mathrm{win}$ is the cutoff frequency of the low-pass filter, hence a 60 mHz low-pass filtering is equivalent to averaging time $T_\mathrm{win} \simeq 1/0.06 \simeq 17$ s. The condition for the laser linewidth Eq. (\ref{nuTwin}) dictates that $\nu \ll 17 \cdot 10^2/(2 \pi) \simeq 265$ Hz, and this condition was satisfied by the linewidth of the fiber laser used in the study of Ref. \onlinecite{Mazur:24}, which was less than 100 Hz.}

% It is questionable whether the detection of processes with much weaker dynamics that occur over the same timescale or longer, like those associated with tsunami's propagation, would be possible even utilizing lasers with ultra-low linewidth.

The above considerations apply to approaches based on the analysis of temporal traces. Let us now consider the case in which the phase is analyzed using instead a spectrogram, which is a time-frequency representation employing a short-time Fourier transform.\cite{Oppenheim:99} Assume that the phase $\varphi_0(t)$ is characterized by a bandwidth $1/(2T)$.  The instantaneous frequency perturbed by the frequency noise of the laser has the expression
\be \frac{\df \varphi_0(t)}{\df t} = N_0(t) + \frac{\df \varphi(t)}{\df t}, \ee
and where $N_0(t)$ is the frequency noise of a quantum limited laser with linewidth $\nu$, namely a white noise term with correlation function
\be \langle N_0(t) N_0(t')\rangle = D_\varphi \delta(t-t'). \label{whitenoise} \ee
Let us assume that the signal spectrum is entirely contained within a bandwidth $1/(2T)$. Complete signal reconstruction can be accomplished by sampling the signal at intervals of $T$. Let $n \gg 1$ be the number of samples, so that the overall detection window is $nT$. Optimal signal reconstruction is achieved by convolution of the signal plus noise with the matched filter
\be F(t) = (1/T) \operatorname{sinc}(t/T), \label{filt0} \ee
where $\operatorname{sinc}(x) = \sin(\pi x)/(\pi x)$. After filtering, Eq. (\ref{whitenoise}) becomes
\be \frac{\df \varphi_0(t)}{\df t} = N(t) + \frac{\df \varphi(t)}{\df t}, \label{freq0}\ee
where 
\be N(t) = \int_{-\infty}^\infty F(t-t') N_0(t') \df t', \label{filt} \ee
while the second term at right-hand side of Eq. (\ref{freq0}), its spectrum being entirely contained within the filter bandwidth, remains unaltered by filtering. Let us define the spectra
\be \tilde N(\Omega_k) = \frac 1 n \, \sum_{h =0}^{n-1} N(h T) \exp\left(i \Omega_k h T\right), \label{NOmega} \ee
\be \tilde \varphi(\Omega_k) = \frac 1 n \, \sum_{h =0}^{n-1} \varphi(h T) \exp\left(i \Omega_k h T\right), \label{phiOmega} \ee
and likewise, the spectrum of $\varphi_0(t)$. Here, the angular frequencies have the discrete values (assuming $n$ even)
\be \Omega_k = \frac{2 \pi k}{n T}, \quad k = -n/2+1, \ldots, n/2. \label{Omegak} \ee
The normalization factor $1/n$ in the Fourier transform definition ensures that the spectral amplitude of a pure sinusoidal modulation $\varphi(t) = A \cos(\Omega_{\overline k} t)$ is $\tilde \varphi(\Omega_{\overline k}) = (A/2) (\delta_{k, -\overline{k}} + \delta_{k, \overline{k}})$ and does not depend on the number of samples $n$. Fourier transforming both sides of Eq. (\ref{freq0}) yields
\be i \Omega_k \tilde \varphi_0(\Omega_k) = \tilde N(\Omega_k) + i \Omega_k \tilde \varphi(\Omega_k). \label{spectral}\ee
Using Eq. (\ref{whitenoise}) into Eq. (\ref{filt}), we obtain 
\be \langle N(h T) N(k T) \rangle = \frac{2 \pi \nu}{T} \delta_{h,k}. \label{Nhk} \ee
Taking the square of Eq. (\ref{NOmega}), averaging and using Eq. (\ref{Nhk}) yields 
\be \langle \tilde N^*(\Omega_h) \tilde N(\Omega_k) \rangle = \frac {2 \pi \nu} {n T} \delta_{h,k}. \label{NOmegahk} \ee
A faithful detection of the phase modulation at angular frequency $\Omega_k$ requires that the signal in Eq. (\ref{spectral}) is much larger than the noise,
\be \langle  |\tilde N(\Omega_k)|^2 \rangle \ll |\Omega_k|^2 |\tilde \varphi(\Omega_k)|^2. \label{cond1} \ee
By using Eq. (\ref{NOmegahk}) for $h = k$ into Eq. (\ref{cond1}), solving for $\nu$ and defining $T_\mathrm{win} = n T$ as the amplitude of the time window of the Fourier transform and $f_k = \Omega_k/(2 \pi)$ as the frequency in hertz, Eq. (\ref{cond1}) yields
\be \nu \ll 2 \pi |f_k|^2 |\tilde \varphi(2 \pi f_k)|^2 T_\mathrm{win}. \label{null} \ee
This equation gains clarity when we define $T_k = 1/f_k$ as the period of the spectral component with frequency $f_k$, becoming
\be \nu \ll 2 \pi |f_k| |\tilde \varphi(2 \pi f_k)|^2 \frac{T_\mathrm{win}}{T_k}. \label{null1} \ee
The conditions stated by Eqs. (\ref{null}) and (\ref{null1}) are independent of the sampling time $T$ and the number of samples $n$, provided that $f_k \le 1/(2 T)$, which results from the condition $|\Omega_k| \le |\Omega_{n/2}|$ in Eq. (\ref{Omegak}). Specifically, Eq. (\ref{null1}) illustrates that the maximum tolerable linewidth is directly proportional to the modulation frequency, the amplitude square of the phase modulation and the ratio $T_\mathrm{win}/T_k$, which is the number of temporal periods contained within the time window of the Fourier transform $T_\mathrm{win} = nT$. Therefore, widening the time window of the short-time Fourier transform $T_\mathrm{win}$ can alleviate the requirements on the laser linewidth compared to the detection of transient phase deviations discussed earlier in this section. However, this can only be achieved by sacrificing temporal resolution, which is determined by the amplitude of the time window $T_\mathrm{win}$. Consequently, the necessity of using ultra-stable lasers for detecting low intensity sub-hertz signals persists even with approaches based on the short-time Fourier transform. 

The independence of Eqs. (\ref{null}) and (\ref{null1}) on the sampling time $T$ and the number of samples $n$ suggests that these equations are also valid in the continuous limit where $n$ tends to infinity and $T$ to zero, with their product $T_\mathrm{win} = nT$ finite. The derivation of Eqs. (\ref{null}) and (\ref{null1}) in the continuous case is presented in appendix \ref{AppendixB}.

{ Equation (\ref{null}) applies to all measurements using phase as a sensing probe, including distributed acoustic sensing. It does not however prevent the detection of millihertz signals if the oscillations have sufficient amplitude. For instance, Eq. (\ref{null}) for the 1 mHz oscillations reported in Fig. 4 c) of Ref. \onlinecite{Lindsey:19}, obtained with $T_\mathrm{win} = 7200$ s, yields $\nu \ll 0.045 \, |\tilde \varphi|^2$. If the amplitude of the 1000 s period oscillations corresponds to a change of the optical path by $\pm 10$ wavelengths, corresponding to a frequency shift of $\pm 10$ mHz, we have $|\tilde \varphi| \simeq 20 \pi$, we obtain $\nu \ll 714$ Hz, which is a condition satisfied by the linewidth of good quality fiber lasers.\cite{Mazur:24}}

{ As a final remark, it is worth noting that the above analysis does not apply to self-referenced schemes, like those employing self-homodyne or self-heterodyne detection, in which the same laser is used as the probe and the local oscillator. This case is discussed in appendix \ref{AppendixC}.}

\section{Polarization-based sensing} \label{Pbs}

Now, let us explore the potential of the use of polarization for sensing. The prominent advantage of polarization compared to phase is that polarization is unaffected by laser phase noise. This characteristic makes polarization the preferred choice for detecting environmental processes with very low frequencies. { Of course, the use of polarization has also many disadvantages, which include the non-deterministic dependence of the probe signal on the perturbation and a problematic localization capability.} Let us now analyze how polarization can be used for sensing with the help of the theory established in the previous sections.

The solution of (\ref{dsdz1}) is the concatenation of infinitesimal rotations around the axes $\vec \beta(z) \df z$. For sensing, we are interested into the change of the state of polarization induced by small time-dependent changes $\Delta \vec \beta(z,t)$ of $\vec \beta$
%
% \be \frac{\df \vec s}{\df z} = [\vec \beta(z) + \Delta \vec \beta(z,t)] \times \vec s, \ee
%
%
\be \frac{\df \mathbf{U}(z,t)}{\df z} = \frac{i}{2} \, [\vec \beta(z) + \Delta \vec \beta(z,t)] \cdot \vec \sigma \, \mathbf{U}(z,t). \label{dUdz} \ee
Let us now use an approach is similar to the interaction picture in quantum mechanics, separating the ``free'' static evolution from the ``interaction'' time-dependent part.\cite{sakurai:17} To this purpose, let us represent $\mathbf U(z,t)$ as the concatenation of two unitary matrices
\be \mathbf U(z,t) = \mathbf U_0(z) \mathbf U_1(z,t). \label{U0} \ee
The matrix $\mathbf U(z,t)$ is solution of Eq. (\ref{dUdz}) if $\mathbf U_0(z,t)$ and $\mathbf U_1(z,t)$ satisfy the equations
\be \frac{\df \mathbf{U}_0(z)}{\df z} = \frac{i}{2} \, \vec \beta(z) \cdot \vec \sigma \, \mathbf{U}_0(z), \ee
\be \frac{\df \mathbf{U}_1(z,t)}{\df z} = \frac{i}{2} \, \left\{\mathbf{U}_0^{-1}(z) \left[\Delta \vec \beta(z,t) \cdot \vec \sigma \right] \, \mathbf{U}_0(z) \right\} \, \mathbf{U}_1(z,t). \label{U11} \ee
Equation (\ref{U11}) is equivalent to
\be \frac{\df \mathbf{U}_1(z,t)}{\df z} = \frac{i}{2} \, \left\{ \left[\mathbf{R}_0^{-1}(z) \Delta \vec \beta(z,t)\right] \cdot \vec \sigma \right\} \, \mathbf{U}_1(z,t), \ee
where $\mathbf{R}_0(z)$ is the rotation operator in Stokes space corresponding to the unitary operator $\mathbf{U}_0(z)$ in Jones space by the relation $\mathbf{R}_0(z) \vec \sigma = \mathbf{U}_0^{-1}(z) \vec \sigma \mathbf{U}_0(z)$. By doing so, we employ a frame that rotates with the static birefringence $\vec \beta(z)$, effectively eliminating the static, $z$-dependent rotations induced by $\vec{\beta}(z)$.\cite{Gordon:05} In this reference frame, the state of polarization becomes in terms of the original one $\vec s' = \mathbf{R}_0^{-1}(z) \vec s$, and the evolution of the polarization vector $\vec s'$ is described by the equation
\be \frac{\df \vec s'}{\df z} =  \Delta \vec \beta' (z,t) \times \vec s', \label{Delta_s1} \ee
where 
\be \Delta \vec \beta' (z,t) = \mathbf{R}_0^{-1}(z) \Delta \vec \beta (z,t). \label{Deltabetaprime} \ee
This equation shows that when perturbations are absent and therefore $\Delta \vec \beta'(z,t) \equiv 0$, we have $\df \vec s' /\df z \equiv 0$ and hence $\vec s' \equiv \vec s_0$, where $\vec s_0$ is the input Stokes vector (which is identical in the rotating frame and in the original frame). If the fluctuations of the birefringence are sufficiently small such that their impact on $\vec s'$ is linear -- a prerequisite for ensuring linearity of the sensing mechanism -- a perturbative approach can be applied around the unperturbed solution $\vec s' = \vec s_0$. This involves setting $\vec s(z,t) = \vec s_0 + \Delta \vec s(z,t)$ with $\Delta \vec s(z,t)$ of the same order of $\Delta \vec \beta' (z,t)$, leading to
\be \frac{\df \Delta \vec s'}{\df z} =  \Delta \vec \beta'(z,t) \times \vec s_0. \ee
Integration of the above gives
\be \Delta \vec s'(z,t) = \int_0^z \Delta \vec \beta'(z',t) \df z' \times \vec s_0, \label{Delta_s2}\ee
where $\Delta \vec s'(z,t) = \vec s'(z,t) - \vec s_0$ is the deviation of the rotated state of polarization from the static position. { This is the main quantity of interest for polarization sensing because it is proportional to the integrated perturbations of the birefringence, which are in turn proportional to the external perturbations.} 

Let us now use the general relations that we have just derived first considering the case in which the perturbation is caused by strain. In this case we have
\be \Delta \vec s'(z,t) = \xi \int_0^z \epsilon(z',t)  \vec \beta_\perp'(z') \df z', \label{Delta_s} \ee
where we have defined  $\Delta \vec \beta_\perp'(z) = \Delta \vec \beta'(z,t) \times \vec s_0$, the component of $\Delta \vec \beta'(z,t)$ perpendicular to $\vec s_0$ and rotated around $\vec s_0$ by 90$^\circ$. Similarly to phase deviations, the deviation of the state of polarization from its steady-state value is proportional to $\epsilon(z',t)$, ensuring the linearity of the sensing probe. However, the behavior differs when it comes to signal accumulation. Unlike the deterministic nature of the polarization-averaged birefringence vector, which is uniform in space in the absence of perturbations, in standard single mode fibers, the preexisting birefringence varies randomly over a scale of tens of meters. Fortunately, however, a simplification arises from the fact that external perturbations generally occur on orders of magnitude larger scales, making it appropriate to perform statistical averaging over the short-scale variation of the birefringence. As we will demonstrate below, this results in the accumulation of the deviation of the state of polarization along the fiber being incoherent, contrasting with the coherent accumulation of phase deviations.

A realistic birefringence correlation function is
\be \langle \vec \beta(z) \cdot \vec \beta(z') \rangle = \langle \beta^2 \rangle \exp(-|z-z'|/L_\mathrm{f}), \ee
where $L_\mathrm{f}$, the birefringence correlation length, is of the order of meters.\cite{Galtarossa:01} Assuming that the scale of variation of the external perturbation is much longer, it is appropriate to replace right-hand side of the above with a Dirac delta function of equal area
\be \langle \vec \beta(z) \cdot \vec \beta(z') \rangle = 2 L_\mathrm{f} \langle \beta^2 \rangle \delta(z-z'). \label{deltaapp} \ee
The fiber birefringence $\vec \beta(z)$ represents linear birefringence and therefore belongs to the equatorial plane of the Stokes space. Conversely, in the rotated reference frame the rotated birefringence $\vec \beta'(z) = \mathbf{R}_0^{-1}(z) \vec \beta(z)$ is instead isotropically distributed because $\mathbf{R}_0^{-1}(z)$ it is the concatenation of rotations with axes $\vec \beta(z)$ that vary over a length scale of few meters, so that its components are uncorrelated to each other and each one has a variance one third of the total. Consequently we have
\be \langle \vec \beta_\perp' (z) \cdot \vec \beta_\perp' (z') \rangle = \frac 2 3 \langle \vec \beta(z) \cdot \vec \beta(z') \rangle. \ee
Using now Eq. (\ref{Delta_s}) we obtain 
\be \langle \Delta \vec s(z,t) \cdot \Delta \vec s(z,t') \rangle = \frac 4 3 \langle \beta^2 \rangle L_\mathrm{f} \xi^2 \int_0^z \epsilon(z',t)  \epsilon(z',t') \, \df z'. \ee 
The contribution of a fiber section to the fluctuations of the polarization are proportional to the strength of the local static birefringence, which is a quantity well characterized in optical fibers because the fiber's polarization mode dispersion depends on it. The fiber polarization mode dispersion is related to $\langle \beta^2 \rangle L_\mathrm{f}$ by
\be \langle \tau^2 \rangle = \frac {1}{\omega_0^2} \, 2 L_\mathrm{f} \langle \beta^2 \rangle z, \ee
with $\langle \tau \rangle^2 = 8 \langle \tau^2 \rangle/(3 \pi)$ the mean polarization mode dispersion square.\cite{Galtarossa:01} If we define as $\kappa^2 = \langle \tau \rangle^2/z$ the averaged square polarization mode dispersion of the fiber in ps/$\sqrt{\mathrm{km}}$, we may eliminate $2 L_\mathrm{f} \langle \beta^2 \rangle$ in the correlation functions by using
\be 2 L_\mathrm{f} \langle \beta^2 \rangle = \frac{3 \pi \omega_0^2 \kappa^2} 8. \label{beta2} \ee
After using $\omega_0 = 2 \pi c/\lambda$, we obtain
\be \langle \Delta \vec s(z,t) \cdot \Delta \vec s(z,t') \rangle = \frac{\pi}{4} \left(\frac{2 \pi c}{\lambda} \right)^2 \kappa^2 \xi^2 \int_0^z  \epsilon(z',t)  \epsilon(z',t') \, \df z', \label{corr} \ee

Let us now analyze twist. Combining Eqs. (\ref{Deltabetaalpha}), (\ref{Deltabetaprime}) and (\ref{Delta_s2}), yields
\be \Delta \vec s'(z,t) = \int_0^z 2 \, \alpha(z',t)  \mathbf{R}_0^{-1}(z') \left[\hat e_3 \times \vec \beta(z') \right] \df z' \times \vec s_0. \ee
The outcome of $\hat e_3 \times \vec \beta(z)$ yields a vector with the same magnitude as $\vec \beta(z)$ but rotated by 90$^\circ$ in the equatorial plane. Considering that $\vec \beta(z)$ lies within the equatorial plane and is distributed isotropically, the result of this rotation is statistically equivalent to $\vec \beta(z)$. Consequently, we can substitute $\mathbf{R}_0^{-1}(z') [\hat e_3 \times \vec \beta(z') ]$ with $\vec \beta'(z)$ and proceed along the same route of the analysis for strain. The final result is
\be \langle \Delta \vec s(z,t) \cdot \Delta \vec s(z,t') \rangle = \frac 4 3 \langle \beta^2 \rangle L_\mathrm{f}\int_0^z \, 4 \, \alpha(z',t)\alpha(z',t' ) \, \df z', \ee
and, using Eq. (\ref{beta2})
\be \langle \Delta \vec s(z,t) \cdot \Delta \vec s(z,t') \rangle = \frac{\pi}{4} \left(\frac{2 \pi c}{\lambda} \right)^2 \kappa^2 \int_0^z \, 4 \, \alpha(z',t)\alpha(z',t' ) \, \df z'. \ee
As discussed in Ref. \onlinecite{Mecozzi:21}, strain is most likely the predominant source of perturbation in submarine systems employing jelly-filled cables, while twist has a dominant role in aerial cables exposed to wind.\cite{Wuttke:03}

It is important to note that, in both cases of strain and twist, the temporal correlation functions of the deviations of the state of polarization are proportional to the temporal correlation functions of the perturbations, integrated over the entire link length. This property insures that the spectrum of the fluctuations of the state of polarization faithfully reproduces the spectrum of the integrated strain and twist. Additionally, in both cases, the sensitivity is proportional to the polarization mode dispersion coefficient of the fiber.

With polarization, the spectra are obtained in terms of ensemble averages. It is well established\cite{Shtaif:00} that polarization fluctuations exhibit ergodic behavior in frequency, meaning that ensemble averages can be effectively replaced by averages over frequency, if taken over a bandwidth containing a sufficient number of principal state of polarization bandwidths. In this particular scenario, however, measurement data over a sufficiently large frequency span are not easily accessible, but fortunately, they are also unnecessary. This is because the length scale of the environmental perturbations, of the order of tens of kilometers, is much greater than the length scale of the spatial variations of the birefringence, of the order of tens of meters. Consequently, significant self averaging of the birefringence fluctuations occurs over the length scale of the environmental perturbations. This issue will be further discussed and the effectiveness of self averaging validated against experimental data in Sec. \ref{exp}, where we analyze a few cases of earthquakes and sea swell detection.

\section{Comparison between phase and polarization accumulation} \label{Comp}

Let us now compare the signal accumulation between polarization and phase deviations, limiting ourselves to the case in which the perturbation is due to strain because twist is uncoupled to phase. If we compare Eq. (\ref{corr}) and that obtained from Eq. (\ref{phase})
\bea \Delta \varphi(z,t)\Delta \varphi(z,t') &=& \left(\frac{2 \pi}{\lambda}\right)^2 n_0^2 \xi^2 \nonumber \\
&& \int_0^z \int_0^z \epsilon(z',t) \epsilon(z'',t') \df z' \df z'', \eea
we notice that, although both $\langle \Delta \vec s(z,t) \cdot \Delta \vec s(z,t') \rangle$ and $\Delta \varphi(z,t)\Delta \varphi(z,t')$ are proportional to the \textit{temporal} correlation function of the strain, the accumulation of the strain contributions along the fiber is different in the two cases. For polarization, sections with positive and negative strain give the same contribution to signal strength, because $\langle |\Delta \vec s(z,t)|^2 \rangle$ depends only on $\epsilon(z,t)^2$. On the contrary, Eq. (\ref{phase}) reveals that sections subjected to positive strain yield a positive contribution to the phase deviations, while sections experiencing negative strain contribute negatively, thus partially offsetting each other's effects.

To get an order of magnitude estimate of the effect that averaging produces on phase measurements, let us assume a seismic wave of wavelength $\Lambda$ whose amplitude is modulated by the envelope $\epsilon_0(z,t)$, namely $\epsilon(z,t) = \sin(2\pi z/\Lambda)\epsilon_0(z,t)$. With this assumption, Eq. (\ref{phase}) becomes
\be \Delta \varphi(z,t) = \frac{2 \pi}{\lambda} n_0 \int_0^z \xi \cos(2\pi z/\Lambda) \epsilon_0(z',t) \df z', \ee
that is
\be \Delta \varphi(z,t) = \frac{2 \pi}{\lambda} n_0 \, \xi \, \tilde \epsilon_0(2 \pi/\Lambda, t), \ee
where
\be \tilde \epsilon_0(K, t) = \frac 1 2 \operatorname{Re} \left[\int_{-\infty}^\infty \exp(-i K z) \epsilon_0(z',t) \df z'\right], \ee
is the spatial Fourier transform of the strain perturbation (which we assume zero outside the fiber length) calculated at the spatial wavevector $K=2\pi z/\Lambda$. Assuming for the envelope of the perturbation $\epsilon_0(z',t)$ a Gaussian distribution of root mean square $L_0$ entirely contained into the fiber length
\be \epsilon_0(z,t) = \epsilon_0 \exp\left(- \frac{z^2}{2 L_0^2}\right), \ee
we obtain
\be \tilde \epsilon_0(K, t) = \frac{\sqrt{2 \pi}}{2} L_0 \epsilon_0 \exp\left(- \frac{2 \pi^2 L_0^2}{\Lambda^2}\right). \ee
The wavelength of seismic waves $\Lambda_0$ is of the order of 100 km, but if we assume a plane seismic wave incident on the local direction of the cable with an angle $\vartheta$, the spatial periodicity is $\Lambda = \Lambda_0/\cos \vartheta$, so it is in general larger than $\Lambda_0$ and equal only if the wavevector of the seismic wave is parallel to the direction of the cable. With this simplified assumption we obtain
\be \Delta \varphi(z,t) = \frac{2 \pi}{\lambda} n_0 \, \xi \, \frac{\sqrt{2 \pi}}{2} L_0 \epsilon_0 \exp\left(- \cos^2 \vartheta \frac{2 \pi^2 L_0^2}{\Lambda_0^2}\right). \ee
{ In the above expression, the right-hand side is independent of the link length $z$. This is a consequence of the spatial averaging and of the fact that we assumed that the perturbation is all contained within the fiber length. Notice the dependence on $\cos^2 \vartheta$ at the exponent, suggesting a larger sensitivity for seismic wave approximately orthogonal to the cable, consistent with experimental observations detailed in Ref. \onlinecite{Donadello:24}.} For comparison, with polarization
\be \langle |\Delta \vec s(z,t)|^2 \rangle^{1/2} = \frac{\sqrt{\pi}}{2} \frac{2 \pi c}{\lambda} \kappa \xi  \left[\int_0^z \epsilon(z,t)^2 \, \df z'\right]^{1/2}, \ee 
that is, with the Gaussian assumption and assuming $\Lambda \ll L_0$ so that we may replace the cosine square with its average one half,
\be \langle |\Delta \vec s(z,t)|^2 \rangle^{1/2} = \frac{\sqrt{\pi}}{2} \frac{2 \pi c}{\lambda} \kappa \xi  \epsilon_0 \left(\frac{\sqrt{2 \pi} L_0}{4}  \right)^{1/2}. \ee 
Comparing the two signals, we have
\bea \Delta \varphi(z,t) &=& \left(\frac{32}{\pi} \right)^{1/4} \frac{n_0 \sqrt{L_0}}{\kappa \, c}\exp\left(- \cos^2 \vartheta \frac{2 \pi^2 L_0^2}{\Lambda_0^2}\right) \nonumber \\ && \langle |\Delta \vec s(z,t)|^2 \rangle^{1/2}.\eea
Introducing the full-width at half maximum of a Gaussian $L_\mathrm{f} = 2 \sqrt{2 \ln(2)} L_0$
\bea \Delta \varphi(z,t) &=& 4 \left(\frac{\ln 2}{\pi} \right)^{1/4} \frac{n_0 \sqrt{L_\mathrm{f}}}{\kappa \, c} \exp\left(- \cos^2 \vartheta \frac{\pi^2 \, L_\mathrm{f}^2}{4 \Lambda_0^2 \ln 2 }\right) \nonumber \\ && \langle |\Delta \vec s(z,t)|^2 \rangle^{1/2}.\eea
Assuming $\kappa = 0.03$ ps/$\sqrt{\mathrm{km}}$ and $n_0 = 1.5$, we obtain for $L_\mathrm{f} = 2 \Lambda_0/\cos \vartheta$ and $L_f = 400$ km,  $\Delta \varphi(z,t) \simeq 6000 \, \langle |\Delta \vec s(z,t)|^2 \rangle^{1/2}$ rad.  When the ratio $L_\mathrm{f}/(\Lambda/\cos \vartheta)$ becomes larger, the efficiency of the phase modulation drops very rapidly as a consequence of the Gaussian profile and the fact that we assumed $\Lambda$ as a constant, so that the expression may become meaningless in this limit.

The root mean square deviations of the signal polarization being much smaller than the amplitude of the phase deviations ensures that the polarization deviations fall within the linear range even for very large external perturbations, like those applied by earthquakes of high magnitudes, as we will show in Sec. \ref{exp}.

\section{The Jones matrix in the rotating frame} \label{Rot}

In the rotating frame in which the static birefringence is removed, the Jones matrix can be represented, similarly to Eq. (\ref{Uz}), as
\be \mathbf{U}_1(t) = \exp\left[\frac i 2 \Delta \vec \varphi'(t) \cdot \vec \sigma \right]. \label{U1}\ee
In the same frame, the output Stokes vector is
\be \vec s'(t) = \langle \vec s_0|\mathbf{U}_1^{\dagger}(t) \, \vec \sigma \, \mathbf{U}_1(t)| \vec s_0 \rangle = \exp \left[ \Delta \vec \varphi'(t) \times \right] \vec s_0. \label{s0U}\ee
Being in the cases of interest for sensing $|\Delta \vec \varphi'(t)| \ll \pi$, because otherwise the probe does not depend linearly on the perturbations, we may expand the exponential in the last member of Eq. (\ref{s0U}) to first order obtaining for $\Delta \vec s(t) = \vec s(t)-\vec s_0$
\be \Delta \vec s'(t) = \Delta \vec \varphi(t) \times \vec s_0. \label{deltasp} \ee
If we use the expression of $\Delta \vec s'(t)$ given by Eq. (\ref{Delta_s2}) we get\cite{Mecozzi:23}
\be \Delta \vec \varphi(t) = \int_0^z \Delta \vec \beta'(z',t) \df z'. \ee
Comparison of Eqs. (\ref{U1}) with Eq. (\ref{deltasp}) reveals that the detection of the fluctuations of the output polarization from the average value, $\Delta \vec s'(t)$, with a fixed input enables the characterization of the fluctuations of two out of the three parameters that identify the Jones matrix of the link, the two components of $\Delta \vec \varphi(t)$ orthogonal to $\vec s_0$.\cite{Mecozzi:23} 

The three components of $\Delta \vec \varphi(t)$ and hence a complete characterization of the fluctuations of the rotation vector can be obtained from $\mathbf U_1(t)$ with the following procedure. A coherent receiver reconstructs the Jones matrix in the original frame, $\mathbf U(t)$. Using Eq. (\ref{U0}), we find that $\mathbf U_1(t)$ is related to $\mathbf U(t)$ by the following equation
\be \mathbf U_1(t) = \mathbf U_0^{-1} \mathbf U(t). \label{U1_1} \ee
In practical terms, the Jones matrix in the rotating frame can be extracted by left-multiplying the Jones matrix directly obtained from the receiver, $\mathbf U(t)$, by the inverse of the ``static'' Jones matrix, $\mathbf U_0$.\cite{Mecozzi:23} The latter is obtained by averaging the Jones matrix $\mathbf U(t)$ over a sufficiently long time window. The duration of the averaging time window sets the lower limit on the bandwidth of $\Delta \vec \varphi(t)$. Once $\mathbf U_1(t)$ is extracted, we obtain
\be \Delta \vec \varphi(t) = \operatorname{trace}\left\{-i \log \left[\mathbf U_1(t)\right] \vec \sigma\right\}. \ee
Whether the polarization state corresponding to a fixed input is measured, or the Jones matrix is reconstructed from the receiver, to achieve a linear dependency of the measured quantity on the applied strain both approaches require employing of a frame that rotates with the static birefringence, and both approaches yield identical results. When utilizing a fixed polarization at the input, as done in refs. \onlinecite{Mecozzi:21,Zhan:21}, one extracts the two components of $\int_0^z \df z' \Delta \vec{\beta}'(z',t)$ orthogonal to the input polarization $\vec{s}_0$.\cite{Mecozzi:23} Conversely, knowledge of the full Jones matrix provides access to all three components of $\int_0^z \df z' \Delta \vec{\beta}'(z',t)$.

%To summarize, the characterization of the full Jones matrix allows the calculation of the output polarization with any polarization at input, hence its characterization permits the extraction of the Jones matrix in the rotating frame and consequently the determination of $\Delta \vec \varphi(t)$, which as we have shown, is the quantity linearly dependent on the applied perturbation. Alternatively, one can compute using the Jones matrix the output Stokes vector corresponding to a given polarization at input, and compute the deviation of the state of polarization from the static output caused by the perturbation. This is equivalent to the use of the deviation from the North pole of the Poincar\'e sphere of the first two components of the Stokes space once the Stokes space is rotated such that the time averaged polarization is set to the North pole. On the contrary, partial information on the Jones matrix, for instance the modulus of their eigenvalues \cite{Costa:23}, does not allow in general the extraction of quantities linearly related to the strength of the perturbation. 

\section{Localization with polarization} \label{Loc}

The integral in $\Delta \vec \varphi(t)$ can be readily obtained in transmission experiments \cite{Mecozzi:21,Mecozzi:23}. However, in such experiments, only the perturbation accumulated over the entire link can be extracted. Below, we outline a procedure demonstrating that in experiments utilizing time-resolved backscattering, such as in distributed acoustic sensing, or in experiments employing high-loss loopbacks as in Ref. \onlinecite{Mazur:22a,Mazur:23,Yaman:23,Costa:23}, it is possible to extract one of the three components of $\Delta \vec \varphi(t)$ specific to a section located at any position along the link.

Assume that $\mathbf U_\mathrm{f}$ is the Jones matrix that describes the evolution of the polarization of a single mode fiber up to a given distance $z$ and $\Delta \mathbf U$ the unitary matrix describing the evolution in a section that goes from $z$ to $z+\Delta z$ along the same fiber, which we will refer to as the section of interest in the following. Then, the backscattered field is either rerouted through a different fiber with Jones matrix $\mathbf U_\mathrm{b}$ or transmitted back through the same fiber, in which case $\mathbf U_\mathrm{b} = \mathbf U_\mathrm{f}$. The Jones matrix $\mathbf U_\mathrm{rt}(1)$ describing the round-trip propagation from $0$ to $z$ and back, and the Jones matrix $\mathbf U_\mathrm{rt}(2)$ describing the round-trip propagation from $0$ to $z+\Delta z$ and back, are\cite{Galtarossa:08}
\be \mathbf U_\mathrm{rt}(1) = \mathbf U_\mathrm{b}^T \mathbf U_\mathrm{f}, \ee
\be \mathbf U_\mathrm{rt}(2) = \mathbf U_\mathrm{b}^T \left(\Delta \mathbf U^T \Delta \mathbf U\right) \, \mathbf U_\mathrm{f}, \ee
where the superscript $T$ stands for transpose. Our aim is to characterize the unitary matrix 
\be \Delta \mathbf U_\mathrm{rt} = \Delta \mathbf U^T \Delta \mathbf U, \ee
from a measurement of $\mathbf U_\mathrm{rt}(1)$ and $\mathbf U_\mathrm{rt}(2)$. Left multiplying by $\mathbf U_\mathrm{rt}^{-1}(1)$ both sides of the above, we obtain
\be \mathbf U_\mathrm{meas}  = \mathbf U_\mathrm{f}^{-1}  \Delta \mathbf U_\mathrm{rt} \, \mathbf{U}_\mathrm{f}, \label{Um1U2} \ee
where we have defined the unitary matrix
\be \mathbf U_\mathrm{meas} = \mathbf U_\mathrm{rt}(1)^{-1} \mathbf{U}_\mathrm{rt}(2). \ee
From now on, we will assume that $\mathbf U_\mathrm{meas}$ has been characterized experimentally and is known. Let us now define
\be \mathbf U_\mathrm{meas} = \exp\left(i \vec \varphi_\mathrm{meas}/2 \cdot \vec \sigma \right), \label{Umeas} \ee
\be \mathbf{U}_\mathrm{f} = \exp\left(i \vec \varphi_\mathrm{f}/2 \cdot \vec \sigma \right), \ee
and
\be  \Delta \mathbf U_\mathrm{rt} = \exp\left[ (\Delta \vec \varphi_{0}+\Delta \vec \varphi_{1})/2 \cdot \vec \sigma\right), \ee
where $\Delta \vec \varphi_{0}$ is the contribution of the static birefringence in the section of interest and $\Delta \vec \varphi_{1}$ of the perturbation.  Entering the above definitions into Eq. (\ref{Um1U2}) we obtain
\be \vec \varphi_\mathrm{meas} = \exp(-\vec \varphi_\mathrm{f} \times) (\Delta \vec \varphi_{0}+\Delta \vec \varphi_{1}). \label{varphimeas} \ee
The vector $\vec \varphi_\mathrm{meas}$ can be extracted from the measurement data using
\be \vec \varphi_\mathrm{meas} = \operatorname{trace}\left[-i \log (\mathbf U_\mathrm{meas}) \vec \sigma \right]. \ee
Taking the modulus square of both terms of (\ref{varphimeas}) yields
\be |\vec \varphi_\mathrm{meas}|^2 =|\Delta \vec \varphi_{0}+\Delta \vec \varphi_{1}|^2 = |\Delta \vec \varphi_{0}|^2 + |\Delta \vec \varphi_{1}|^2 + 2 \Delta \vec \varphi_{0} \cdot \Delta \vec \varphi_{1}. \ee
In this equation $\Delta \vec \varphi_{0}$ is time independent whereas $\Delta \vec \varphi_{1}$ is time-dependent. The technique used in Ref. \onlinecite{Costa:23,Yaman:23} was based on the analysis of the temporal modulation of $|\vec \varphi_\mathrm{meas}|^2$ or its square root. In both cases, this approach fails to return the spectrum of the perturbation because the term $|\Delta \vec \varphi_{1}|^2$ is nonlinear in the perturbation and is generally not negligible compared to the linear term $\Delta \vec \varphi_{0} \cdot \Delta \vec \varphi_{1}$. This is because, although the birefringence is much larger than its fluctuations, being $\Delta \varphi_{0} \leq \pi$ for the periodicity of rotations, the vectors $\Delta \vec \varphi_{0}$ and $\Delta \vec \varphi_{1}$ have comparable magnitudes. Consequently, the temporal variations of the length of the rotation vector are generally not proportional to any of the components of $\Delta \vec \varphi_{1}$, which are the quantities of interest because related to the fiber strain in the section of interest. However, we will show in the remainder of this section that further manipulations of the above equations enable the establishment of a procedure for extracting one of the three components of $\Delta \vec{\varphi}_1$.

If we insert into Eq. (\ref{Um1U2}) the decomposition $\mathbf{U}_\mathrm{f} = \mathbf{U}_{0,\mathrm{f}} \mathbf{U}_{1,\mathrm{f}}$, where $\mathbf{U}_{0,\mathrm{f}}$ is the static contribution and $\mathbf{U}_{1,\mathrm{f}}$ the contribution of the perturbations of the forward propagation, Eq. (\ref{Um1U2}) becomes
\be \mathbf U_\mathrm{meas}  = \mathbf U_{1,\mathrm{f}}^{-1}  \Delta \mathbf U_\mathrm{rt}' \mathbf{U}_{1,\mathrm{f}}, \label{U12} \ee
where
\be \Delta \mathbf U_\mathrm{rt}' = \mathbf U_{0,\mathrm{f}}^{-1} \Delta \mathbf U_\mathrm{rt} \, \mathbf{U}_{0,\mathrm{f}} \ee
is the Jones matrix of the roundtrip propagation through the section under test rotated by the \textit{static} birefringence of the forward propagation. Let us define, using the prime for the quantities rotated by the static birefringence, 
\be \Delta \mathbf U_\mathrm{rt}' = \exp\left[i (\Delta \vec \varphi_{0}' + \Delta \vec \varphi_{1}')/2 \cdot \vec \sigma \right], \ee
%
%
%\be \Delta \mathbf U^T \Delta \mathbf U = \exp\left[i (\Delta \vec \varphi_{0} + \Delta \vec \varphi_{1})/2 \cdot \vec \sigma \right], \ee
%
where $\Delta \vec \varphi_{0}'$ is the contribution of the static birefringence in the section under test and $\Delta \vec \varphi_{1}'$ of the perturbations, rotated by the static birefringence of the forward propagation, and represent $\mathbf{U}_{1,\mathrm{f}}$ as
\be \mathbf{U}_{1,\mathrm{f}} = \exp\left(i \vec \varphi_{1,\mathrm{f}}'/2 \cdot \vec \sigma \right). \ee
If we now use Eq. (\ref{U12}), we may express the matrix $\mathbf U_\mathrm{meas}$ as\cite{Gordon:00}
%
%\be \Delta \mathbf U_\mathrm{rt} = \exp\left[i (\Delta \vec \varphi_{0} + \Delta \vec \varphi_{1})/2 \cdot \vec \sigma \right], \ee
%
% The matrix $\Delta \mathbf U_\mathrm{rt}$ can be expressed as\cite{Gordon:00}
%
\be \mathbf U_\mathrm{meas} = \exp\left[i \mathbf R_{1,\mathrm{f}} (\Delta \vec \varphi_{0}' + \Delta \vec \varphi_{1}')/2 \cdot \vec \sigma \right], \label{Umeas1} \ee
where
\be \mathbf R_{1,\mathrm{f}} = \exp\left(-\vec \varphi_{1,\mathrm{f}} \times \right) \simeq \mathds 1 - \vec \varphi_{1,\mathrm{f}} \times \label{R1} \ee
where we used that $\vec \varphi_{1,\mathrm{f}} \ll \pi$ because it is produced by the small perturbations in the forward propagation. Comparing Eq. (\ref{Umeas}) with Eq. (\ref{Umeas1}) we obtain
\be \vec \varphi_\mathrm{meas} = \Delta \vec \varphi_{0}' + \Delta \vec \varphi_{1}' - \vec \varphi_{1,\mathrm{f}} \times\Delta \vec \varphi_{0}' - \vec \varphi_{1,\mathrm{f}} \times \Delta \vec \varphi_{1}'. \ee
The terms $\Delta \vec \varphi_{1}'$ and $\vec \varphi_{1,\mathrm{f}}$ are small time-dependent perturbations in the section of interest and in the forward propagation. If we average them over a sufficiently long time interval these terms vanish, so that we have
\be \mathds{E} \left( \vec \varphi_\mathrm{meas} \right) = \Delta \vec \varphi_{0}'. \label{deltaphi0} \ee
If we now define
\be \Delta \vec \varphi_\mathrm{meas} = \vec \varphi_\mathrm{meas} - \mathds{E} \left( \vec \varphi_\mathrm{meas} \right), \ee
we have 
\be \Delta \vec \varphi_\mathrm{meas} = \Delta \vec \varphi_{1}' - \vec \varphi_{1,\mathrm{f}} \times\Delta \vec \varphi_{0}' - \vec \varphi_{1,\mathrm{f}} \times \Delta \vec \varphi_{1}'. \ee
Being the term $\vec \varphi_{1,\mathrm{f}}' \times \Delta \vec \varphi_1'$ the product of perturbation terms which we may assume much smaller than one, it can be neglected with respect to linear terms, so that the equation above, solved for $\Delta \vec \varphi_1'$, gives
\be \Delta \vec \varphi_1' \simeq  \Delta \vec \varphi_\mathrm{meas} - \vec \varphi_{1,\mathrm{f}}' \times \Delta \vec \varphi_0'. \label{deltaphimeas}\ee
In Eq. (\ref{deltaphimeas}), $\Delta \vec \varphi_0'$ is the known, time independent, rotation vector given by Eq. (\ref{deltaphi0}), accounting for the effect of the static birefringence in the round-trip through the section under test, $\vec \varphi_{1,\mathrm{f}}'$ is the (unknown) time-dependent result of the birefringence perturbations in the forward propagation up to $z$, and
\be \Delta \vec \varphi_1' = \int_z^{z+\Delta z} \Delta \vec \beta_f'(z',t) \df z' + \int_{z+\Delta z}^z \Delta \vec \beta_b'(z',t) \df z' \ee
is the time-dependent birefrigence accumulated over the roudtrip from $z$ to $z+\Delta z$ rotated by the static birefringence of the forward propagation up to $z$. 

Equation (\ref{deltaphimeas}) states that the three components of $\Delta \vec \varphi_1'$ are equal to the known vector $\Delta \vec \varphi_\mathrm{meas}$ corrupted by an extra term involving the unknown time-dependent vector $\vec \varphi_{1,\mathrm{f}}'$. One of the three components of $\Delta \vec \varphi_1'$ is however unaffected by the extra term. Specifically, if we project $\Delta \vec \varphi_\mathrm{meas}$ over the direction parallel to $\hat e_0 = \Delta \vec \varphi_0'/\Delta \varphi_0'$, we obtain
\be \Delta \vec \varphi_1' \cdot \hat e_0 = \Delta \vec \varphi_\mathrm{meas} \cdot \hat e_0. \ee
Projection over the direction of $\Delta \vec \varphi_0'$ allows the separation of the time-dependent fluctuations of the polarization caused by the birefringence perturbations in the section of interest from those originating from the light propagation before the section of interest. Although only a single component of the integrated birefringence fluctuations can be extracted by this procedure, this is sufficient to derive the strain perturbation spectrum due to the isotropic nature of $\Delta \vec \varphi_1'$. The isotropy of the polarization fluctuations will be corroborated through the analysis of experimental data in the following section.

As a final comment, we notice that the procedure that we have just described is equivalent to subtract the steady state value of the rotation matrix by a procedure similar to that described and experimentally validated in Ref. \onlinecite{Mecozzi:23}. That approach, in turn, was equivalent to the rotation used in  Ref. \onlinecite{Mecozzi:21} to align the average of the output polarization with the north pole of the Poincar\'e sphere. This alignment was crucial for establishing a correspondence between the spectrum of polarization fluctuations and the spectra of earthquakes and microseisms in Ref. \onlinecite{Zhan:21}. 

\section{Environmental sensing using the state of polarization} \label{exp}

In this section, we explore the potential of utilizing the state of polarization as a sensing probe for detecting earthquakes and sea swells in proximity to an optical cable. We will make extensive use of the database of Ref. \onlinecite{zhongwenzhan_2020}, demonstrating how the analysis of the correlation between measurements obtained from two wavelength division multiplexing channels offers valuable insights, potentially enabling a coarse localization of earthquake events.

Fibers embedded in loose tube gel-filled cables laid on the seafloor experience a static pressure of about $10^5$ Pa every 10 meters of depth. For a 4-kilometer depth (the average depth of the Curie cable\cite{Zhan:21}), the pressure is approximately $400 \times 10^5$ Pa, roughly 400 times atmospheric pressure. This static pressure induces strain due to the Poisson effect. During an oceanic earthquake, the movement of the seafloor causes the water column above the cable to oscillate. The inertia of this water column leads to fluctuations in the pressure acting on the cable around its static value. Consequently, these pressure fluctuations induce oscillations in strain, affecting both the polarization-averaged propagation constant and the birefringence of the fiber.

We will analyze data from two channels, denominated channel 1 and channel 2 in Ref. \onlinecite{zhongwenzhan_2020}, belonging to the Curie submarine cable system, connecting Los Angeles, California, to Valparaiso, Chile, for a total length of 10,500 km. In the measurement campaign, both channels were looped back in Valparaiso, so that the signals were transmitted and received in Los Angeles, covering a total round-trip distance of 21,000 km. Channel 1 operated at a frequency centered at 193.5805 THz, while channel 2 operated at 193.6570 THz, resulting in a frequency separation of $\Delta f = 76.5$ GHz. { These polarization data were extracted from the transceiver of the Curie submarine system while transmitting data. The operation of the Curie system, like that of the vast majority of long-haul coherent systems in use nowadays, requires that the receiver estimates the phase and polarization variations that an optical field undergoes when propagating through an optical fiber. The Curie receiver is designed to reconstruct two messages that the transmitter injects on the $x$ and $y$ linear polarizations of the fiber. To do so, it produces by various means an estimate of the Jones matrix of the fiber. Since the reconstruction of the polarization of an optical field transmitted over the $x$ or $y$ polarization does not require the recovery of the phase of the optical field, it is immune to the phase and frequency noise of the transmit and local oscillator lasers, and is not affected by their linewidth and frequency mismatch. The data in Ref. \onlinecite{zhongwenzhan_2020} were obtained using the Jones matrix, reconstructed by the receiver of the Curie system, to calculate the output polarization for an input linear polarization parallel to the $x$-axis. This approach emulates an experiment in which an $x$-polarized optical field is transmitted through the Curie cable, and its output polarization is measured. Being polarization immune to the phase noise of the laser, the noise on the output polarization calculated by this procedure is equal to the noise in an ideal polarization interferometer, even if the transmit and local oscillator lasers used in the Curie system had typical telecom-grade linewidths. For further details, the reader should refer to Ref. \onlinecite{Zhan:21} and the supplementary material therein. The use of coherent receivers for sensing is also discussed in Ref. \onlinecite{Mecozzi:23}.}

The database of Ref. \onlinecite{zhongwenzhan_2020} provides the three components of the Stokes vector rotated such that the Stokes vectors averaged over a 200 s time window coincide with the north pole of the Poincar\'e sphere. From now on, we will refer to these components as if they were those in the frame rotating with the static fiber birefringence. Although in the field experiment the input to the fiber was a fixed linear polarization,\cite{Mecozzi:23} hence a point on the equator of the sphere, this equivalence is justified by the fact that the constant rotation from the fixed point on the equator to the north pole does not impact the time-dependent deviations from the average polarization. It should be noted, however, that the rotation of the sphere that aligns the average polarization to a fixed point of the sphere does not specify the orientation of the sphere around the average point. The rotating frame is therefore undetermined for a rotation of the cloud of polarization points around the average. This indetermination does not affect accuracy when polarization dependent loss is negligible as assumed in our analysis, because in this case the fluctuations of the deviations of the Stokes vector from the average Stokes vector are isotropic. %Consequently, we also need to use the assumption of isotropy of the fluctuations of the deviations of the Stokes vector from the average Stokes vector, which is valid if polarization dependent loss is negligible.%, because the rotation leaves a rotation of the cloud of points around the north pole of the sphere undetermined.

\begin{figure}
\includegraphics[width=\columnwidth]{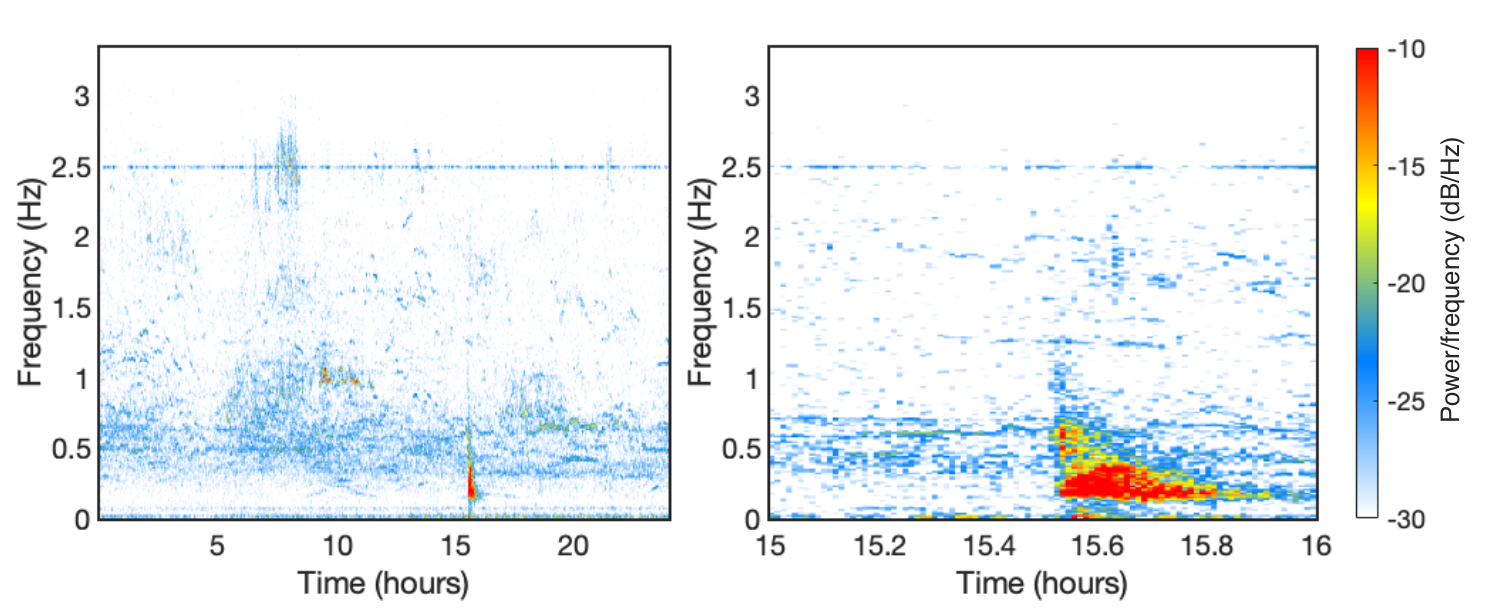}% Here is how to import EPS art
\caption{\label{fig1} Sum of the spectrograms of the two components of $\Delta \vec s$ orthogonal to $\vec s_0$ of channel 1, referring to an M7.3 earthquake that occurred in Oaxaca on June 23, 2020, at UTC time 15:29:05. The horizontal axis represents UTC time. The left panel depicts the entire day of the earthquake, while the right panel zooms in on a one-hour time window around the earthquake event.}
\end{figure}

Figure \ref{fig1} displays the sum of the spectrograms of the two components of $\Delta \vec s$ orthogonal to $\vec s_0$ of channel 1, concerning an M7.3 earthquake that took place in Oaxaca on June 23, 2020, at UTC time 15:29:05. The horizontal axis represents UTC time. The left panel illustrates the entire day of the earthquake, while the right panel zooms in on a one-hour time window around the earthquake occurrence.

%
%Figure \ref{fig2} shows the spectrograms of the two components of the deviations of the Stokes vector separately. It is worth to notice that the same spectral features are displayed by the two spectrograms, although the random nature of the birefringence acts independently on the two components, as confirmed by the fact that the magnitude of the cross-correlation between the two components never rises above 5\%. This is a clear indication that, as we anticipated at the end of section \ref{Pbs}, significant spatial self averaging occurs, and therefore the random nature of the static birefringece is not a limiting factor for the accuracy of the sensing approaches based on the analysis of the modulation of the state of polarization in a fiber with random birefringence.

\begin{figure}
\includegraphics[width=\columnwidth]{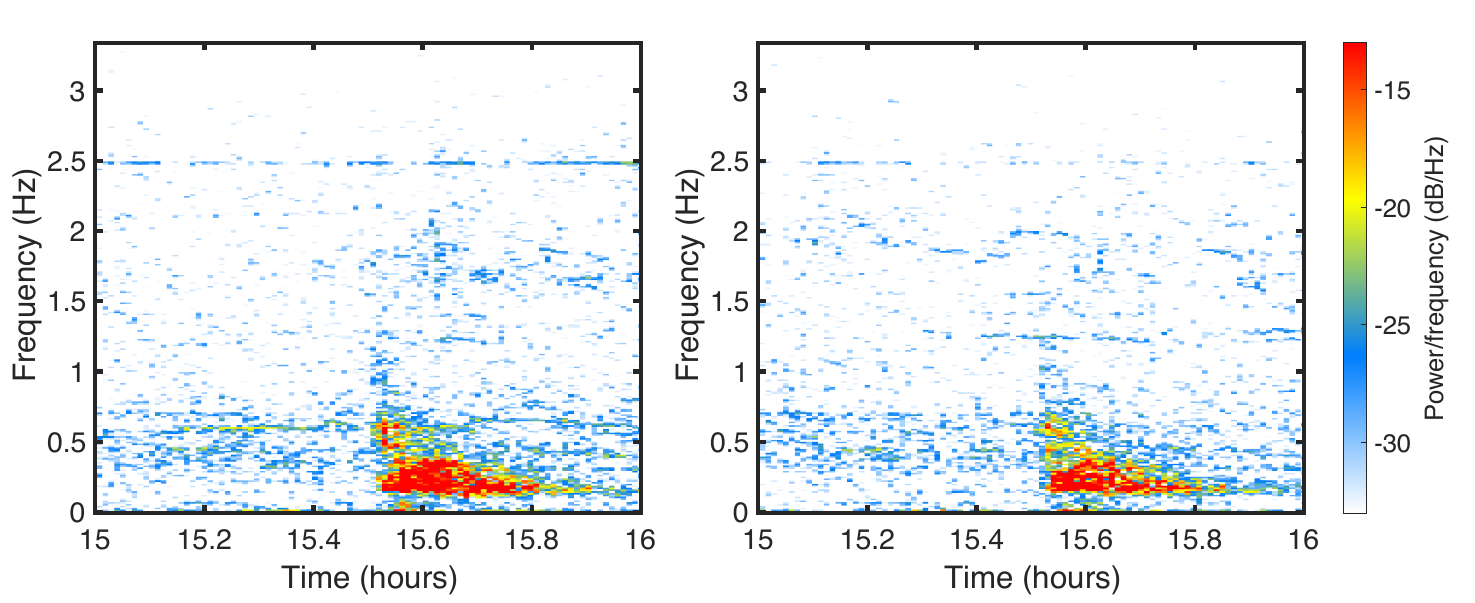}% Here is how to import EPS art
\caption{\label{fig2} Spectrograms of the two components of $\Delta \vec s$ orthogonal to $\vec s_0$ relative to a M7.3 earthquake occurred in Oaxaca on 23 June 2020, UTC time 15:29:05.}
\end{figure}

Figure \ref{fig2} illustrates the spectrograms of the two components of the deviations of the Stokes vector separately. It is noteworthy that both spectrograms display similar spectral features, despite the random nature of the birefringence acting independently on each component. This independence is confirmed by the observation that the magnitude of the cross-correlation between the two components does not exceed 5\%. This similarity strongly suggests that, as we anticipated at the end of section \ref{Pbs}, significant spatial self averaging occurs, and therefore the random nature of the static birefringence does not impose a significant limitation on the accuracy of sensing approaches based on the analysis of the modulation of the state of polarization at the output of a long fiber with random birefringence.

\begin{figure}
\includegraphics[width=\columnwidth]{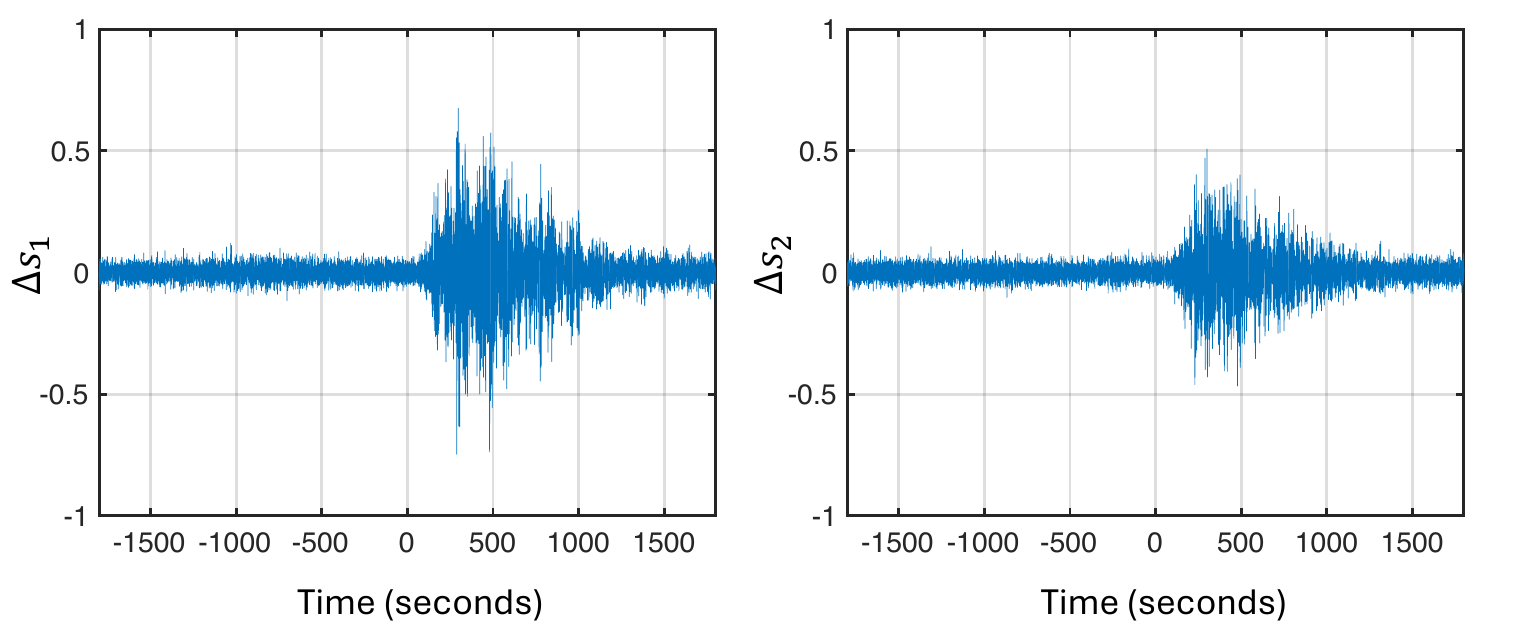}% Here is how to import EPS art
\caption{\label{fig3} Temporal traces of the two components of $\Delta \vec s$ orthogonal to $\vec s_0$ relative to a M7.3 earthquake occurred in Oaxaca on 23 June 2020, UTC time 15:29:05. The traces are filtered in the band 0.1 to 1.5 Hz. The origin of the temporal axis is set to the time of the earthquake.}
\end{figure}

Figure \ref{fig3} depicts the temporal traces of the two components of $\Delta \vec s$ orthogonal to $\vec s_0$, namely $\Delta s_1$ and $\Delta s_2$, extracted from the receiver of channel 1 during the Oaxaca earthquake, filtered in the band 0.1 to 1.5 Hz. 

\begin{figure}
\includegraphics[width=\columnwidth]{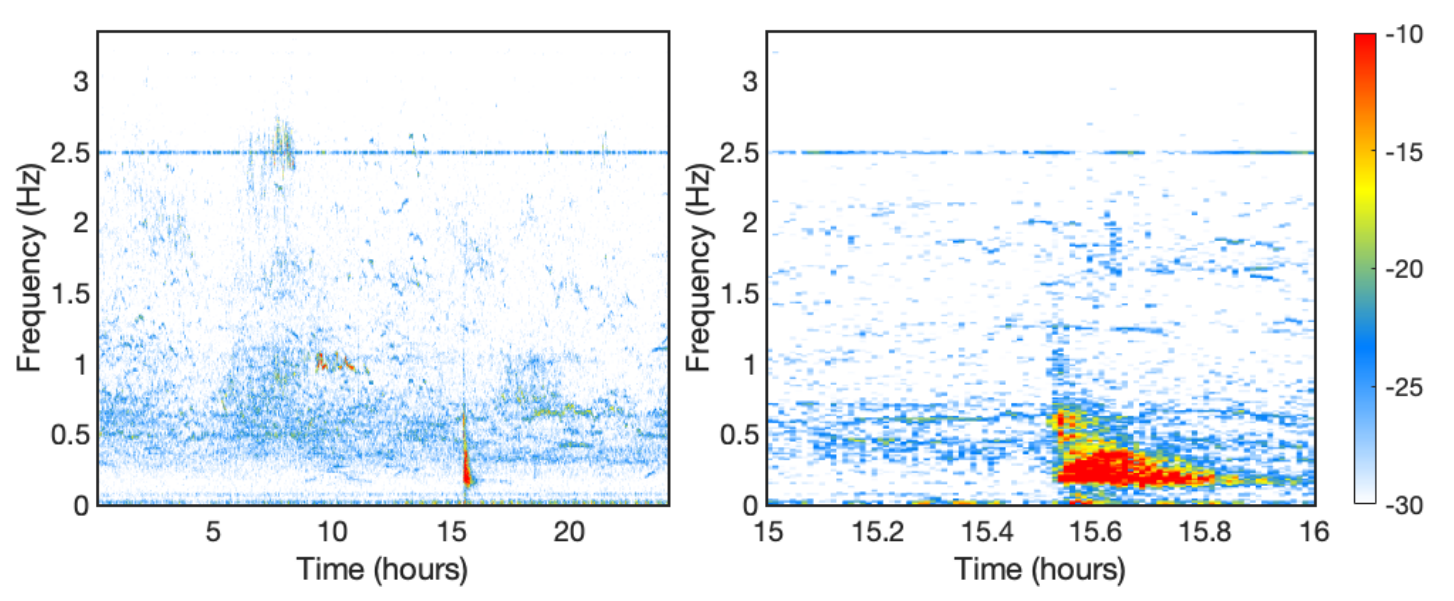}% Here is how to import EPS art
\caption{\label{fig4} Same of Fig. \ref{fig1} on a channel separated by 76.5 GHz (Channel 1 is centered at 193.5805 THz and channel 2 at 193.6570 THz).}
\end{figure}

\begin{figure}
\includegraphics[width=\columnwidth]{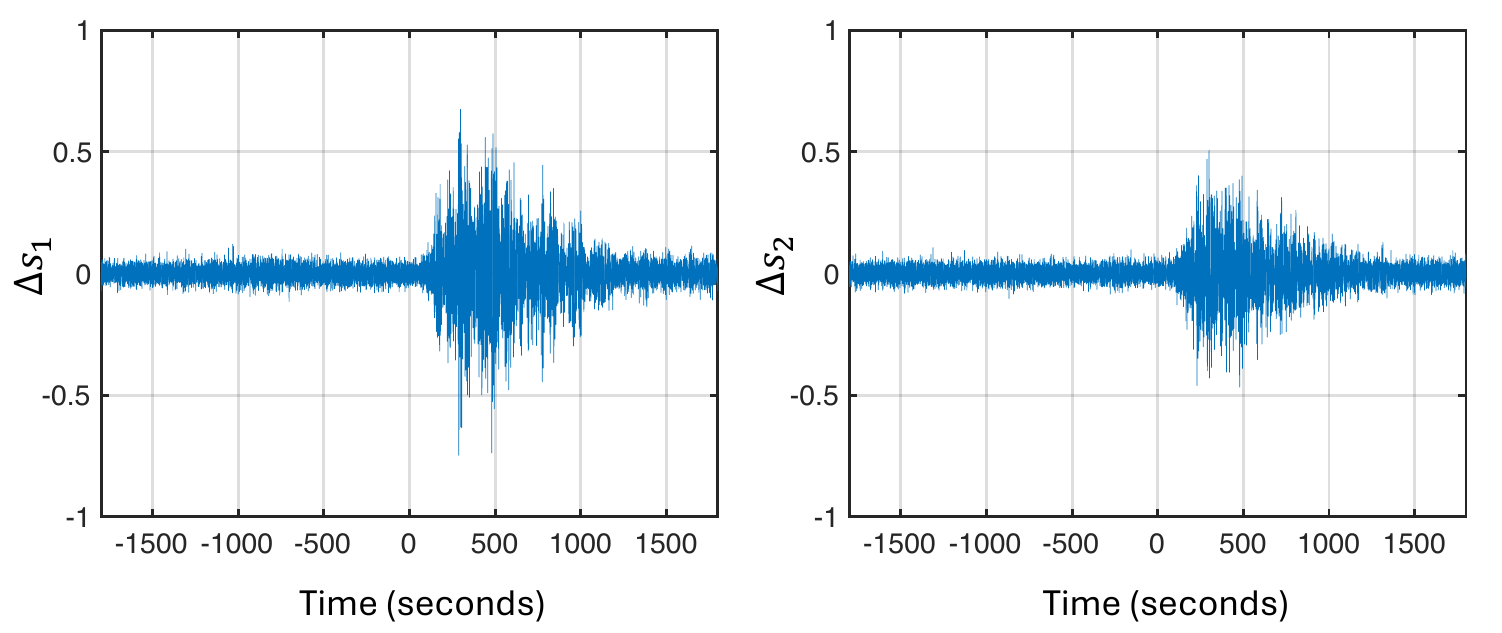}% Here is how to import EPS art
\caption{\label{fig5} Same of Fig. \ref{fig3} for the same channel of Fig. \ref{fig4}.}
\end{figure}

Figure \ref{fig4} displays the sum of the spectrograms of the two components of the Stokes vector orthogonal to the input polarization $\vec s_0$ for channel 2, which we remind is spaced 76.5 GHz from channel 1. The spectrogram closely resembles that of Fig. \ref{fig1}. Figure \ref{fig5} shows the temporal traces of the two components extracted from the same channel. A remarkable similarity between the traces of the two components of the Stokes vector extracted from channel 1 and channel 2 emerges from the comparison between Fig. \ref{fig3} and Fig. \ref{fig5}. The two traces nearly overlap when the trace of channel 2 is delayed by 552.35 s. The time misalignment between the two channels can be attributed to the clock of channel 2 being out of synchronization and experiencing a slow drift, leading to lags that could accumulate to several minutes from the UTC time (for further details, refer to the supplementary material of Ref. \onlinecite{Zhan:21}).


It is valuable at this point to develop a quantitative framework for estimating the correlations between the two wavelength division multiplexing channels. Let us consider two frequencies, $\omega_1$ and $\omega_2 = \omega_1 + \Delta \omega$ corresponding to the center frequencies of the two wavelength division multiplexing channels. We have shown in the previous analysis that the process of correlating the polarization fluctuations to the spectrum of the perturbations involves a rotation of the Poincaré sphere. This rotation aligns, for each channel, the average Stokes vector with a fixed point of the Poincar\'e sphere, which we have arbitrarily chosen as the north pole. Subsequently, we analyze the deviations of the Stokes vector from this average. We have already pointed out that, apart from an immaterial time-independent rotation, this is equivalent to the use of the fluctuations of the Stokes vectors in a frame rotating with the static birefringence. Since the birefringence is frequency-dependent, the rotating frame of the two channels is different. In a frame rotating with the static birefringence at the generic frequency $\omega$ and within a first order approximation, the fluctuations of the Stokes vector at frequency $\omega$ are 
\be \Delta \vec s_{\omega}'(z,t) = \xi \int_0^z \df z' \epsilon(z',t)  \mathbf{R}_{\omega}^{-1}(z') \vec \beta (z') \times \vec s_0, \ee
where we used Eqs. (\ref{Delta_s2}), (\ref{Deltabetaprime}) and (\ref{deltavecbeta}), and we neglect, as customarily done in the theory of polarization mode dispersion, the (weak) dependence on frequency of the birefringence, but not the effect of the frequency dependence on the rotation operators.  The correlation function of the fluctuations of the rotated Stokes vector at the two frequencies is 
\bea \langle \Delta \vec s_{\omega_2}'(z,t) \cdot \Delta \vec s_{\omega_1}'(z,t) \rangle &=& \xi^2 \int_0^z \df z' \int_0^z \df z'' \epsilon(z',t) \epsilon(z'',t) \nonumber \\ && \hspace{-3.5cm} \left\langle \left[\mathbf{R}_{\omega_2}^{-1}(z') \vec \beta(z') \times \vec s_0 \right] \cdot \left[\mathbf{R}_{\omega_1}^{-1}(z'') \vec \beta(z'') \times \vec s_0 \right] \right\rangle. \label{xcorrelation0} \eea
The result of the average is (see appendix \ref{appendixA} for the detailed derivation) 
\bea \langle \Delta \vec s_{\omega_2}'(z,t')  \cdot \Delta \vec s_{\omega_1}'(z,t) \rangle =  \nonumber \\
&& \hspace{-3.8cm} \frac{\pi}{4} \left(\frac{2 \pi c}{\lambda} \right)^2 \kappa^2  \int_0^z \df z' \epsilon(z',t') \epsilon(z',t) F(z'), \label{xcorrelation1} \eea
where
\be F(z) = \exp\left(-\frac{\pi \Delta \omega^2 \kappa^2 z}  8 \right). \ee
Expressing $F(z) = \exp(-z/\Delta Z)$ with $\Delta Z = 2 /(\pi^3 \Delta f^2 \kappa^2)$, and using the values of the systems of Ref. \onlinecite{zhongwenzhan_2020}, $\kappa = 0.03$ ps/$\sqrt{\mathrm{km}}$ and $\Delta f = 76.5$ GHz, we obtain $\Delta Z \simeq 12,247$ km. In the presence of a localized perturbation centered at $z = z_p$ over a spatial extension $\Delta z \ll \Delta Z$, we may replace $F(z')$ with $F(z_p)$ in the integral at right-hand side of Eq. (\ref{xcorrelation1}), obtaining
\bea \langle \Delta \vec s_{\omega_2}'(z,t')  \cdot \Delta \vec s_{\omega_1}'(z,t) \rangle &\simeq&  \nonumber \\
&& \hspace{-3.8cm} \frac{\pi}{4} \left(\frac{2 \pi c}{\lambda} \right)^2 \xi^2 \kappa^2 F(z_p) \int_0^z \df z' \epsilon(z',t') \epsilon(z',t), \eea
that is
\be \langle \Delta \vec s_{\omega_2}'(z,t')  \cdot \Delta \vec s_{\omega_1}'(z,t) \rangle \simeq \langle \Delta \vec s_{\omega}'(z,t')  \cdot \Delta \vec s_{\omega}'(z,t) \rangle \, F\left(z_p\right), \label{xcorrelation} \ee
where at right-hand side $\omega$ is either $\omega_1$ or $\omega_2$. Since the fluctuations of the Stokes vector $\Delta \vec s$ are isotropic on the tangent plane of the Poincar\'e sphere centered on the tip of $\vec s_0$, the above equations in terms of the components on a canonical basis of the rotated Stokes space $\vec e_i$ become
\bea && \langle \left[\Delta \vec s_{\omega_2}'(z,t') \cdot \vec e_i\right] \left[\Delta \vec s_{\omega_1}'(z,t) \cdot \vec e_j\right] \rangle \nonumber \\ 
&=& \frac 1 2 \langle \Delta \vec s_{\omega_2}'(z,t')  \cdot \Delta \vec s_{\omega_1}'(z,t) \rangle \delta_{i,j}, \quad i,j = 1,2, \eea
and 
\bea && \langle \left[\Delta \vec s_{\omega}'(z,t')  \cdot \vec e_i \right] \left[\Delta \vec s_{\omega}'(z,t) \cdot \vec e_j\right] \rangle \nonumber \\ 
&=& \frac 1 2 \langle \Delta \vec s_{\omega}'(z,t')  \cdot \Delta \vec s_{\omega}'(z,t) \rangle \delta_{i,j}, \quad i,j = 1,2, \eea
with the fluctuations of the third components zero to first order. 

When a single localized perturbation is dominant over the others, the cross-correlation is proportional to the autocorrelation of the two channels, with the scaling factor $F(z_p)$. This is the case of earthquakes strongly coupled to the fiber. When, on the other hand, there are multiple perturbations scattered at different positions along the cable, the proportionality cannot be established and cross-correlation and  autocorrelations have different shapes. For a given $z_p$, the width of the function $F(z_p)$ is the bandwidth over which the polarization of two channels at different frequencies decorrelates, which is related (but is not equal) to the bandwidth of the principal states of polarization of the fiber section from the transmitter to the position $z_p$ along the fiber.\cite{Shtaif:00} This bandwidth is directly proportional to the inverse of the polarization mode dispersion coefficient, $\kappa$, multiplied by the square root of the distance from transmitter.

When a narrowband perturbation affects all channels equally at a specific point in the fiber, it might seem somehow obvious that the perturbation is fully correlated if the perturbation is applied at the receiver, that is $z_p = z$, and that the distance for considering the decorrelation of the perturbation imprinted on the channels is the distance from the point where the perturbation is applied to the receiver. It may therefore appear counterintuitive the result stated by Eq. (\ref{xcorrelation}) that the distance affecting the depolarization is the distance of the perturbation point from the transmitter $z_p$. This produces the seemingly paradoxical results that a perturbation at the cable input produces a fully correlated polarization perturbation between distant channels. The explanation of this result stems from the use of a frame rotating with the static birefringence of each channel. The use of this reference frame is equivalent to the application of a backpropagation that rotates back the output polarization under the action of the static birefringence only. The output polarization of two channels does get decorrelated proportionally to the distance of the section that goes from the perturbation point to the receiver, but the backpropagation from the receiver to the transmitter compensates exactly the decorrelation that occurs in the section from the receiver to the point where the perturbation is applied, leaving uncompensated only the propagation from the point of perturbation to the transmitter. 

\begin{figure}
\includegraphics[width=\columnwidth]{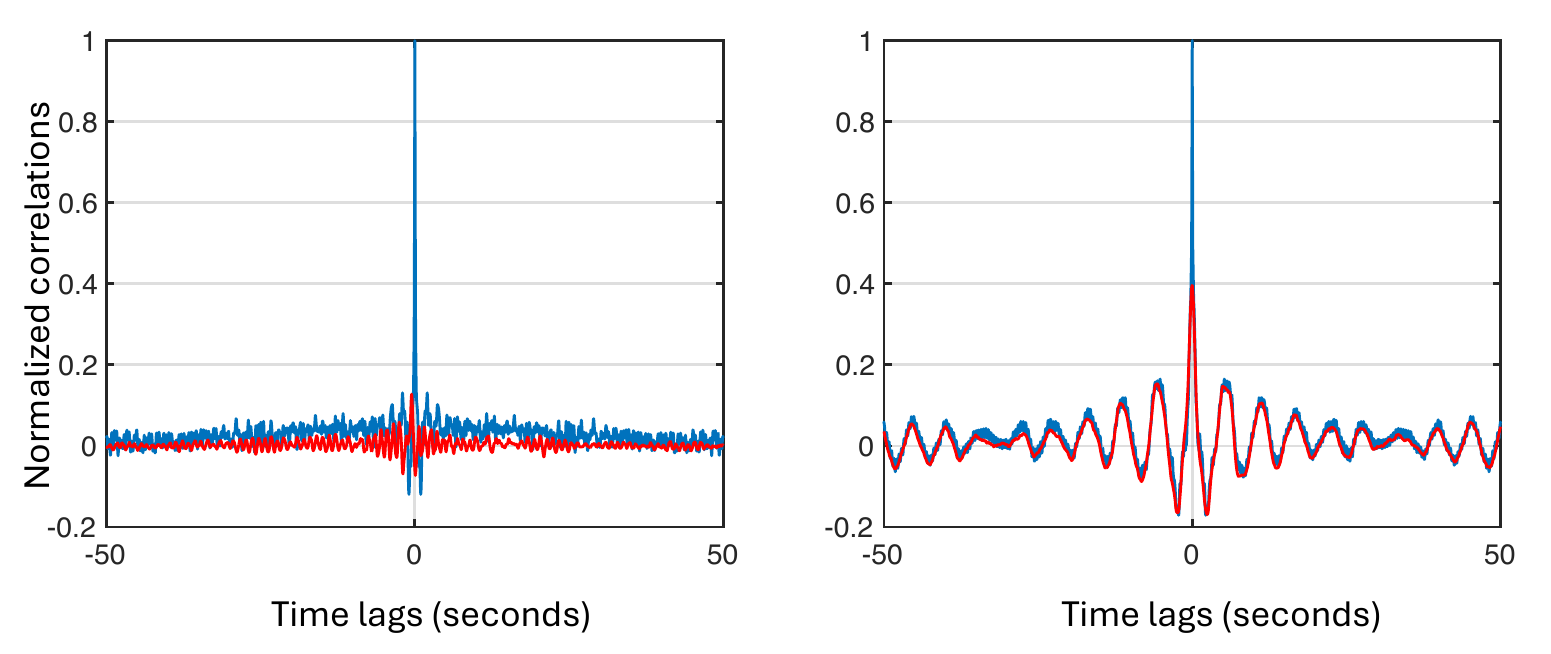}% Here is how to import EPS art
\caption{\label{fig9} Autocorrelation of $\Delta s_1$ for channel 1 (blue) with overlapped cross-correlation between channel 1 and channel 2 delayed by 552.35 s, from 0 to 12 UTC time of 23 June 2020 (left panel), and from 12 to 24 UTC time of the same day, which include the time of the earthquake (right panel).}
\end{figure}
\begin{figure}
\includegraphics[width=\columnwidth]{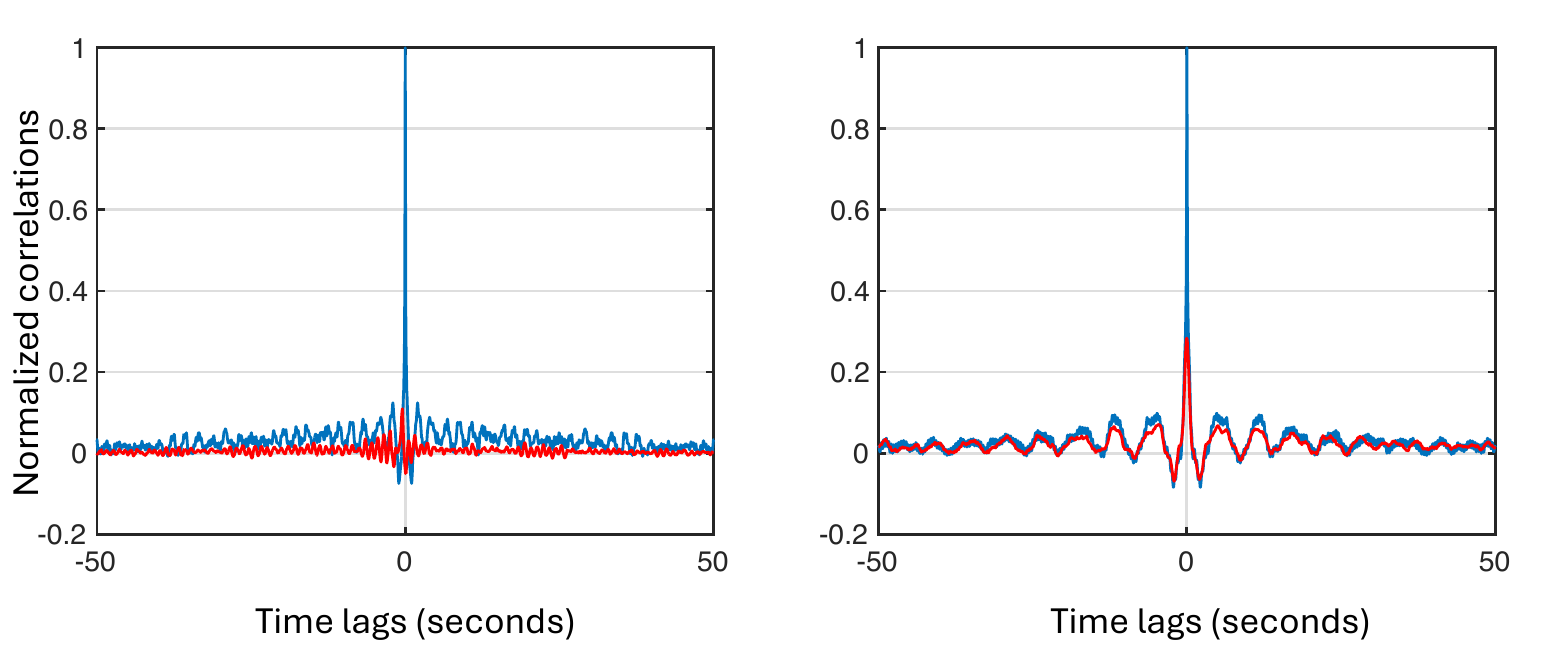}% Here is how to import EPS art
\caption{\label{fig10} Autocorrelation of $\Delta s_2$ for channel 1 (blue) with overlapped cross-correlation between channel 1 and channel 2 delayed by 552.35 s, from 0 to 12 UTC time of 23 June 2020 (left panel), and from 12 to 24 UTC time of the same day, which include the time of the earthquake (right panel). The cross-correlation is displayed with a sign inverted (see text).}
\end{figure}

Let us now compare the expressions that we have just derived with the data provided in Ref. \onlinecite{zhongwenzhan_2020}. Since the cable perturbation is inherently a nonstationary process, its correlation functions do not depend on $t-t'$ only. For a better visual representation, we decided to compare autocorrelation and cross-correlation plotting the normalized correlation of the polarization traces as a function of the time difference $t-t'$, averaged over a suitable time window and normalized such that the autocorrelation for $t = t'$ is one. The time window for the average will be chosen on a case-by-case basis to highlight specific features of the process. 

Figure \ref{fig9} illustrates the autocorrelation of $\Delta s_1$ for channel 1 (blue), with overlapped the cross-correlation between $\Delta s_1$ extracted from channel 1 and channel 2 delayed by 552.35 s (red). The cross correlation was calculated spanning the time interval from 0 to 12 UTC time of 23 June 2020 (left panel), and the time interval, containing the time of the earthquake, from 12 to 24 UTC time of the same day (right panel). Figure \ref{fig10} depicts the same curves for $\Delta s_2$, with the cross-correlation displayed with the sign inverted. Notably, achieving consistency between the data extracted from channel 1 and channel 2 always necessitates inverting one of the components (not always the same) of the Stokes vector of one of the two channels, that we arbitrarily choose as being channel 2. A plausible explanation is that the transmission matrix, from which the output polarization of channel 2 (or channel 1) is derived, is the transpose of that pertaining to the propagation direction of channel 1 (or channel 2). The transposition of a matrix entails the inversion of the third component of the rotation vector in Stokes space, effectively resulting in an improper rotation of the Stokes space. The resulting change of the frame parity (from right-handed to left-handed) in Stokes space implies that after rotation, one of the two components of the fluctuations of the rotated Stokes vector changes its sign.
\begin{figure}
\includegraphics[width=\columnwidth]{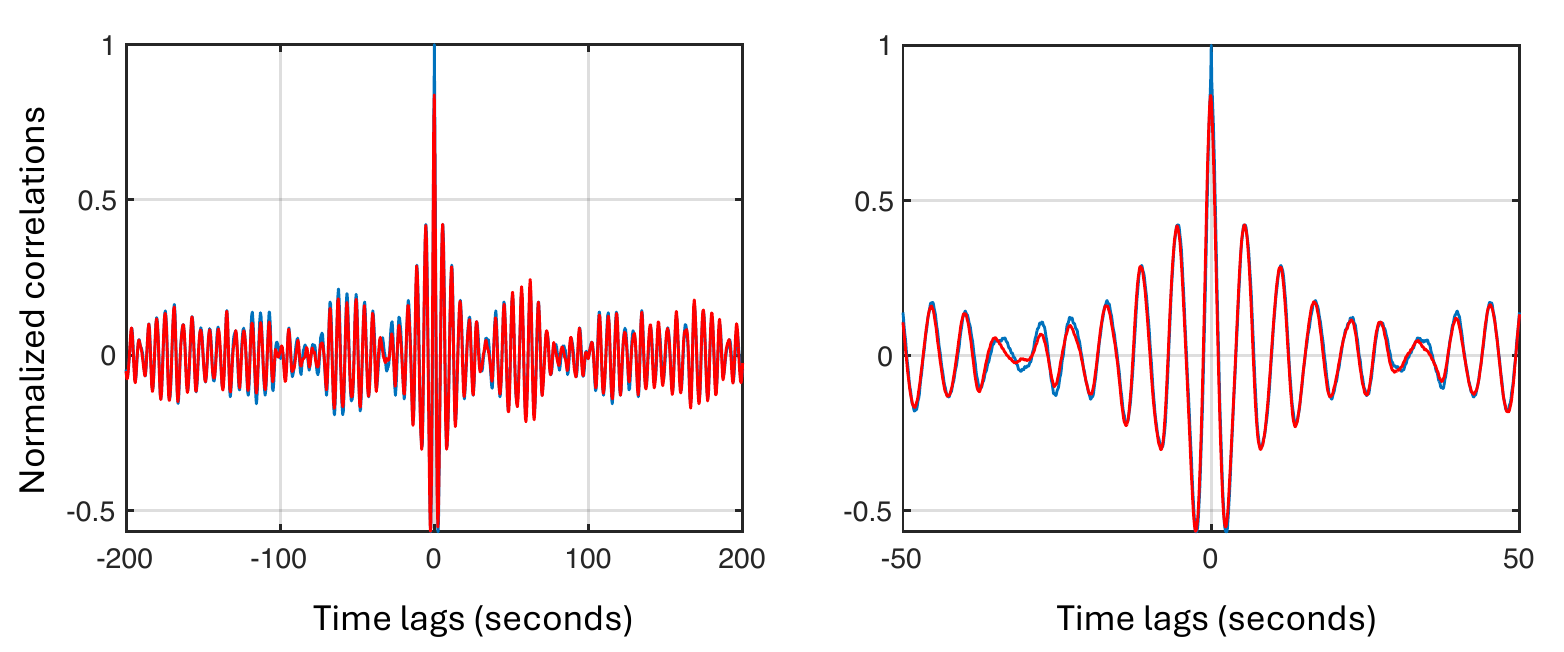}% Here is how to import EPS art
\caption{\label{fig13} Autocorrelation of $\Delta s_1$ for channel 1 (blue) with overlapped cross-correlation between channel 1 and channel 2 delayed by 552.35 s, from UTC time $t_0 - 1000$ s to $t_0 + 2000$ s where $t_0$ is the UTC time of the earthquake 23 June 2020 15:29:05 plotted from $- 200$ s to $200$ s (left panel),  and a zoom from $-50$ s to $50$ s (right panel).}
\end{figure}
\begin{figure}
\includegraphics[width=\columnwidth]{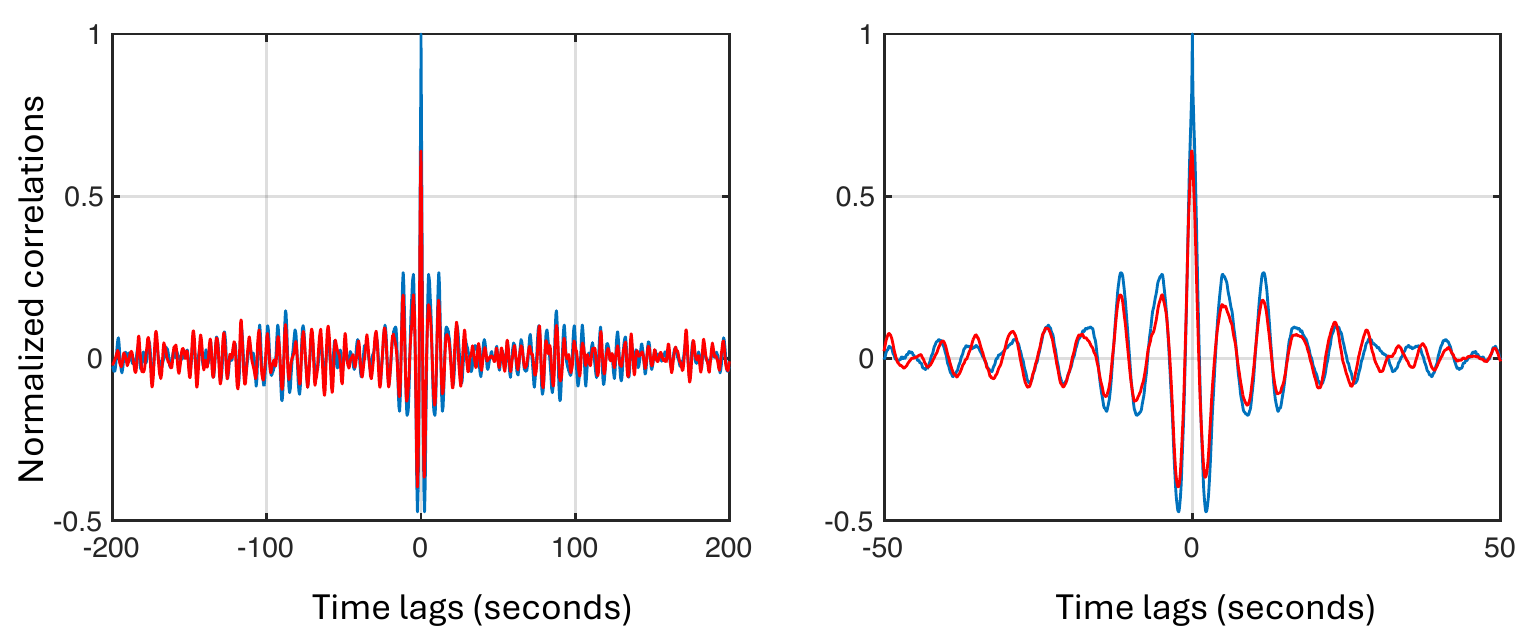}% Here is how to import EPS art
\caption{\label{fig14} Autocorrelation of $\Delta s_2$ for channel 1 (blue) with overlapped cross-correlation between channel 1 and channel 2 delayed by 552.35 s, from UTC time $t_0 - 1000$ s to $t_0 + 2000$ s where $t_0$ is the UTC time of the earthquake on 23 June 2020, 15:29:05 plotted from $- 200$ s to $200$ s (left panel), and a zoom from $-50$ s to $50$ s (right panel).}
\end{figure}

The right panels of Figs. \ref{fig9} and \ref{fig10} demonstrate, in agreement with our theoretical predictions, that in the presence of a dominant localized earthquake perturbation, the autocorrelation and the cross-correlation exhibit approximate proportionality, whereas they manifest different shapes when the perturbations are numerous and distributed, as in the left panels of the same figures. However, while this proportionality becomes evident for nonzero time lags, the value when the time lag equals zero appears reduced compared to the expected value. This reduction may stem from either the residual presence of distributed perturbations or a slow drift of the clock of channel 2, resulting in a misalignment of the polarization traces. This misalignment effectively produces a dilatation of the time axis of channel 2, thus reducing the narrow cross-correlation peak. Both of these effects become less pronounced if we narrow down the time window of the cross-correlation around the time of the earthquake. This is corroborated by the analysis of Figs. \ref{fig13} and \ref{fig14}, where the right panels depict the same autocorrelations and cross-correlations as the right panels of Figs. \ref{fig9} and \ref{fig10}, but calculated over a time window of 3000 s starting from 1000 s before the UTC time of the earthquake. The left panels represent the same correlations as the right panels, with enlarged windows for the time lags. The rise of the central peak is evident, along with the near overlap of the two correlation traces at non-zero time lags.

\begin{figure}
\includegraphics[width=\columnwidth]{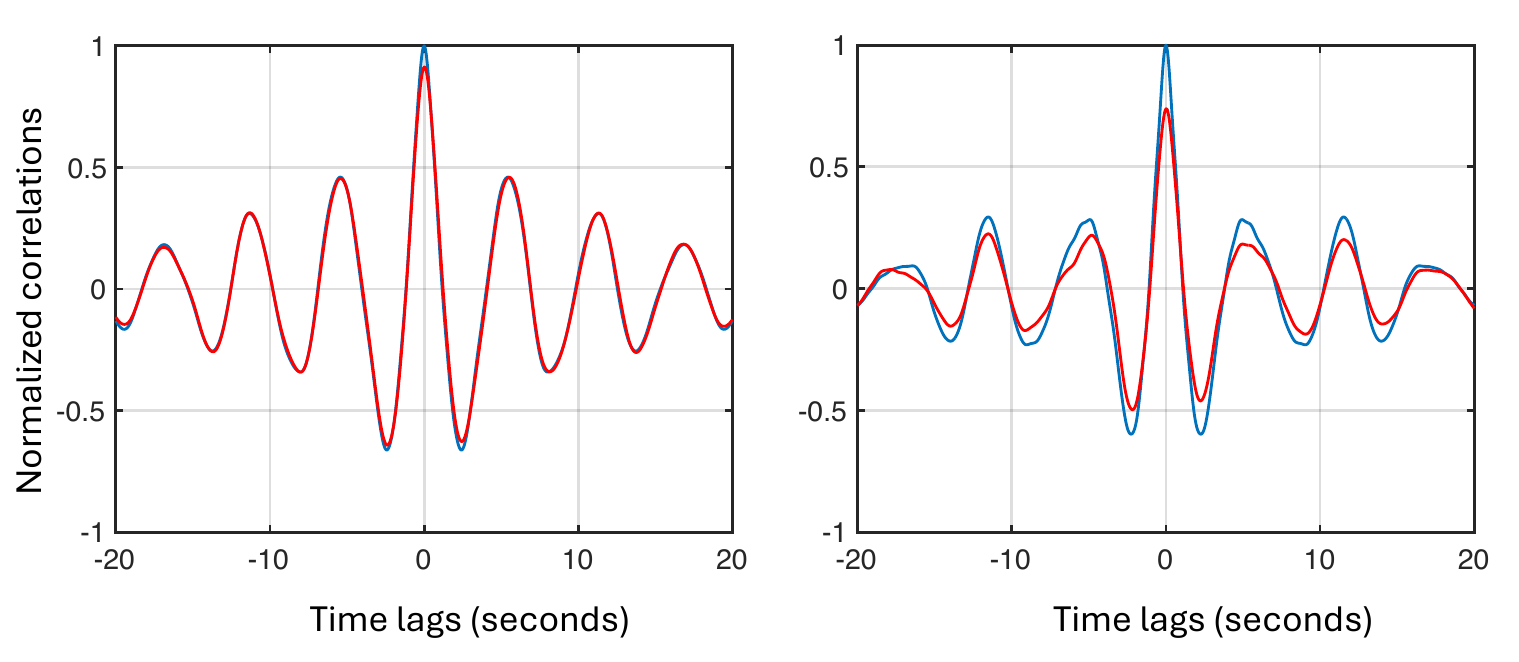}% Here is how to import EPS art
\caption{\label{fig200} In blue: autocorrelations of $\Delta s_1$ (left panel), and of $\Delta s_2$ (right panel) for channel 1. In red: cross-correlations between $\Delta s_1$ obtained from channel 1 and channel 2 (left panel) and $\Delta s_2$ obtained from channel 1 and channel 2, respectively, delayed by 552.35 s. Before processing, the signals were filtered within the frequency range of 0.1 to 1.5 Hz, corresponding to the earthquake perturbation frequency range. The time window from UTC time $t_0 - 1000$ s to $t_0 + 2000$ s where $t_0$ is the UTC time of the earthquake on 23 June 2020, 15:29:05.}
\end{figure}

Further confirmation of the strong correlation between the polarization fluctuations of the two channels is evident when the Stokes vector deviations are filtered within the frequency range of 0.1 to 1.5 Hz, which approximately corresponds to the earthquake perturbation frequency range. This is shown in Fig. \ref{fig200}, which in the left panel illustrates in blue the autocorrelations of $\Delta s_1$, while in the right panel the autocorrelations of $\Delta s_2$, both for channel 1. Overlaid in red are the cross-correlations between $\Delta s_1$ and $\Delta s_2$, respectively, obtained from the two channels, delayed by 552.35 s. Before processing, the signals were filtered within the frequency range of 0.1 to 1.5 Hz. For the component with more pronounced dynamics, $\Delta s_1$, the autocorrelation and cross-correlation curves essentially coincide. For the component exhibiting smaller amplitude fluctuations (possibly because of some residual polarization dependent loss), the two curves are more distinct. This difference might be attributed to the rotation of the sphere aimed at aligning the average output Stokes vector with the north pole. The rotation could have been different for the two channels, potentially causing a slight misalignment of the cluster of polarization points due to a small rotation around the third axis of the Stokes vector.

\begin{figure}
\includegraphics[width=\columnwidth]{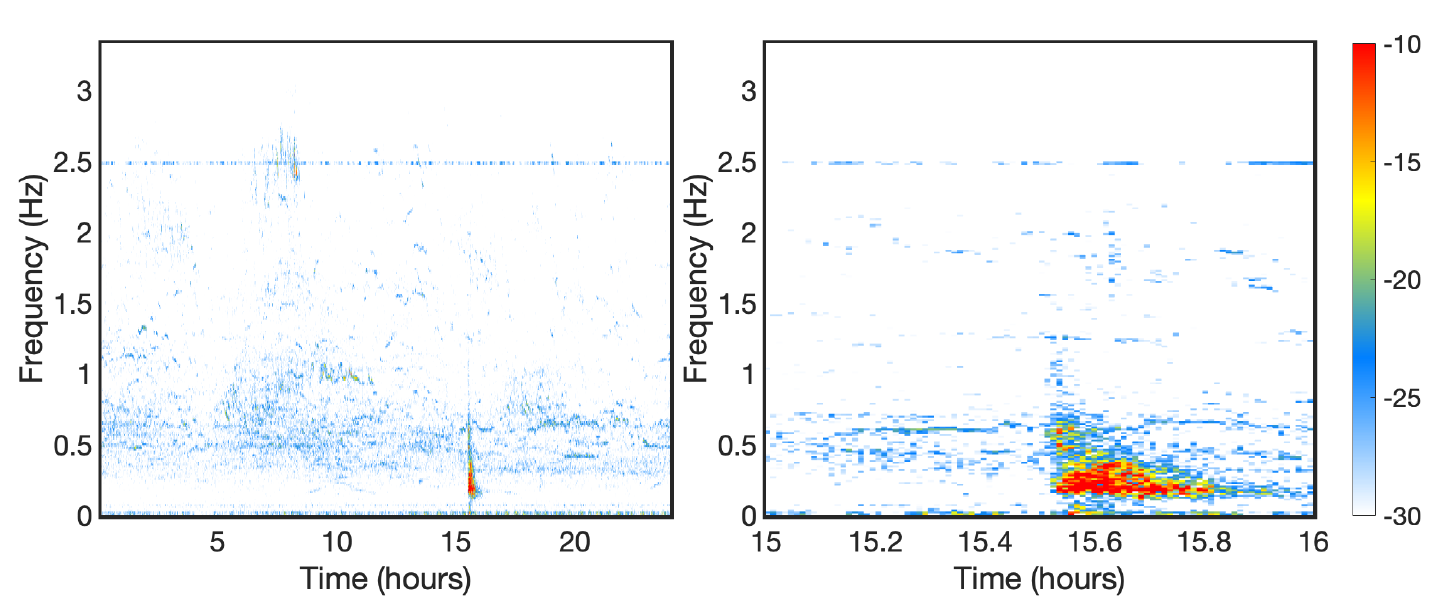}% Here is how to import EPS art
\caption{\label{fig310} Sum of the cross-spectrograms of the $\Delta s_1$ and $\Delta s_2$ of channel 1 and channel 2, relative to a M7.3 earthquake occurred in Oaxaca on 23 June 2020, UTC time 15:29:05. Left panel shows the entire day of the earthquake, the right panel a zoom of the hour of the earthquake.}
\end{figure}

\begin{figure}
\includegraphics[width=\columnwidth]{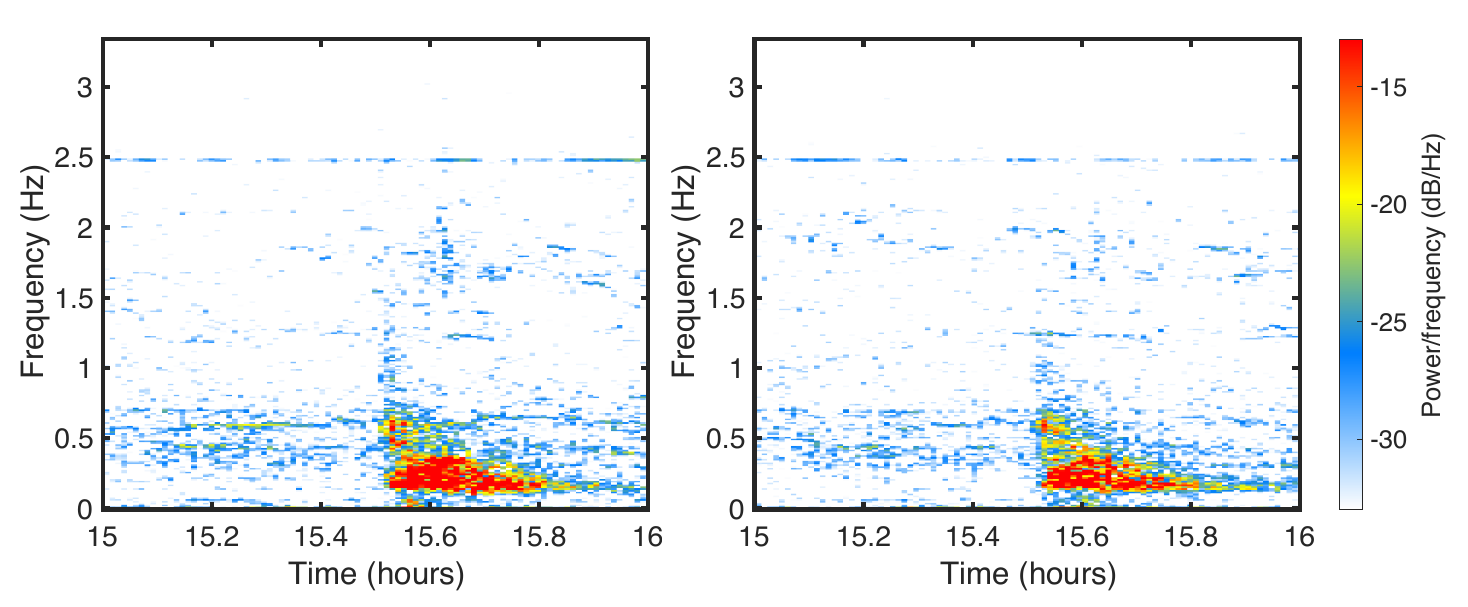}% Here is how to import EPS art
\caption{\label{fig300} Cross-spectrograms of the $\Delta s_1$ and $\Delta s_2$ of channel 1 and channel 2, relative to a M7.3 earthquake occurred in Oaxaca on 23 June 2020, UTC time 15:29:05.}
\end{figure}

Cross-spectrograms,\cite{Oppenheim:99} which are the time-frequency representation of the products of the short-time Fourier transforms of the polarization traces of the two channels (or equivalently, of the modulus of the short-time Fourier transform of the cross-correlations), can also provide a useful visualization of the correlations between channels. Figure \ref{fig310} shows the sum of the cross-spectrograms of $\Delta s_1$ and $\Delta s_2$ of channel 1 and channel 2 for the entire day of the earthquake (left panel) and its magnification in the hour of the earthquake (right panel). The trace of channel 2 has been delayed by 552.35 s. The strong similarity with Fig. \ref{fig1}, representing the sum of the spectrograms of $\Delta s_1$ and $\Delta s_2$ of channel 1, is self-evident. It is also evident that the cross-spectrograms show an earthquake signature practically unaltered from the spectrograms of Fig. \ref{fig1}, while the background features appear more diffuse, suggesting a less localized origin. Figure \ref{fig300} shows separately the cross-spectrogram of $\Delta s_1$ (left panel) and $\Delta s_2$ (right panel) of channel 1 and channel 2. Again, the cross-spectrograms displayed here show features very similar to those in Fig. \ref{fig2}, which presents separately the spectrograms of $\Delta s_1$ and $\Delta s_2$ of channel 1. 

%

% Let us now analyze another earthquake of magnitude 6.8 with epicenter 200 km east of the city of Antofagasta, in Chile, close to the Valparaiso terminal of the Curie cable, on 3 June 2020, UTC time 07:35:33. Figure \ref{fig7} shows the sum of the spectrograms of the two components of the rotation vector orthogonal to the input Stokes vector for channel 1. The left panel reports the entire day of the earthquake while the right panel the zoom of one hour time window around the earthquake time. The spectral signature of this earthquake is much weaker than that of the Oaxaca earthquake, an indication of a possible weaker coupling of the cable with the environment.

\begin{figure}
\includegraphics[width=\columnwidth]{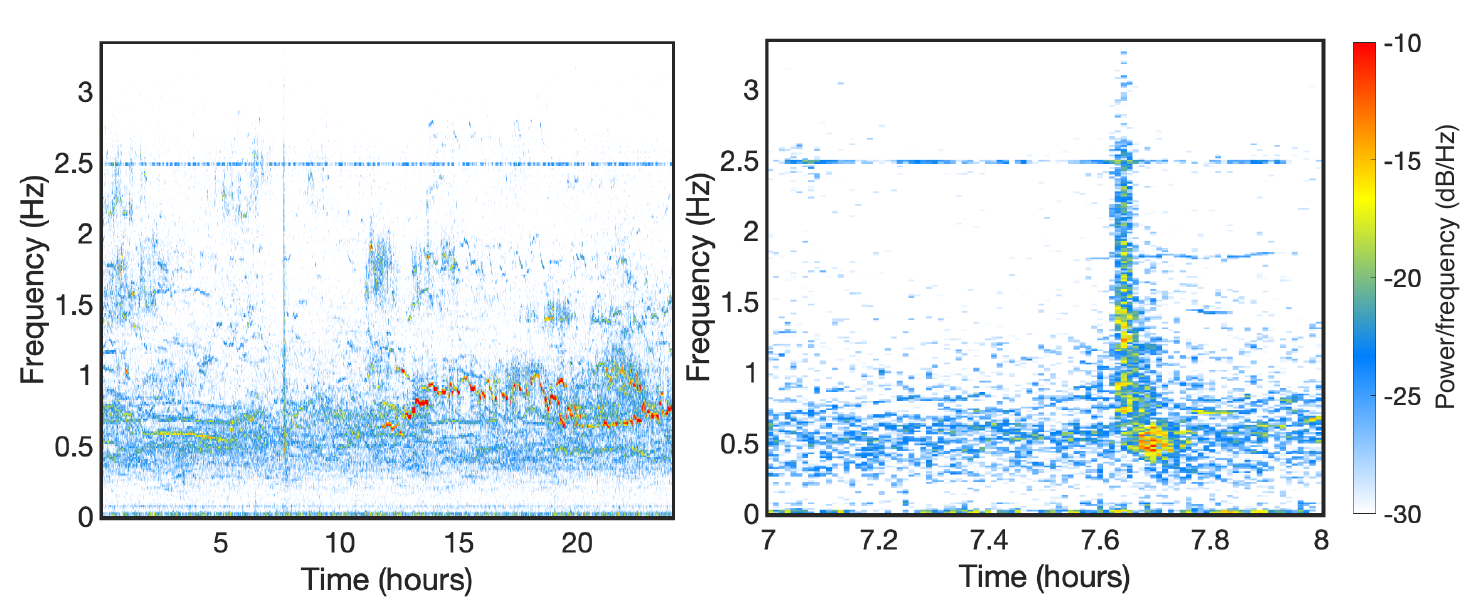}% Here is how to import EPS art
\caption{\label{fig7} Sum of the spectrograms of the two components of $\Delta \vec s$ orthogonal to $\vec s_0$ relative to M6.8 earthquake occurred approximately 200 km east of the city of Antofagasta, in Chile, on 3 June 2020, UTC time 07:35:33. The abscissa reports to the UTC time. The left panel reports the entire day of the earthquake while the right panel the zoom of one hour time window around the earthquake time.}
\end{figure}

Let us now examine another earthquake, namely a magnitude 6.8 event with its epicenter located 200 km east of the city of Antofagasta, Chile, near the Valparaiso terminal of the Curie cable. This earthquake occurred on June 3, 2020, at UTC time 07:35:33. Figure \ref{fig7} displays the sum of the spectrograms of the two components of the output Stokes vectors orthogonal to the input Stokes vector for channel 1. The left panel illustrates the entire day of the earthquake, while the right panel zooms in on a one-hour time window around the earthquake event. Compared to the Oaxaca earthquake, the spectral signature of this event appears much weaker, suggesting a potentially lower coupling of the cable with the surrounding environment.

\begin{figure}
\includegraphics[width=\columnwidth]{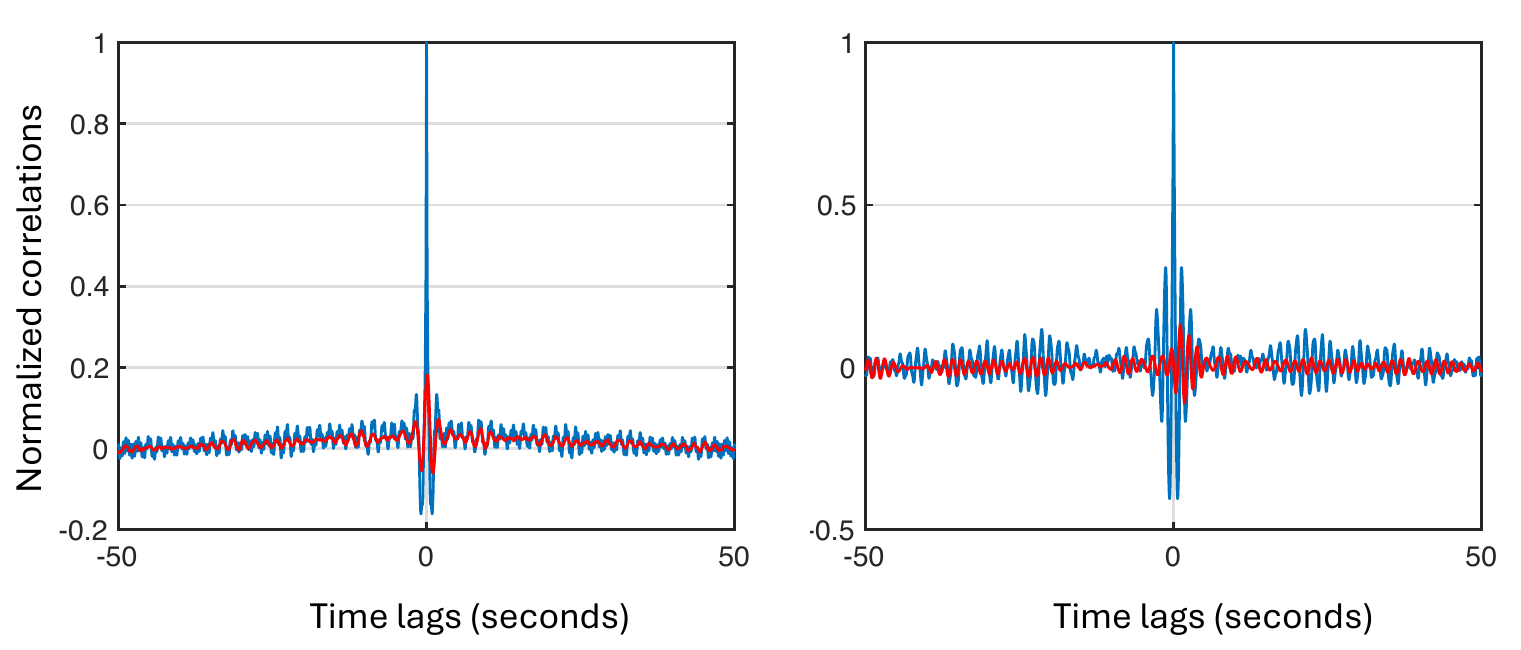}% Here is how to import EPS art
\caption{\label{fig11} Autocorrelation (blue) of $\Delta s_1$ of channel 1 and cross-correlation (red) between $\Delta s_1$ of the channel 1 and channel 2, from 0 to 12 UTC time of 3 June 2020, which include the time of the earthquake (left panel), and from 12 to 24 UTC time of the same day (right panel). The cross-correlation is displayed with a sign inverted (see text)}
\end{figure}

\begin{figure}
\includegraphics[width=\columnwidth]{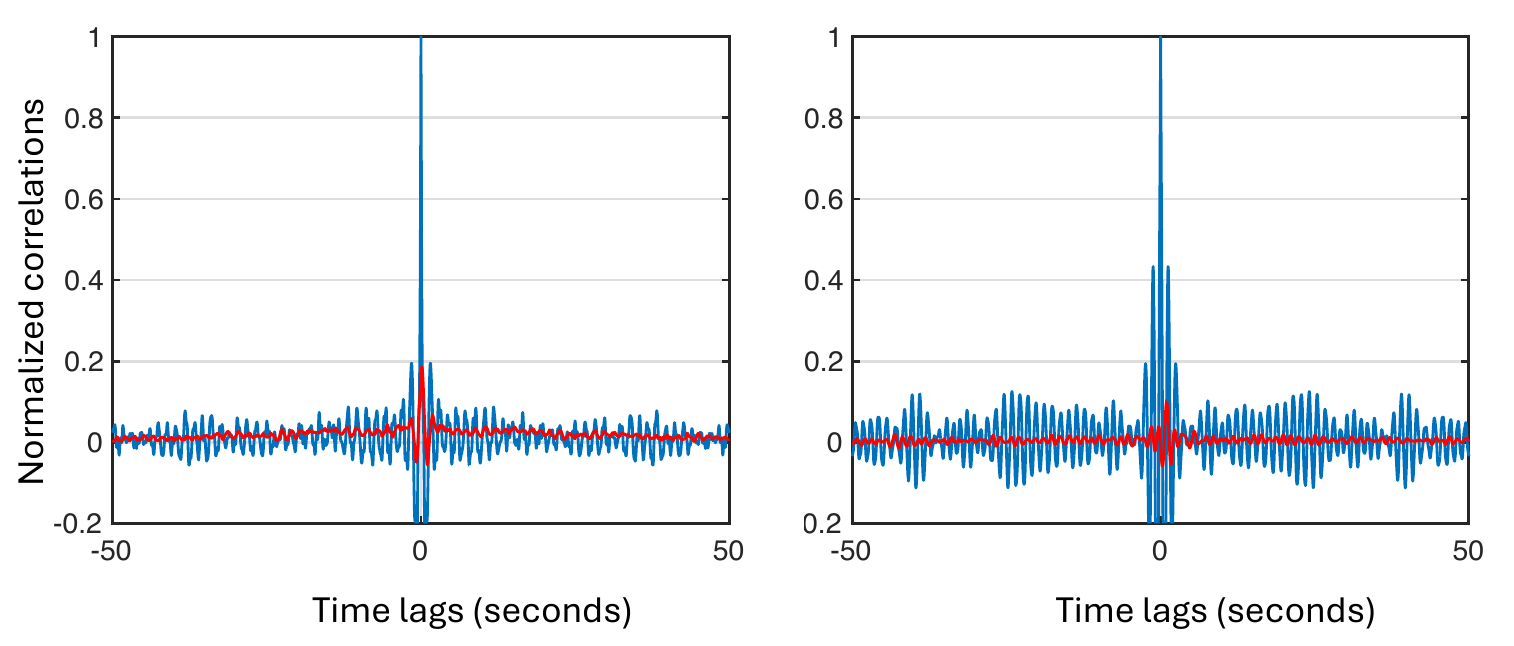}% Here is how to import EPS art
\caption{\label{fig12} Autocorrelation of $\Delta s_2$ for channel 1 (blue) with overlapped cross-correlation between channel 1 and channel 2 delayed by 519.9 s (red), from 0 to 12 UTC time of 3 June 2020, (left panel), and from 12 to 24 UTC time of the same day (right panel).}
\end{figure}

For this earthquake as well the database of Ref. \onlinecite{zhongwenzhan_2020} includes state of polarization data extracted from channel 1 and 2. Figure \ref{fig11} displays the autocorrelation of $\Delta s_1$ of channel 1 in blue and the cross-correlation between $\Delta s_1$ of channel 1 and channel 2 in red. The left panel covers the time interval from 0 to 12 UTC time on June 3, 2020, an interval including the time of the earthquake, while the right panel covers the time interval from 12 to 24 UTC time of the same day. In both plots, $\Delta s_1$ of channel 2 was inverted. Figure \ref{fig12} shows the same quantities for $\Delta s_2$. Notice that, differently from the state of polarization data of channel 2 relative to the Oaxaca earhquake where, to make the data compatible with those of channel 1, $\Delta s_1$ was left unchanged and $\Delta s_2$ inverted, in this case $\Delta s_1$ was inverted and $\Delta s_2$ left unchanged. As we discussed previously, the change of sign of only one of the components of the Stokes vector of channel 2 is indicative of a change of parity. This could potentially result from the unitary matrix used to derive the polarization data for channel 2 being the transpose of that for channel 1, suggesting a reversal in propagation direction.


\begin{figure}
\includegraphics[width=\columnwidth]{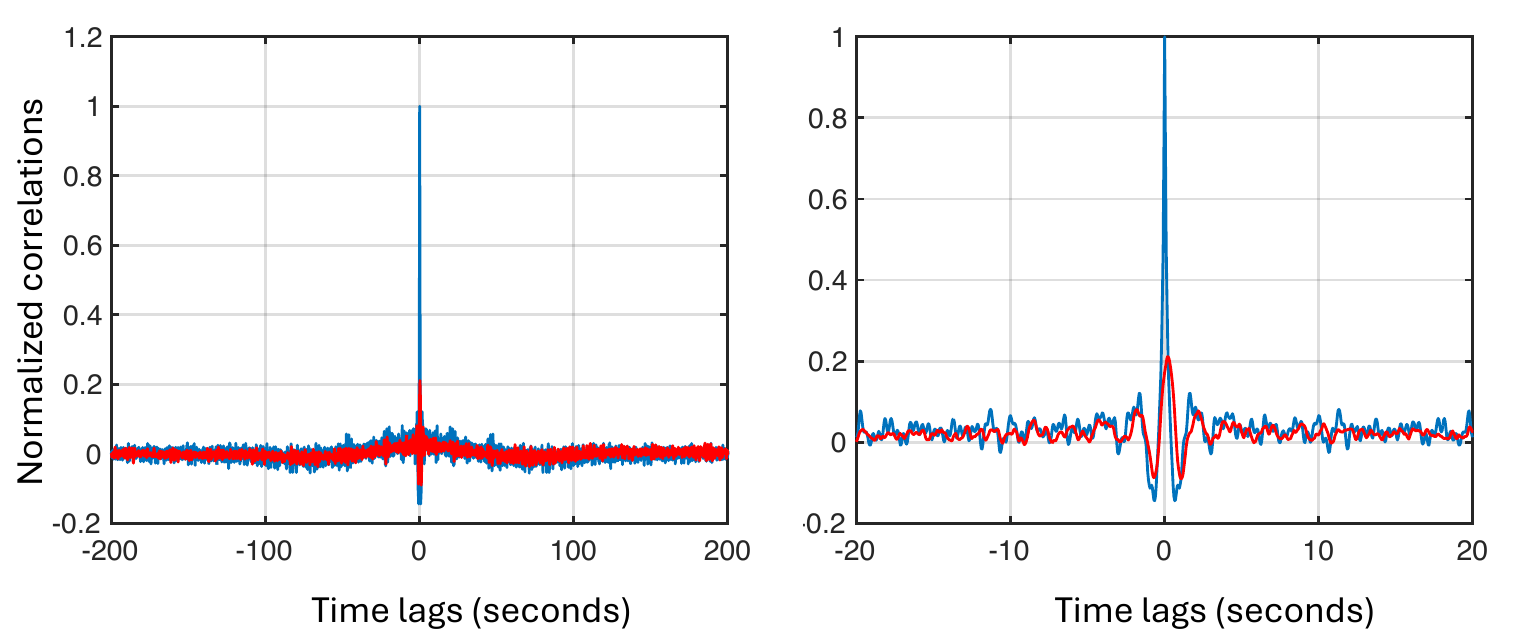}% Here is how to import EPS art
\caption{\label{fig15} Autocorrelation of $\Delta s_1$ for channel 1 (blue) with overlapped cross-correlation between channel 1 and channel 2 delayed by 519.9 s (red), from UTC time $t_0 - 1000$ s to $t_0 + 2000$ s where $t_0$ is the UTC time of the earthquake, 3 June 2020 07:35:33 plotted from $- 200$ s to $200$ s (left panel), and a zoom from $-50$ s to $50$ s (right panel).}
\end{figure}

\begin{figure}
\includegraphics[width=\columnwidth]{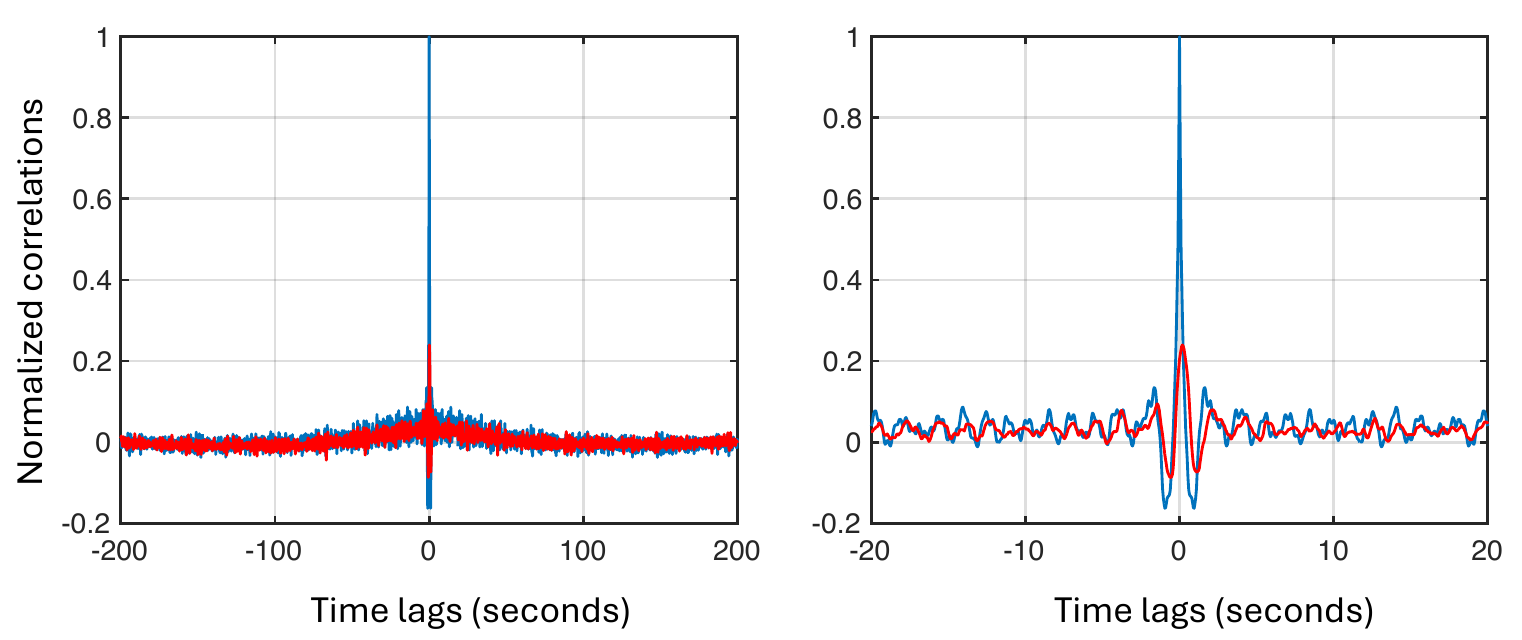}% Here is how to import EPS art
\caption{\label{fig16} Autocorrelation of $\Delta s_2$ for channel 1 (blue) with overlapped cross-correlation between channel 1 and channel 2 delayed by 519.9 s (red), from UTC time $t_0 - 1000$ s to $t_0 + 2000$ s where $t_0$ is the UTC time of the earthquake on 3 June 2020, 07:35:33 plotted from $- 200$ s to $200$ s (left panel), and a zoom from $-50$ s to $50$ s (right panel).}
\end{figure}

Both Figs. \ref{fig11} and \ref{fig12} fail to show a clear proportionality between autocorrelation and cross-correlation, because the earthquake is not the dominant source of perturbation when averaged over 12 hours. If we restrict the autocorrelations and cross-correlation to a neighbour of the earthquake, we expect this proportionality to rise up. This is confirmed by looking at Figs. \ref{fig15} and \ref{fig16} which show the autocorrelations of $\Delta s_1$ and $\Delta s_2$ for channel 1 and the cross-correlations of the same quantities of channel 1 and 2, calculated from UTC time $t_0 - 1000$ s to $t_0 + 2000$ s where $t_0$ is the UTC time of the earthquake, 07:35:33 of 3 June 2020. Autocorrelations and cross-correlations have in this case a distinct similarity, as the theory suggests.

\begin{figure}
\includegraphics[width=\columnwidth]{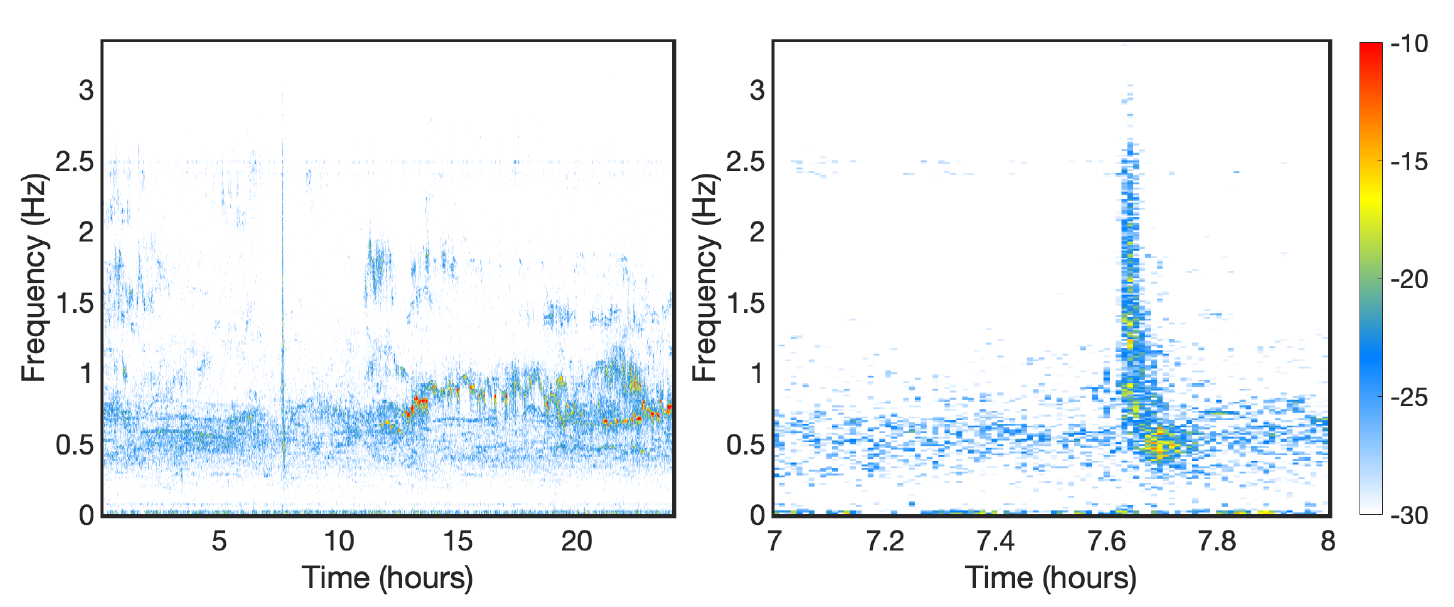}% Here is how to import EPS art
\caption{\label{fig400} Sum of the cross-spectrograms of $\Delta s_1$ and $\Delta s_2$ of channel 1 and channel 2, relative to M6.8 earthquake occurred approximately 200 km east of the city of Antofagasta, in Chile, on 3 June 2020, UTC time 07:35:33. The trace of channel 2 has been delayed by 519.9 s. The abscissa reports to the UTC time. The left panel reports the entire day of the earthquake while the right panel the zoom of one hour time window around the earthquake time.}
\end{figure}

Figure \ref{fig400} shows the sum of the cross-spectrograms of $\Delta s_1$ and $\Delta s_2$ of channel 1 and channel 2, on the entire day of 3 June 2020, and one hour around UTC time 07:35:33. The trace of channel 2 has been delayed by 519.9 s. Once again the similarity with the autocorrelation of \ref{fig7} is self evident, although the amplitude of the cross-spectrogram appears smaller, compared with the spectrogram, when compared to the similar figures for the Oxaca earthquake. 

Let us delve deeper into the Oaxaca and Chile earthquakes through the lens of the theory established earlier in this section. Regarding the Oaxaca earthquake, the cable point closest to the epicenter of the earthquake lies approximately 2,000 km from Los Angeles. Given the loop-back configuration of the system, the transmitted signal encounters the earthquake's perturbation twice during a round-trip. Assuming the perturbation is small hence it affects the polarization linearly, we can infer that the effects of the two perturbations add up. With the total length of the link from Los Angeles to Valparaiso being 10,500 km, when the signal encounters the earthquake for the first time, the propagation from the transmitter is approximately $z_p(1) \simeq 2,000$ km, resulting in $F(z_p(1)) \simeq 0.85$. When the signal encounters the earthquake for the second time, the distance traveled from the transmitter is $z_p(2) \simeq 21,000-2,000 = 19,000$ km, leading to $F(z_p(2)) \simeq  0.21$. Assuming equal efficiency of modulation in both passes, the ratio between cross-correlation and autocorrelation would be approximately $0.5 \cdot 0.85 + 0.5 \cdot 0.21 = 0.531$, yet cross-correlation and autocorrelation displayed in Figs. \ref{fig13} and \ref{fig14} appear nearly identical, at least for nonzero time lags where the effect of the clock misalignment is negligible. This suggests that the coupling with the earthquake is likely to be much stronger in the forward direction that in the backward. We speculate that a possible reason could be that the polarization modulation is imprinted in the fiber from Los Angeles to Valparaiso on a clean signal, whereas in the fiber from Valparaiso to Los Angeles on a signal strongly depolarized by the amplified spontaneous emission of the inline amplifiers. Notice that the cable was designed for one-way operation, and hence the loop-back arrangement makes the amplified emission noise power close to the receiver in Los Angeles approximately double the system's nominal value. Since the receiver can faithfully decode the signal modulation even in the loop-back configuration, it can also detect the additional polarization modulation imprinted by environmental perturbations on a polarized optical field near the transmitter. On the contrary, polarization modulation may be less efficient on the return fiber because it is applied on a signal considerably depolarized by the amplified emission noise power generated by nearly twice the number of amplifiers specified in the system's design.

Concerning the Chile earthquake, we can assume that the cable is perturbed around $z_p \simeq 10,000$ km for both passes, resulting in $F(z_p) \simeq 0.44$. Again, a significant, although smaller than the previous case, correlation between the two traces is expected and confirmed by the experimental traces in Figs. \ref{fig15} and \ref{fig16}.

These observations suggest that, in principle, within a single-pass configuration of an operational transmission system, cross-correlation between two closely spaced channels, whose spacing can be optimized for maximum accuracy, may permit the localization of the position on the link where an earthquake occurs. The optimization involves the choice of a frequency spacing maximizing the sensitivity of $F(z_p)$ on a $z_p$ ranging from $0$ to the link span $z$. A good recipe may be setting $\Delta z \simeq z$. 

To conclude the analysis, we will use the ``continuous'' data provided in Ref. \onlinecite{zhongwenzhan_2020} to highlight the impressive sensitivity at sub-hertz frequencies of environmental sensing achieved through the detection of the light polarization. Furthermore, we will experimentally confirm that the state of polarization is primarily sensitive to variations of the hydrostatic pressure rather than to mechanical vibrations in the environment. Environmental vibrations are likely to be decoupled from the fiber due to the loose-tube configuration of the cable and the presence of petroleum jelly in which the fibers are immersed. 

\begin{figure}
\includegraphics[width=7cm]{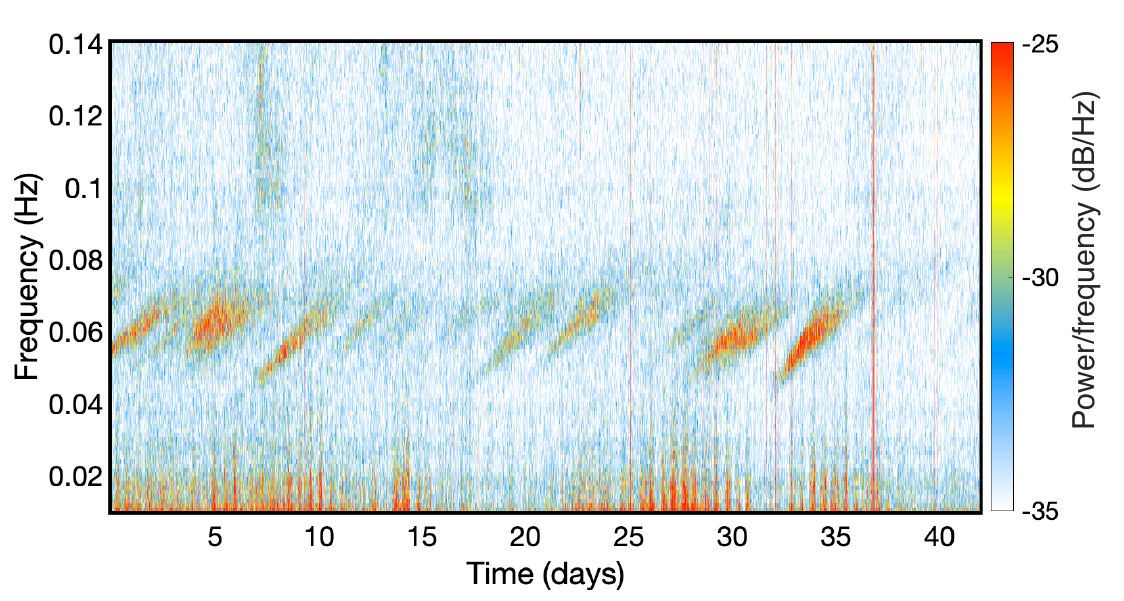}% Here is how to import EPS art
\caption{\label{fig100} Sum of the spectrograms of the two components of $\Delta \vec s$ orthogonal to $\vec s_0$, acquired from channel 1, between June 1, 2022, and July 12, 2022. The abscissa represents the number of days elapsed since June 1, 2022.}
\end{figure}

\begin{figure}
\includegraphics[width=7cm]{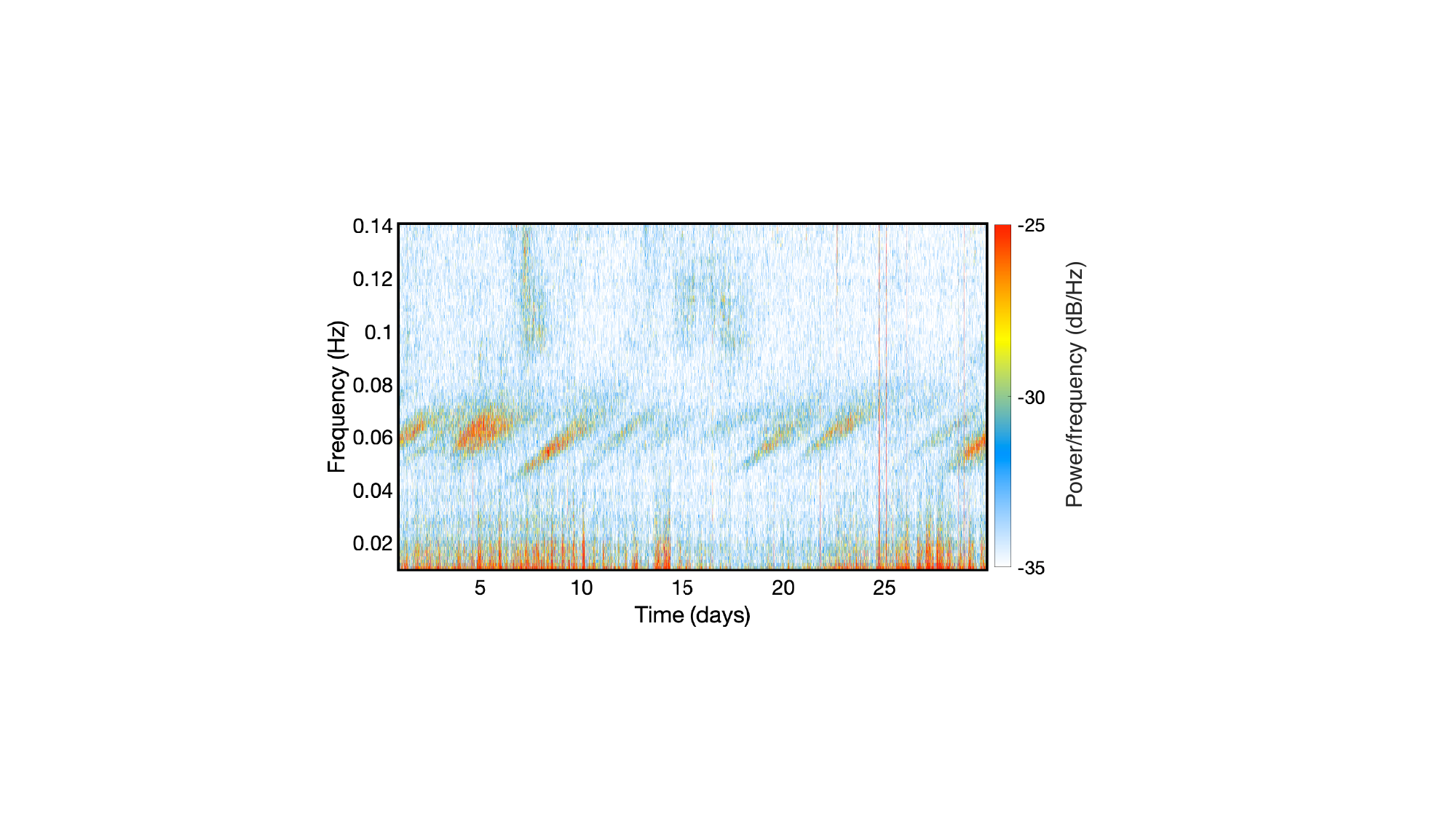}% Here is how to import EPS art
\caption{\label{fig101} Sum of the spectrograms of the two components of $\Delta \vec s$ orthogonal to $\vec s_0$, acquired from channel 2, between June 2, 2022, and June 30, 2022. The abscissa represents the number of days elapsed since June 1, 2022.}
\end{figure}

Figures \ref{fig100} and \ref{fig101} (see also refs. \onlinecite{Zhan:21,Mecozzi:21}) depict the sum of the spectrograms of the two components of the Stokes vector orthogonal to the input state of polarization, recorded from June 1, 2022, to July 12, 2022 for channel 1 (Fig. \ref{fig100}) and from June 2, 2022, to June 30, 2022 for channel 2 (Fig. \ref{fig101}). Two prominent features are observable. First, there are spectral features attributed to ocean swells, which appear as dispersive wave structures. Remarkably, these structures are also clearly visible in spectrograms obtained from onshore seismographs near Los Angeles.\cite{Zhan:21} Notably, these structures are exclusively present in the primary microseism band and lack a corresponding second-harmonic spectral signature in the secondary microseism band. The secondary microseism is the second-harmonic signal generated by the phase-matched excitation by pressure variations on the seafloor of two nearly counterpropagating seismic waves. The seconday microseism is clearly visible in spectrograms from on-shore seismographs (see Fig. 4, panels (B) and (C), of Ref. \onlinecite{Zhan:21}) whereas it is absent in Figs. \ref{fig100} and \ref{fig101}. This observation suggests that the state of polarization is relatively insensitive to vibrations but highly sensitive to strain induced by the direct action of pressure variations caused by ocean swells.

The second spectral feature discerned from the analysis of the spectrogram is a very distinct semidiurnal modulation at around 20 mHz. This feature arises from the modulation of the pressure applied to the fiber caused by ocean tides. It is remarkable that this impressive sensitivity to ultralow frequency perturbations was achieved through the use of a laser with a linewidth in the hundreds of kHz range, a manifestation of the immunity of the laser polarization from phase noise. Achieving comparable sensitivity at such low frequencies would be challenging, if at all possible, using phase even with lasers of ultralow linewidth.

\section{Conclusions}

In this paper, we initially established a theoretical framework for understanding the sensing capabilities of optical fibers. We delineated the advantages and limitations associated with utilizing polarization-averaged optical phase and the light polarization as sensing tools, showing the distinct advantage of polarization over phase to discriminate sub-hertz environmental processes. Subsequently, we proposed a scheme capable of extracting the spectrum of perturbations affecting the cable by detecting the state of polarization of the backreflected light. Exploiting the extensive dataset of Ref. \onlinecite{zhongwenzhan_2020}, we discussed two examples of earthquake detection and the detection of sea swells and ocean tides through the state of polarization reconstructed by the receiver of the Curie cable. Finally, we gave the analytical expression of the polarization cross-correlations between two channels at nearby frequencies and demonstrated how the analysis of these correlations can provide valuable insights into the localization of earthquakes.

\appendix

\section{Continuous limit of Eqs. (\ref{null}) and (\ref{null1})} \label{AppendixB}

In this appendix, we derive the continuous limit of Eqs. (\ref{null}) and (\ref{null1}). After multiplying and dividing by the sampling time $T$, Eqs. (\ref{NOmega}) and (\ref{phiOmega}) become
\be \tilde N(\Omega_k) = \frac 1 {nT} \, \sum_{h =0}^{n-1} N(h T) \exp\left(i \Omega_k h T\right) T, \label{NOmegaT} \ee
\be \tilde \varphi(\Omega_k) = \frac 1 {n T} \, \sum_{h =0}^{n-1} \varphi(h T) \exp\left(i \Omega_k h T\right) T. \label{phiOmegaT} \ee
Letting $T$ tend to zero and $n$ to infinity while maintaining $n T = T_\mathrm{win}$ finite, $T$ becomes $\df t$, $h T$ in the exponents becomes a continuous time $t$ and the sums transform into integrals. Thus, we arrive at the continuous limit of these expressions as
%Equation (\ref{null1}) can be derived in the continuous limit by applying a finite-time Fourier transform to the phase and the noise, using the definitions
%
\be
\tilde N_0(\Omega) = \frac 1 {T_\mathrm{win}} \int_0^{T_\mathrm{win}} \exp(i \Omega t) N_0(t) \df t, \label{N0omegacont}\ee
and
\be \tilde \varphi(\Omega) = \frac 1 {T_\mathrm{win}} \int_0^{T_\mathrm{win}} \exp(i \Omega t) \varphi(t) \df t. \ee
Strictly speaking, Eq. (\ref{Omegak}) dictates that $\Omega$ takes on the discrete values $\Omega = 2 \pi k/(nT) = 2 \pi k/T_\mathrm{win}$, with $k \in \mathbb Z$. We removed the dependence on the integer $k$ because we assume that $\Omega$ is analytically continued over the entire real axis. Again, like $1/n$ in the discrete case, the normalization factor $1/T_\mathrm{win}$ ensures that the peak amplitude of the Fourier transform of a sinusoidal modulation is independent of the time window $T_\mathrm{win}$. %The normalization factor $1/T_\mathrm{win}$ plays the role of the normalization factor $1/n$ in Eqs. (\ref{NOmega}) and (\ref{phiOmega}). It ensures that the peak amplitude of the Fourier transform of a pure harmonic modulation is independent of the observation time $T_\mathrm{win}$. 
In the continuous limit, the laser phase noise is negligible over the signal if 
\be |\tilde N_0(\Omega)|^2 \ll |\Omega|^2 |\tilde \varphi(\Omega)|^2. \label{condcont} \ee
Inserting into $|\tilde N_0(\Omega)|^2$ the expression (\ref{N0omegacont}), averaging the result and using Eq. (\ref{whitenoise}) produces the equality
\be \langle |\tilde N_0(\Omega)|^2 \rangle = \frac{2 \pi \nu}{T_\mathrm{win}}.  \label{N0app} \ee
Inserting this expression into Eq. (\ref{condcont}) and using the definition $f = \Omega/(2 \pi)$ yields
\be \nu \ll 2 \pi f^2 |\tilde \varphi(2 \pi f)|^2 T_\mathrm{win}, \label{nullcont0} \ee
which is the continuous limit of Eq. (\ref{null}). Using now the definition $f = 1/T_f$, we obtain the continuous limit of Eq. (\ref{null1}), namely
\be \nu \ll 2 \pi |f| |\tilde \varphi(2 \pi f)|^2 \frac{T_\mathrm{win}}{T_f}. \label{nullcont} \ee

{ 
\section{Self-referenced schemes} \label{AppendixC}

In this appendix, we examine the qualitatively different situation that arises with self-referenced phase measurements, such as those using self-homodyne or self-heterodyne detection.\cite{Donadello:24} These schemes are characterized by the same laser serving as both the probe and the local oscillator, a configuration commonly used in roundtrip scenarios where transmitter and receiver are colocated. If $\tau$ is the roundtrip time of the probe across the fiber, the detected phase noise is the difference between the noise of the laser used as local oscillator and the noise of the same laser used as the probe, launched through the fiber a time $\tau$ earlier, that is
\be \varphi_\mathrm{noise}(t) = \varphi_\mathrm{laser}(t) - \varphi_\mathrm{laser}(t-\tau), \label{varphidiff} \ee
If $D_\varphi \tau \gg 2 \pi$, that is $\tau \gg 1/\nu$, the roundtrip time exceeds the coherence time of the laser and hence the two processes $\varphi_\mathrm{laser}(t)$ and $\varphi_\mathrm{laser}(t-\tau)$ are uncorrelated. In this case, the phase noise is the same as if the probe and local oscillator were generated by independent lasers. Conversely, if $\tau < 1/\nu$, the phase noise is becomes a stationary process, whose correlation function is
\be \langle \varphi_\mathrm{noise}(t) \varphi_\mathrm{noise}(t') \rangle = D_\varphi \tau \operatorname{tri}\left(\frac{t-t'} \tau\right), \label{varphiapp} \ee
where $\operatorname{tri}(u)$ is the triangular function equal to $1-|u|$ for $|u| \le 1$ and zero elsewhere. 

Let us first consider the direct estimation of the frequency deviation obtained by using a single phase increment. A direct quantification of the noise in this estimate can be gained using for the noise the estimate (\ref{dotvarphi}) of the frequency deviations
\be \dot \varphi (T) = \frac {\Delta \varphi(T)} {T}  =  \frac{\varphi_\mathrm{noise}(t+T) -  \varphi_\mathrm{noise}(t)}{T}, \ee
squaring, averaging and using that $\varphi_\mathrm{noise}(t+T)^2 = \varphi_\mathrm{noise}(t)^2$, yields for $\dot \varphi_\mathrm{rms}(T)^2 = \langle \dot \varphi^2 (T) \rangle$ 
\be \dot \varphi_\mathrm{rms}(T)^2 = \frac{2 \langle \varphi_\mathrm{noise}(t)^2 \rangle - 2 \langle \varphi_\mathrm{noise}(t)\varphi_\mathrm{noise}(t+T)\rangle }{T^2}. \label{dotvarirms} \ee
Using now the correlation function (\ref{varphiapp}) into Eq. (\ref{dotvarirms}), we obtain for the mean square of the frequency noise the following expression
\be \dot \varphi_\mathrm{rms}(T)^2 = \frac{2 D_\varphi \tau}{T^2} \left[1-\operatorname{tri}\left(\frac{T} \tau\right)\right]. \ee
For $T\le\tau$, the above expression yields $\dot \varphi_\mathrm{rms}(T)^2 = 2 D_\varphi /T$ indicating that the noise is twice the mean square noise $\dot \varphi_\mathrm{rms}(T)^2$ expressed by Eq. (\ref{dotvarphi1}). This occurs because for $T \le \tau$ the increments over the time $T$ of the two terms on the right-hand side of Eq. (\ref{varphidiff}) are independent,  and this leads to their variances adding up. In contrast, for $T >\tau$ the expression becomes 
\be \dot \varphi_\mathrm{rms}(T)^2 = \frac{2 D_\varphi} {T} \left(\frac{\tau} {T}\right), \quad T >\tau. \label{nuapp1} \ee
This is because only a fraction $\tau/T$ of the phase increments pertaining to the two terms are independent, while the others are equal and cancel each other out.

Let us now focus the analysis on low-frequency fluctuations. Low-pass filtering over a bandwidth of $1/T_\mathrm{win}$ can be accomplished by averaging the frequency estimations obtained with time increments of duration $T$ contained in a time window of amplitude $T_\mathrm{win}$ 
\be \dot \varphi_\mathrm{ave}(T_\mathrm{win}) = \frac 1 {T_\mathrm{win}} \int_0^{T_\mathrm{win}} \frac{\varphi_\mathrm{noise}(t+T)- \varphi_\mathrm{noise}(t)}{T}\df t . \label{Omega0} \ee
The values of $\varphi_\mathrm{noise}(t)$ for $t \in [T, T_\mathrm{win})$ cancel out, and hence
\be \dot \varphi_\mathrm{ave}(T_\mathrm{win}) = \frac 1 {T} \int_0^T \frac{\varphi_\mathrm{noise}(t+T_\mathrm{win}) - \varphi_\mathrm{noise}(t)}{T_\mathrm{win}}\df t . \label{Omega00} \ee
After squaring Eq. (\ref{Omega00}), we obtain a double integral with an integrand consisting of four terms. For $T_\mathrm{win} \ge T + \tau$, the two terms of the form $\varphi_\mathrm{noise}(t+T_\mathrm{win}) \varphi_\mathrm{noise}(t')$ vanish after averaging because they are the product of  statistically independent terms with zero mean. The only non-zero contributions are the terms $\langle \varphi_\mathrm{noise}(t'+T_\mathrm{win}) \varphi_\mathrm{noise}(t+T_\mathrm{win})\rangle$ and $\langle \varphi_\mathrm{noise}(t)\varphi_\mathrm{noise}(t') \rangle$, which are equal.  With this in mind, it is easy to show that 
\be \langle \dot \varphi_\mathrm{ave}^2(T_\mathrm{win}) \rangle = \frac 2 {T^2} \int_0^T \df t \int_0^T \df t' \frac{\langle \varphi_\mathrm{noise}(t)\varphi_\mathrm{noise}(t') \rangle}{T_\mathrm{win}^2}. \label{dotvariint} \ee
Inserting Eq. (\ref{varphiapp}) into the Eq. (\ref{dotvariint}) and performing the integral, we obtain, for $T \ge \tau$
\be \langle \dot \varphi_\mathrm{ave}^2(T_\mathrm{win}) \rangle =  \frac {2 D_\varphi \tau^2} {T T_\mathrm{win}^2} \left( 1 - \frac{\tau}{3 T} \right), \quad T \ge \tau, \quad T_\mathrm{win} \ge T + \tau, \label{edgefinal} \ee
and for $T \le \tau$
\be \langle \dot \varphi_\mathrm{ave}^2(T_\mathrm{win}) \rangle =  \frac {2 D_\varphi \tau} {T_\mathrm{win}^2} \left( 1 - \frac{T}{3 \tau} \right), \quad T \le \tau, \quad T_\mathrm{win} \ge T + \tau. \label{edgefinal1} \ee
The condition $\dot \varphi^2 \gg \langle \dot \varphi_\mathrm{ave}^2(T_\mathrm{win}) \rangle$ becomes, using $D_\varphi = 2 \pi \nu$
\be \nu \ll \frac{\dot \varphi^2 T}{4 \pi} \, \left(\frac{T_\mathrm{win}}{\tau}\right)^2 \frac{3 T}{3T-\tau}, \quad T \ge \tau, \quad T_\mathrm{win} \ge T + \tau, \label{nuapp4} \ee
and
\be \nu \ll \frac{\dot \varphi^2 \tau}{4 \pi} \, \left(\frac{T_\mathrm{win}}{\tau}\right)^2 \frac{3 \tau}{3\tau-T}, \quad T \le \tau, \quad T_\mathrm{win} \ge T + \tau. \label{nuapp5} \ee
The parameter $T$ has the meaning of the time interval used to estimate the frequency deviations. Using a small $T$ increases the effective noise until it clamps for $T < \tau$. Being the effect of the noise minimum for maximum $T$, and because of the condition $T_\mathrm{win} \ge T + \tau$, the optimal choice is $T = T_\mathrm{win}-\tau$. 

Assume now that temporal averaging of the frequency is accomplished using a (sliding) window function more general than the simple averaging employed in Eq. (\ref{Omega0}), that is by using the following expression
\be \dot \varphi_\mathrm{ave}(t) = \int_{-\infty}^{\infty} h(t-t') \frac{\varphi_\mathrm{noise}(t'+T)- \varphi_\mathrm{noise}(t')}{T}\df t', \label{dotacca} \ee
where the window function is normalized such that 
\be \int_{-\infty}^{\infty} h(t) \df t = 1. \label{unitnor} \ee
Multiplication by a sliding window and subsequent integration is equivalent to the application of a low-pass filter. The quantity $\dot \varphi_\mathrm{ave}(T_\mathrm{win})$ in Eq. (\ref{Omega0}) is a particular case of Eq. (\ref{dotacca}), corresponding to $h(t) = 1/T_\mathrm{win}$ for $t\in[0,T_\mathrm{win})$ and zero elsewhere. To simplify the analysis, let us make the additional assumption that $\tau$ is much smaller than both $T$ and the width of $h(t)$ so that we may approximate Eq. (\ref{varphiapp}) as
\be \langle \varphi_\mathrm{noise}(t) \varphi_\mathrm{noise}(t') \rangle = D_\varphi \tau^2 \delta(t-t'). \label{varphiappdelta} \ee
Squaring and averaging Eq. (\ref{dotacca}) we obtain
%
%\bea \langle \dot \varphi_\mathrm{ave}(t)^2 \rangle &=& \frac{2 D_\varphi \tau^2}{T^2} \int_{-\infty}^{\infty} \left[h(t-t')^2 \right.\nonumber \\
%&& \left. -h(t-t')h(t-t'-T)\right] \df t'. \eea
%
%
\be \langle \dot \varphi_\mathrm{ave}(t)^2 \rangle = \frac{2 D_\varphi \tau^2}{T^2} \int_{-\infty}^{\infty} \left[h(t')^2 -h(t')h(t'-T)\right] \df t', \label{dotvarapp}  \ee
where we used the substitution $t-t' \to t'$. This expression is consistent with the result obtained with simple averaging for $\tau \ll T$. This can be shown using $h(t) = 1/T_\mathrm{win}$ for $t\in[0,T_\mathrm{win})$ and zero elsewhere. With this substitution, we obtain $\langle \dot \varphi_\mathrm{ave}(T_\mathrm{win})^2 \rangle = 2 D_\varphi \tau^2 /(T T_\mathrm{win}^2)$,  which coincides for $\tau \ll T$ with Eq. (\ref{edgefinal}). 

Assume now that $h(t)$ is a smooth function, and that $T_\mathrm{win}$ is the scale of variation of $h(t)$. This condition is not satisfied by simple averaging, because in this case $h(t)$ is discontinuous. If $T \ll T_\mathrm{win}$, we may perform the approximation 
%
%\be h(t-t'-T) \simeq h(t-t') + \frac{\partial h(t-t')}{\partial t'}T + \frac 1 2 \frac{\partial^2 h(t-t')}{\partial t'^2} T^2, \ee
%
%
\be h(t'-T) \simeq h(t') - \frac{\partial h(t')}{\partial t'}T + \frac 1 2 \frac{\partial^2 h(t')}{\partial t'^2} T^2, \ee
and use this approximation into Eq. (\ref{dotvarapp}), obtaining
%
%\bea \langle \dot \varphi_\mathrm{ave}(t)^2 \rangle &\simeq& \frac{2 D_\varphi \tau^2}{T^2} \int_{-\infty}^{\infty} h(t-t') \nonumber \\ && \left[-\frac{\partial h(t-t')}{\partial t'} T - \frac 1 2 \frac{\partial^2 h(t-t')}{\partial t'^2} T^2\right] \df t'. \nonumber \\ && \eea
%
%
\be \langle \dot \varphi_\mathrm{ave}(t)^2 \rangle \simeq \frac{2 D_\varphi \tau^2}{T^2} \int_{-\infty}^{\infty} h(t') \left[\frac{\partial h(t')}{\partial t'} T - \frac 1 2 \frac{\partial^2 h(t')}{\partial t'^2} T^2\right] \df t'. \ee
Being $h(t) \partial h(t) / \partial t = (1/2) \partial h(t)^2/\partial t$, the term proportional to $T$ can be explicitly integrated and, being $h(t) \to 0$ for $|t| \to \infty$, it goes to zero. Integration by parts of the second term then gives
\be \langle \dot \varphi_\mathrm{ave}(t)^2 \rangle \simeq D_\varphi \tau^2 \int_{-\infty}^{\infty} \left[\frac{\partial h(t')}{\partial t'}\right]^2 \df t'. \label{genfilfin} \ee
It is interesting to note that, in contrast to the case of a square window, for smooth window functions $\langle \dot \varphi_\mathrm{ave}(t)^2 \rangle$ is independent of $T$. As an example, consider a triangular window of width $2 T_\mathrm{win}$ and height $1/T_\mathrm{win}$. This window has twice the width and the same height and area of a rectangular window of width $T_\mathrm{win}$.\footnote{A regularized version of a triangular window is $h(t) = (1/T_\mathrm{win})[1-\sqrt{\epsilon^2+(t/T_\mathrm{win})^2}]$ for $|T/T_\mathrm{win}|\le \sqrt{1-\epsilon^2}$, and zero elsewhere, with $\epsilon \ll 1$}. In this case we have $|\partial h(t)/\partial t| = 1/T_\mathrm{win}^2$ for $t \in (0, 2 T_\mathrm{win})$ and hence we get
\be \langle \dot \varphi_\mathrm{ave}(2 T_\mathrm{win})^2 \rangle \simeq \frac{2 D_\varphi \tau^2 }{T_\mathrm{win}^3}, \ee
which is the same result obtained with a square window of amplitude $T_\mathrm{win}$ with the choice $T = T_\mathrm{win} -\tau \simeq T_\mathrm{win}$, which was previously shown to minimize the variance. 
The condition $\dot \varphi^2 \gg \langle \dot \varphi_\mathrm{ave}^2(T_\mathrm{win}) \rangle$ gives the following condition on the minimum linewidth when a general window function is used, which is parallel to Eq. (\ref{nuapp4})
%
%\be \nu  \ll \frac{\dot \varphi^2}{2 \pi \tau^2} \left\{ \int_{-\infty}^{\infty} \left[\frac{\partial h(t')}{\partial t'}\right]^2 \df t' \right\}^{-1}. \ee
%
\be \nu  \ll \frac{\dot \varphi^2}{2 \pi \tau^2 \int_{-\infty}^{\infty} \left[\partial h(t')/\partial t'\right]^2 \df t'}. \ee
Let us now move to the consideration of the case in which the analysis is performed using a short time Fourier transform. We will first consider the case where a square window of duration $T_\mathrm{win}$ is used. Let us assume, as before, that the frequency is calculated dividing by $T$ the phase changes occurring over the time interval $T$. This is the most convenient approach because it is generally difficult to track the absolute phase whereas frequency shifts are easy to follow.\cite{Mazur:24} The short time Fourier transform of the frequency noise is in this case
\be \tilde N(\Omega) = \frac 1 {T_\mathrm{win}} \int_0^{T_\mathrm{win}} \exp\left(i \Omega t\right) \frac{\varphi_\mathrm{noise}(t+T)- \varphi_\mathrm{noise}(t)}{T}\df t . \label{Omega000} \ee
Simple algebraic manipulation yields
\be \tilde N(\Omega) = \frac {\exp\left(-i \Omega T\right) - 1} {T} \tilde \varphi_\mathrm{noise}(\Omega) + \tilde R(\Omega), \label{sumN} \ee
where
\be \tilde \varphi_\mathrm{noise}(\Omega) = \frac 1 {T_\mathrm{win}} \int_T^{T_\mathrm{win}} \exp\left(i \Omega t\right) \varphi_\mathrm{noise}(t)\df t, \ee
and the rest
\bea R(\Omega) &=& \frac{1}{T_\mathrm{win} T} \int_0^T \exp\left(i \Omega t\right) \big[\exp(i \Omega T_\mathrm{win}) \varphi_\mathrm{noise}(t+T_\mathrm{win}) \nonumber \\
&& - \varphi_\mathrm{noise}(t) \big]\df t . \label{ROmega}\eea
Using Eq. (\ref{varphiapp}) for $T_\mathrm{win} \gg \tau$ and $T_\mathrm{win} \gg T$, we obtain
\be \langle |\tilde \varphi_\mathrm{noise}(\Omega)|^2 \rangle = \frac{D_\varphi \tau^2} {T_\mathrm{win}} \operatorname{sinc}^2\left(\frac{\Omega \tau} 2\right), \ee
where $\operatorname{sinc}(u) = \sin(x)/x$. Using the independence of the two terms at right-hand side of Eq. (\ref{sumN}) we obtain for the spectrum of the noise of the (angular) frequency (the time derivative of the phase) the expression
\be \langle |\tilde N(\Omega)|^2 \rangle = \frac{4 \sin^2(\Omega T/2)}{T^2} \langle |\tilde \varphi_\mathrm{noise}(\Omega)|^2 \rangle + \langle |\tilde R(\Omega)|^2 \rangle, \ee 
that is
%
%\be \langle |\tilde N(\Omega)|^2 \rangle = \frac{4 \sin^2(\Omega T/2)}{T^2} \frac{\tau^2 D_\varphi}{T_\mathrm{win}} \operatorname{sinc}^2\left(\frac{\Omega \tau} 2\right) + \langle |\tilde R(\Omega)|^2 \rangle, \ee
%
%
\be \langle |\tilde N(\Omega)|^2 \rangle = \frac{\Omega^2 \tau^2 D_\varphi}{T_\mathrm{win}} \operatorname{sinc}^2\left(\frac{\Omega T} 2 \right) \operatorname{sinc}^2\left(\frac{\Omega \tau} 2\right) + \langle |\tilde R(\Omega)|^2 \rangle. \ee
We are interested here to low frequencies, so that we may assume $\Omega T \ll 2 \pi$, and the equation above becomes
\bea \langle |\tilde N(\Omega)|^2 \rangle &=& \frac{\Omega^2 \tau^2 2 \pi \nu}{T_\mathrm{win}} \operatorname{sinc}^2\left(\frac{\Omega \tau} 2\right) + \langle |\tilde R(\Omega)|^2 \rangle \nonumber \\
&=& 4 \sin^2\left(\frac{\Omega \tau} 2\right) \frac{2 \pi \nu}{T_\mathrm{win}} + \langle |\tilde R(\Omega)|^2 \rangle, \label{noisepowerspectrum}\eea
where we used that $D_\varphi = 2 \pi \nu$. The first term in Eq. (\ref{noisepowerspectrum}) is generally dominant over the second, which is however the only one giving a nonzero contribution at $\Omega=0$. Let us analyze the first term first.

The first term of the noise power spectrum in Eq. (\ref{noisepowerspectrum}) is periodic in $f = \Omega/(2 \pi)$ with a period of $f_\tau = 1/\tau$. Averaging this term over one period yields $4 \pi \nu /T_\mathrm{win}$, which is twice the value given by Eq. (\ref{N0app}). This demonstrates that the main effect of the use of a delayed probe as local oscillator is the redistribution of the noise across the spectrum around an average noise level equal to the scenario where the probe and the local oscillator are generated by two independent laser sources. When the frequency noise contributions of the probe and local oscillator is out of phase, which occurs when the frequency $f = \Omega/(2 \pi)$ is zero or a multiple of $1/\tau$, they cancel each other out. When they are in phase, their contribution is four times that of a single source. %This result can be generalized: with a probe laser characterized by the generic frequency noise power spectrum $N_0(\Omega)$, the frequency noise affecting the phase measurement when the same laser delayed by $\tau$ is used as a local oscillator is $N(\Omega) = 4 \sin^2(\Omega \tau/2) N_0(\Omega)$.

In the limit where $\Omega \tau \ll 2 \pi$, the frequency noise spectrum becomes
\be \langle |\tilde N(\Omega)|^2 \rangle = \frac{\Omega^2 \tau^2 2 \pi \nu}{T_\mathrm{win}} + \langle |\tilde R(\Omega)|^2 \rangle. \label{nOmegaapp} \ee
For $\Omega \to 0$, the first term vanish and only the second contribute to the spectrum. It is immediate to see by setting $\Omega = 0$ into Eq. (\ref{ROmega}) and comparing with Eq. (\ref{Omega00}) that $\tilde R(0) = \dot \varphi_\mathrm{ave}(T_\mathrm{win})$, so that $\langle |\tilde R(0)|^2 \rangle = \langle \dot \varphi_\mathrm{ave}^2(T_\mathrm{win}) \rangle$. Being $\langle |\tilde R(\Omega)|^2 \rangle$ significant only for $\Omega \to 0$, we may approximate $\langle |\tilde R(\Omega)|^2 \rangle$ with its value at $\Omega =0$ without significant error. Then, the use of Eq. (\ref{edgefinal}) for $T \ge \tau$ gives 
\be \langle |\tilde N(\Omega)|^2 \rangle = \frac{\Omega^2 \tau^2 2 \pi \nu}{T_\mathrm{win}} + \frac {4 \pi \nu \tau^2} {T T_\mathrm{win}^2} \left( 1 - \frac{\tau}{3 T} \right).  \label{nOmegaapp1}\ee
The residual value of $\langle |\tilde N(\Omega)|^2 \rangle$ at $\Omega = 0$ is proportional to $\tau^2$. This indicates that the origin of this residual value is the imperfect cancellation of the laser noise at the edges of the temporal integral defining the short-time Fourier transform, caused by a non zero time delay. It is also noteworthy that this residual value is inversely proportional to $T$.

The inverse proportionality with $T$ disappears with the use of a non-square window functions, that is by using
\be \tilde N(\Omega) =\int_{-\infty}^{\infty} \exp\left(i \Omega t\right) h(t-t') \frac{\varphi_\mathrm{noise}(t'+T)- \varphi_\mathrm{noise}(t')}{T}\df t'. \label{Omega0000} \ee
We omitted in (\ref{Omega0000}) the dependence of $\tilde N(\Omega)$ on time $t$, because the statistical properties of $\tilde N(\Omega)$ are independent of $t$. Squaring Eq. (\ref{Omega0000}), using once again Eq. (\ref{varphiappdelta}) for $\tau \ll T$, and performing one of the two nested integrals with the help of the delta functions, we obtain
\be \langle |\tilde N(\Omega)|^2 \rangle = \frac{2 D_\varphi \tau^2}{T^2} \int_{-\infty}^{\infty} \left[h(t')^2 - \cos(\Omega T) h(t')h(t'-T)\right] \df t',  \ee
and after substituting $\cos(\Omega T) = 1 - 2 \sin^2(\Omega T/2)$,
\bea \langle |\tilde N(\Omega)|^2 \rangle &=& D_\varphi \tau^2 \Omega^2 \operatorname{sinc}^2\left(\frac{\Omega T} 2 \right) \int_{-\infty}^{\infty} h(t')h(t'-T) \df t' \nonumber \\
&& + \frac{2 D_\varphi \tau^2}{T^2} \int_{-\infty}^{\infty} \left[h(t')^2 - h(t')h(t'-T)\right] \df t. \nonumber \\ && \eea
%
%and after expanding the right-hand side to second order in $T$, we arrive for smooth window functions at the final result
Retaining only the lowest order of the expansion in the variable $T$ of the right-hand side, we arrive for smooth window functions at the final result
\be \langle |\tilde N(\Omega)|^2 \rangle = 2 \pi \nu \tau^2 \left\{\Omega^2 \int_{-\infty}^{\infty} h(t')^2 \df t' + \int_{-\infty}^{\infty} \left[\frac{\partial h(t')}{\partial t'}\right]^2 \df t'\right\}.\label{nOmegaapp2}\ee 
If we define as $f_0$ the frequency $f=\Omega/(2 \pi)$ when the cross-over of the two contributions to $\langle |\tilde N(\Omega)|^2 \rangle$ occurs, then we have for a square window 
\be f_0^2  = \frac 1 {2 \pi^2 T T_\mathrm{win}}, \ee
whereas for a smooth window
\be f_0^2  = \frac{\int_{-\infty}^{\infty} \left[\partial h(t')/\partial t'\right]^2 \df t'}{\int_{-\infty}^{\infty} h(t')^2 \df t'}. \ee
For $f \gg f_0$, then, the condition that the noise is much smaller then the signal, that is\footnote{The inclusion of a window function modifies the Fourier transform of the time derivative of a tone of amplitude $\varphi(\bar \Omega)$ and frequency $\bar \Omega$ into $i \bar \Omega \tilde h(\Omega-\bar \Omega) \tilde \varphi(\bar \Omega)$, where $\tilde h(\Omega)$ is the Fourier transform of the window function. Being $\tilde h(0) = 1$ for the normalization condition (\ref{unitnor}), the amplitude of the spectrum of the time derivative at frequency $\bar \Omega$ is not affected by the window function.} $\langle |\tilde N(\Omega)|^2 \rangle \ll (2 \pi f)^2 |\tilde \varphi(2 \pi f)|^2$, becomes for square window
\be \nu \ll \frac{|\tilde \varphi(2 \pi f)|^2 T_\mathrm{win}}{2 \pi \tau^2}, \label{nuappfin} \ee
and for a smooth window function
\be \nu \ll \frac{|\tilde \varphi(2 \pi f)|^2}{2 \pi \tau^2 \int_{-\infty}^{\infty} h(t')^2 \df t'}. \label{nuappfinsmooth} \ee
For these intermediate frequencies, the required laser linewidth is independent of $f$, provided that of course $f \ll 1/\tau$ and $\tau <1/\nu$, the last condition meaning that the rountrip time should be within the laser coherence time. 

%For an optimal choice of the Fourier window or the time $T$ (either choosing $T \simeq T_\mathrm{win}$ for a square window or using a smooth window function like the triangular window discussed above), we have $f_0 \simeq 2/(2 \pi T_\mathrm{win})$. This implies that the condition for the window to be wide enough to affect the signal spectrum negligibly is $f > f_0$. 

%If the window (or the time $T$ for square windows) is not chosen optimally, we may consider $f \ll f_0$. In this case, the first term in Eq. (\ref{nOmegaapp1}) can be neglected, and only the frequency-independent contribution to $\langle |\tilde N(\Omega)|^2 \rangle$ remains. In this limit, the noise is frequency independent, similar to when an independent noiseless local oscillator is employed. The condition that the noise is much smaller than the signal becomes for a square window
For smaller frequencies, namely for $f \ll f_0$, the first term in Eq. (\ref{nOmegaapp1}) can be neglected, and only the frequency-independent contribution to $\langle |\tilde N(\Omega)|^2 \rangle$ remains. In this limit, the noise is frequency independent, similar to when an independent noiseless local oscillator is employed. The condition that the noise is much smaller than the signal becomes for a square window
\be \nu \ll 2\pi f^2 |\tilde \varphi(2 \pi f)|^2 T_\mathrm{win} \frac {T T_\mathrm{win}} {2 \tau^2} \frac{3 T}{3T -\tau}, \ee
whereas for smooth windows
%
%\be \nu  \ll \frac {2\pi f^2 |\tilde \varphi(2 \pi f)|^2}{\tau^2} \left\{\int_{-\infty}^{\infty} \left[\frac{\partial h(t')}{\partial t'}\right]^2 \df t' \right\}^{-1}. \ee
%
\be \nu  \ll \frac {2\pi f^2 |\tilde \varphi(2 \pi f)|^2}{\tau^2 \int_{-\infty}^{\infty} \left[\partial h(t')/\partial t' \right]^2 \df t'}. \ee
At very low frequencies, the required laser linewidth is still proportional to the frequency square like in the case of an independent noiseless local oscillator. Yet, the effect of self referencing is beneficial, because the minimum linewidth condition increases by a factor typically much larger than one.

}

\section{Derivation of Eq. (\ref{xcorrelation1})} \label{appendixA}

In this appendix, we detail the derivation of Eq. (\ref{xcorrelation1}). Equation (\ref{xcorrelation0}) can be rewritten as
\be \langle \Delta \vec s_{\omega_2}'(z,t) \Delta \vec s_{\omega_1}'(z,t) \rangle = \xi^2 \int_0^z \df z' \int_0^z \df z'' \epsilon(z',t) \epsilon(z'',t) A(z',z''). \label{xcorrfin}  \ee
where 
\be A(z',z'') = \left\langle \left[\mathbf{R}_{\omega_2}^{-1}(z') \vec \beta(z') \times \vec s_0 \right] \cdot \left[\mathbf{R}_{\omega_1}^{-1}(z'') \vec \beta(z'') \times \vec s_0 \right] \right\rangle. \label{A} \ee
If we define the auxiliary process $\vec \beta_0(z) = \mathbf{R}_{\omega_1}^{-1}(z) \vec \beta(z)$, Eq. (\ref{A}) becomes 
\be A(z',z'') = \left\langle \left[\mathbf{R}_{\Delta \omega}^{-1}(z') \vec \beta_0(z') \times \vec s_0 \right] \cdot \left[\vec \beta_0(z'') \times \vec s_0 \right] \right\rangle, \ee
with 
\be \mathbf{R}_{\Delta \omega}^{-1}(z) = \mathbf{R}_{\omega_2}^{-1}(z) \mathbf{R}_{\omega_1}(z) . \label{Rdelta} \ee
Expanding the scalar product inside the integral yields
\be A(z',z'') = A_1(z',z'') - A_2(z',z''), \ee
where
\be A_1(z',z'') = \left\langle \mathbf{R}_{\Delta \omega}^{-1}(z') \vec \beta_0(z') \cdot \vec \beta_0(z'') \right\rangle, \label{A1} \ee
and
\be A_2(z',z'') = \left\langle \left[\mathbf{R}_{\Delta \omega}^{-1}(z') \vec \beta_0(z') \cdot \vec s_0\right] \left[\vec \beta_0(z'') \cdot \vec s_0 \right] \right\rangle. \label{A2} \ee
If we assume the delta function approximation for the correlation function of the birefringence, Eq. (\ref{deltaapp}), the same expression applies to $\vec \beta_0(z)$ because an isotropic rotation does not change the statistics of the isotropic vector $\vec \beta(z)$
\be \langle \vec \beta_0(z') \cdot \vec \beta_0(z'') \rangle = 2 L_\mathrm{f} \langle \beta^2 \rangle \delta(z'-z''). \ee
The delta function correlation permits the consideration of the case $z' = z''$ only. As customarily done in the theory of polarization mode dispersion, we include the dependence of the birefringence on frequency only in the rotation operators. If we set $\vec \beta = \vec \beta(\omega_1)$ as the birefringence at $\omega = \omega_1$, the rotation vector Eq. (\ref{Rdelta}) has the form
\be \mathbf{R}_{\Delta \omega}^{-1}(z) = \prod_{z'=z}^{z'=0} \exp\left[-\df z \vec \beta(\omega_2)(z') \times \right] \prod_{z'=0}^{z'=z} \exp\left[\df z \vec \beta(\omega_1)(z') \times \right], \label{prod} \ee
where the products are ordered from right to left. Assume now the following dependence of the birefringence on frequency
\be \vec \beta(\omega_2) = \vec \beta(\omega_1) + \frac{\vec \beta(\omega_1)}{\omega_0} (\omega_2-\omega_1), \label{beta21} \ee
which implies parallelism between $\vec \beta(\omega_2)$ and $\vec \beta(\omega_1)$. Separating the inner term in the products in Eq. (\ref{prod}) we obtain
\bea \mathbf{R}_{\Delta \omega}^{-1}(z) &=& \prod_{z'=z-\df z}^{z'=0} \exp\left[- \df z \vec \beta(\omega_2)(z') \times \right] \nonumber \\
&& \exp\left[-\df z \vec \beta(\omega_2)(z) \times \right] \exp\left[\df z \vec \beta(\omega_1)(z) \times \right] \nonumber \\
&& \prod_{z'=0}^{z'=z-\df z} \exp\left[\df z \vec \beta(\omega_1)(z') \times \right], \eea
which using the fact that $\vec \beta(\omega_1)(0)$ and $\vec \beta(\omega_2)(0)$ are parallel, becomes
\bea \mathbf{R}_{\Delta \omega}^{-1}(z) &=& \prod_{z'=z-\df z}^{z'=0} \exp\left[- \df z \vec \beta(\omega_2)(z') \times \right] \nonumber \\
&& \exp\left\{- \df z \left[\vec \beta(\omega_2)(z) - \vec \beta(\omega_1)(z) \right] \times \right\} \nonumber \\
&& \prod_{z'=0}^{z'=z-\df z} \exp\left[\df z \vec \beta(\omega_1)(z') \times \right],  \eea
and using Eq. (\ref{beta21})
\bea \mathbf{R}_{\Delta \omega}^{-1}(z) &=& \prod_{z'=z-\df z}^{z'=0} \exp\left[- \df z \vec \beta(\omega_2)(z') \times \right] \nonumber \\
&& \exp\left[- \df z \frac{\vec \beta(\omega_1)}{\omega_0} (\omega_2-\omega_1) \times \right] \nonumber \\
&& \prod_{z'=0}^{z'=z-\df z} \exp\left[\df z \vec \beta(\omega_1)(z') \times \right], \eea
For the independence of the rotations, we can average the inner term separately from the others. Using now the property of Gauassian operators
\be \left \langle \exp\left( \mathbf G \right) \right \rangle = \exp\left( \frac 1 2 \left \langle  \mathbf G^2 \right \rangle \right), \ee
and the property that holds for any isotropic vector $\vec \beta(z')$
\be\left \langle \left\{\df z \left[\vec \beta(z')/\omega_0\right] \times \right \}^2\right \rangle = - \frac 2 {3}  \frac{ \left \langle |\vec \beta (z')|^2 \Delta \omega^2 \right \rangle \df z^2}{\omega_0^2} \mathds 1, \ee
and the equality
\be \left \langle |\vec \beta(z')|^2 \right \rangle \df z^2  = 2 L_\mathrm{f} \langle \beta^2 \rangle \df z, \ee
which is the limit for $z'' \to z'$ of Eq. (\ref{deltaapp}), we obtain after averaging the inner term
\bea \mathbf{R}_{\Delta \omega}^{-1}(z) &=& \exp\left(- \frac{2 L_\mathrm{f} \langle \beta^2 \rangle \Delta \omega^2 \df z}{3 \omega_1^2} \right) \mathds 1 \nonumber \\ 
&& \prod_{z'=z-\df z}^{z'=0} \exp\left[- \df z \vec \beta(\omega_2)(z') \times \right] \nonumber \\
&& \prod_{z'=0}^{z'=z-\df z} \exp\left[\df z \vec \beta(\omega_1)(z') \times \right], \eea
Iterating the procedure, we obtain
\be \left \langle \mathbf{R}_{\Delta \omega}^{-1}(z) \right \rangle = \exp\left(- \frac{2 L_\mathrm{f} \langle \beta^2 \rangle \Delta \omega^2 z}{3 \omega_1^2} \right) \mathds 1. \ee
In Eq. (\ref{A1}) $\vec \beta_0(z)$ differs from $\vec \beta(z)$ for a constant rotation. Since this equation is nonzero only for $z' = z''$ and applying a constant rotation to the two terms of a scalar product does not affect the result, we may replace $\vec \beta_0(z')$ with $\vec \beta(z')$. Noting now that $\mathbf{R}_{\Delta \omega}^{-1}(z')$ contains the birefringence of the fiber segments before the section $z'$ we may perform the average of the rotation independently of $\vec \beta(z')$ obtaining 
\be A_1(z',z'') = 2 L_\mathrm{f} \langle \beta^2 \rangle \exp\left(-\frac{2 L_\mathrm{f} \langle \beta^2 \rangle \Delta \omega^2 z'}{3 \omega_0^2} \right) \delta(z'-z''). \label{A1av}\ee
In Eq. (\ref{A2}) we notice that $\vec \beta_0(z)$ preserves the isotropy of $\vec \beta(z)$, and that the absolute orientation of isotropic vectors is immaterial.  Again for the property that $A_1(z',z'')$ is nonzero only for $z' = z''$, $\vec \beta_0(z')$ and $\vec \beta(z')$ differ only for a constant rotation, and hence we can replace $\vec \beta_0(z')$ with $\vec \beta(z')$. With the same arguments used to derive Eq. (\ref{A1av}), we can also in this case average the rotation operator independently of $\vec \beta(z')$ obtaining
\be A_2(z',z'') = \frac 1 3 2 L_\mathrm{f} \langle \beta^2 \rangle \exp\left(-\frac{2 L_\mathrm{f} \langle \beta^2 \rangle \Delta \omega^2 z'}{3 \omega_0^2} \right) \delta(z'-z''). \ee
where we used that $\left\langle \left[\vec \beta_0(z') \cdot \vec s_0\right] \left[\vec \beta_0(z') \cdot \vec s_0 \right] \right\rangle = \langle \vec \beta_0(z')^2 \rangle/3$ for the isotropy of $\vec \beta_0(z')$. Adding the two contributions we obtain
\be A(z',z'') = \frac{2}{3} 2 L_\mathrm{f} \langle \beta^2 \rangle \exp\left(\frac{2 L_\mathrm{f} \langle \beta^2 \rangle \Delta \omega^2 z}{3 \omega_0^2} \right) \delta(z'-z''). \ee
Using Eq. (\ref{beta2}) we obtain
\be A(z',z'') = \frac{\pi}{4} \omega_0^2 \kappa^2 \exp\left(-\frac{\pi \kappa^2 \Delta \omega^2 z}{8} \right) \delta(z'-z''). \ee
Entering this result into Eq. (\ref{xcorrfin}) we obtain Eq. (\ref{xcorrelation1}).

% \nocite{*}
\bibliography{bibliography}% Produces the bibliography via BibTeX.

%merlin.mbs aipnum4-1.bst 2010-07-25 4.21a (PWD, AO, DPC) hacked
%Control: key (0)
%Control: author (8) initials jnrlst
%Control: editor formatted (1) identically to author
%Control: production of article title (0) allowed
%Control: page (1) range
%Control: year (1) truncated
%Control: production of eprint (0) enabled
\begin{thebibliography}{32}%
\makeatletter
\providecommand \@ifxundefined [1]{%
 \@ifx{#1\undefined}
}%
\providecommand \@ifnum [1]{%
 \ifnum #1\expandafter \@firstoftwo
 \else \expandafter \@secondoftwo
 \fi
}%
\providecommand \@ifx [1]{%
 \ifx #1\expandafter \@firstoftwo
 \else \expandafter \@secondoftwo
 \fi
}%
\providecommand \natexlab [1]{#1}%
\providecommand \enquote  [1]{``#1''}%
\providecommand \bibnamefont  [1]{#1}%
\providecommand \bibfnamefont [1]{#1}%
\providecommand \citenamefont [1]{#1}%
\providecommand \href@noop [0]{\@secondoftwo}%
\providecommand \href [0]{\begingroup \@sanitize@url \@href}%
\providecommand \@href[1]{\@@startlink{#1}\@@href}%
\providecommand \@@href[1]{\endgroup#1\@@endlink}%
\providecommand \@sanitize@url [0]{\catcode `\\12\catcode `\$12\catcode `\&12\catcode `\#12\catcode `\^12\catcode `\_12\catcode `\%12\relax}%
\providecommand \@@startlink[1]{}%
\providecommand \@@endlink[0]{}%
\providecommand \url  [0]{\begingroup\@sanitize@url \@url }%
\providecommand \@url [1]{\endgroup\@href {#1}{\urlprefix }}%
\providecommand \urlprefix  [0]{URL }%
\providecommand \Eprint [0]{\href }%
\providecommand \doibase [0]{http://dx.doi.org/}%
\providecommand \selectlanguage [0]{\@gobble}%
\providecommand \bibinfo  [0]{\@secondoftwo}%
\providecommand \bibfield  [0]{\@secondoftwo}%
\providecommand \translation [1]{[#1]}%
\providecommand \BibitemOpen [0]{}%
\providecommand \bibitemStop [0]{}%
\providecommand \bibitemNoStop [0]{.\EOS\space}%
\providecommand \EOS [0]{\spacefactor3000\relax}%
\providecommand \BibitemShut  [1]{\csname bibitem#1\endcsname}%
\let\auto@bib@innerbib\@empty
%</preamble>
\bibitem [{\citenamefont {Zhan}(2020)}]{zhongwenzhan_2020}%
  \BibitemOpen
  \bibfield  {author} {\bibinfo {author} {\bibfnamefont {Z.}~\bibnamefont {Zhan}},\ }\href {\doibase 10.22002/D1.1668} {\enquote {\bibinfo {title} {{Curie Data - Zhan et al. (2021)}},}\ } (\bibinfo {year} {2020}),\ \Eprint {http://arxiv.org/abs/https://data.caltech.edu/records/50509-xhf30} {https://data.caltech.edu/records/50509-xhf30} \BibitemShut {NoStop}%
\bibitem [{\citenamefont {Marra}\ \emph {et~al.}(2018)\citenamefont {Marra}, \citenamefont {Clivati}, \citenamefont {Luckett}, \citenamefont {Tampellini}, \citenamefont {Kronjäger}, \citenamefont {Wright}, \citenamefont {Mura}, \citenamefont {Levi}, \citenamefont {Robinson}, \citenamefont {Xuereb}, \citenamefont {Baptie},\ and\ \citenamefont {Calonico}}]{Marra:18}%
  \BibitemOpen
  \bibfield  {author} {\bibinfo {author} {\bibfnamefont {G.}~\bibnamefont {Marra}}, \bibinfo {author} {\bibfnamefont {C.}~\bibnamefont {Clivati}}, \bibinfo {author} {\bibfnamefont {R.}~\bibnamefont {Luckett}}, \bibinfo {author} {\bibfnamefont {A.}~\bibnamefont {Tampellini}}, \bibinfo {author} {\bibfnamefont {J.}~\bibnamefont {Kronjäger}}, \bibinfo {author} {\bibfnamefont {L.}~\bibnamefont {Wright}}, \bibinfo {author} {\bibfnamefont {A.}~\bibnamefont {Mura}}, \bibinfo {author} {\bibfnamefont {F.}~\bibnamefont {Levi}}, \bibinfo {author} {\bibfnamefont {S.}~\bibnamefont {Robinson}}, \bibinfo {author} {\bibfnamefont {A.}~\bibnamefont {Xuereb}}, \bibinfo {author} {\bibfnamefont {B.}~\bibnamefont {Baptie}}, \ and\ \bibinfo {author} {\bibfnamefont {D.}~\bibnamefont {Calonico}},\ }\bibfield  {title} {\enquote {\bibinfo {title} {Ultrastable laser interferometry for earthquake detection with terrestrial and submarine cables},}\ }\href {\doibase 10.1126/science.aat4458} {\bibfield  {journal} {\bibinfo  {journal}
  {Science}\ }\textbf {\bibinfo {volume} {361}},\ \bibinfo {pages} {486--490} (\bibinfo {year} {2018})},\ \Eprint {http://arxiv.org/abs/https://www.science.org/doi/pdf/10.1126/science.aat4458} {https://www.science.org/doi/pdf/10.1126/science.aat4458} \BibitemShut {NoStop}%
\bibitem [{\citenamefont {Zhan}\ \emph {et~al.}(2021)\citenamefont {Zhan}, \citenamefont {Cantono}, \citenamefont {Kamalov}, \citenamefont {Mecozzi}, \citenamefont {Müller}, \citenamefont {Yin},\ and\ \citenamefont {Castellanos}}]{Zhan:21}%
  \BibitemOpen
  \bibfield  {author} {\bibinfo {author} {\bibfnamefont {Z.}~\bibnamefont {Zhan}}, \bibinfo {author} {\bibfnamefont {M.}~\bibnamefont {Cantono}}, \bibinfo {author} {\bibfnamefont {V.}~\bibnamefont {Kamalov}}, \bibinfo {author} {\bibfnamefont {A.}~\bibnamefont {Mecozzi}}, \bibinfo {author} {\bibfnamefont {R.}~\bibnamefont {Müller}}, \bibinfo {author} {\bibfnamefont {S.}~\bibnamefont {Yin}}, \ and\ \bibinfo {author} {\bibfnamefont {J.~C.}\ \bibnamefont {Castellanos}},\ }\bibfield  {title} {\enquote {\bibinfo {title} {Optical polarization–based seismic and water wave sensing on transoceanic cables},}\ }\href {\doibase 10.1126/science.abe6648} {\bibfield  {journal} {\bibinfo  {journal} {Science}\ }\textbf {\bibinfo {volume} {371}},\ \bibinfo {pages} {931--936} (\bibinfo {year} {2021})},\ \Eprint {http://arxiv.org/abs/https://www.science.org/doi/pdf/10.1126/science.abe6648} {https://www.science.org/doi/pdf/10.1126/science.abe6648} \BibitemShut {NoStop}%
\bibitem [{\citenamefont {Mecozzi}\ \emph {et~al.}(2021)\citenamefont {Mecozzi}, \citenamefont {Cantono}, \citenamefont {Castellanos}, \citenamefont {Kamalov}, \citenamefont {Muller},\ and\ \citenamefont {Zhan}}]{Mecozzi:21}%
  \BibitemOpen
  \bibfield  {author} {\bibinfo {author} {\bibfnamefont {A.}~\bibnamefont {Mecozzi}}, \bibinfo {author} {\bibfnamefont {M.}~\bibnamefont {Cantono}}, \bibinfo {author} {\bibfnamefont {J.~C.}\ \bibnamefont {Castellanos}}, \bibinfo {author} {\bibfnamefont {V.}~\bibnamefont {Kamalov}}, \bibinfo {author} {\bibfnamefont {R.}~\bibnamefont {Muller}}, \ and\ \bibinfo {author} {\bibfnamefont {Z.}~\bibnamefont {Zhan}},\ }\bibfield  {title} {\enquote {\bibinfo {title} {Polarization sensing using submarine optical cables},}\ }\href {\doibase 10.1364/OPTICA.424307} {\bibfield  {journal} {\bibinfo  {journal} {Optica}\ }\textbf {\bibinfo {volume} {8}},\ \bibinfo {pages} {788--795} (\bibinfo {year} {2021})},\ \Eprint {http://arxiv.org/abs/https://opg.optica.org/optica/abstract.cfm?URI=optica-8-6-788} {https://opg.optica.org/optica/abstract.cfm?URI=optica-8-6-788} \BibitemShut {NoStop}%
\bibitem [{\citenamefont {Lindsey}, \citenamefont {Dawe},\ and\ \citenamefont {Ajo-Franklin}(2019)}]{Lindsey:19}%
  \BibitemOpen
  \bibfield  {author} {\bibinfo {author} {\bibfnamefont {N.~J.}\ \bibnamefont {Lindsey}}, \bibinfo {author} {\bibfnamefont {T.~C.}\ \bibnamefont {Dawe}}, \ and\ \bibinfo {author} {\bibfnamefont {J.~B.}\ \bibnamefont {Ajo-Franklin}},\ }\bibfield  {title} {\enquote {\bibinfo {title} {Illuminating seafloor faults and ocean dynamics with dark fiber distributed acoustic sensing},}\ }\href {\doibase 10.1126/science.aay5881} {\bibfield  {journal} {\bibinfo  {journal} {Science}\ }\textbf {\bibinfo {volume} {366}},\ \bibinfo {pages} {1103--1107} (\bibinfo {year} {2019})},\ \Eprint {http://arxiv.org/abs/https://www.science.org/doi/pdf/10.1126/science.aay5881} {https://www.science.org/doi/pdf/10.1126/science.aay5881} \BibitemShut {NoStop}%
\bibitem [{\citenamefont {Landr{\o}}\ \emph {et~al.}(2022)\citenamefont {Landr{\o}}, \citenamefont {Bouffaut}, \citenamefont {Kriesell}, \citenamefont {Potter}, \citenamefont {R{\o}rstadbotnen}, \citenamefont {Taweesintananon}, \citenamefont {Johansen}, \citenamefont {Brenne}, \citenamefont {Haukanes}, \citenamefont {Schjelderup},\ and\ \citenamefont {Storvik}}]{Landrø2022}%
  \BibitemOpen
  \bibfield  {author} {\bibinfo {author} {\bibfnamefont {M.}~\bibnamefont {Landr{\o}}}, \bibinfo {author} {\bibfnamefont {L.}~\bibnamefont {Bouffaut}}, \bibinfo {author} {\bibfnamefont {H.~J.}\ \bibnamefont {Kriesell}}, \bibinfo {author} {\bibfnamefont {J.~R.}\ \bibnamefont {Potter}}, \bibinfo {author} {\bibfnamefont {R.~A.}\ \bibnamefont {R{\o}rstadbotnen}}, \bibinfo {author} {\bibfnamefont {K.}~\bibnamefont {Taweesintananon}}, \bibinfo {author} {\bibfnamefont {S.~E.}\ \bibnamefont {Johansen}}, \bibinfo {author} {\bibfnamefont {J.~K.}\ \bibnamefont {Brenne}}, \bibinfo {author} {\bibfnamefont {A.}~\bibnamefont {Haukanes}}, \bibinfo {author} {\bibfnamefont {O.}~\bibnamefont {Schjelderup}}, \ and\ \bibinfo {author} {\bibfnamefont {F.}~\bibnamefont {Storvik}},\ }\bibfield  {title} {\enquote {\bibinfo {title} {Sensing whales, storms, ships and earthquakes using an arctic fibre optic cable},}\ }\href {\doibase 10.1038/s41598-022-23606-x} {\bibfield  {journal} {\bibinfo  {journal} {Scientific Reports}\ }\textbf
  {\bibinfo {volume} {12}},\ \bibinfo {pages} {19226} (\bibinfo {year} {2022})},\ \Eprint {http://arxiv.org/abs/https://doi.org/10.1038/s41598-022-23606-x} {https://doi.org/10.1038/s41598-022-23606-x} \BibitemShut {NoStop}%
\bibitem [{\citenamefont {Skarvang}\ \emph {et~al.}(2023)\citenamefont {Skarvang}, \citenamefont {Bj{\o}rnstad}, \citenamefont {R{\o}rstadbotnen}, \citenamefont {Bozorgebrahimi},\ and\ \citenamefont {Hjelme}}]{Skarvang:23}%
  \BibitemOpen
  \bibfield  {author} {\bibinfo {author} {\bibfnamefont {K.~S.~Y.}\ \bibnamefont {Skarvang}}, \bibinfo {author} {\bibfnamefont {S.}~\bibnamefont {Bj{\o}rnstad}}, \bibinfo {author} {\bibfnamefont {R.~A.}\ \bibnamefont {R{\o}rstadbotnen}}, \bibinfo {author} {\bibfnamefont {K.}~\bibnamefont {Bozorgebrahimi}}, \ and\ \bibinfo {author} {\bibfnamefont {D.~R.}\ \bibnamefont {Hjelme}},\ }\bibfield  {title} {\enquote {\bibinfo {title} {Observation of local small magnitude earthquakes using state of polarization monitoring in a 250km passive arctic submarine communication cable},}\ }in\ \href {\doibase 10.1364/OFC.2023.W1J.2} {\emph {\bibinfo {booktitle} {Optical Fiber Communication Conference (OFC) 2023}}}\ (\bibinfo  {publisher} {Optica Publishing Group},\ \bibinfo {year} {2023})\ p.\ \bibinfo {pages} {W1J.2}\BibitemShut {NoStop}%
\bibitem [{\citenamefont {Lu}\ \emph {et~al.}(2010)\citenamefont {Lu}, \citenamefont {Zhu}, \citenamefont {Chen},\ and\ \citenamefont {Bao}}]{Lu:10}%
  \BibitemOpen
  \bibfield  {author} {\bibinfo {author} {\bibfnamefont {Y.}~\bibnamefont {Lu}}, \bibinfo {author} {\bibfnamefont {T.}~\bibnamefont {Zhu}}, \bibinfo {author} {\bibfnamefont {L.}~\bibnamefont {Chen}}, \ and\ \bibinfo {author} {\bibfnamefont {X.}~\bibnamefont {Bao}},\ }\bibfield  {title} {\enquote {\bibinfo {title} {{Distributed Vibration Sensor Based on Coherent Detection of Phase-OTDR}},}\ }\href {https://opg.optica.org/jlt/abstract.cfm?URI=jlt-28-22-3243} {\bibfield  {journal} {\bibinfo  {journal} {J. Lightwave Technol.}\ }\textbf {\bibinfo {volume} {28}},\ \bibinfo {pages} {3243--3249} (\bibinfo {year} {2010})}\BibitemShut {NoStop}%
\bibitem [{\citenamefont {Pastor-Graells}\ \emph {et~al.}(2016)\citenamefont {Pastor-Graells}, \citenamefont {Martins}, \citenamefont {Garcia-Ruiz}, \citenamefont {Martin-Lopez},\ and\ \citenamefont {Gonzalez-Herraez}}]{Pastor-Graells:16}%
  \BibitemOpen
  \bibfield  {author} {\bibinfo {author} {\bibfnamefont {J.}~\bibnamefont {Pastor-Graells}}, \bibinfo {author} {\bibfnamefont {H.~F.}\ \bibnamefont {Martins}}, \bibinfo {author} {\bibfnamefont {A.}~\bibnamefont {Garcia-Ruiz}}, \bibinfo {author} {\bibfnamefont {S.}~\bibnamefont {Martin-Lopez}}, \ and\ \bibinfo {author} {\bibfnamefont {M.}~\bibnamefont {Gonzalez-Herraez}},\ }\bibfield  {title} {\enquote {\bibinfo {title} {{Single-shot distributed temperature and strain tracking using direct detection phase-sensitive OTDR with chirped pulses}},}\ }\href {\doibase 10.1364/OE.24.013121} {\bibfield  {journal} {\bibinfo  {journal} {Opt. Express}\ }\textbf {\bibinfo {volume} {24}},\ \bibinfo {pages} {13121--13133} (\bibinfo {year} {2016})}\BibitemShut {NoStop}%
\bibitem [{\citenamefont {Chen}, \citenamefont {Liu},\ and\ \citenamefont {He}(2017)}]{Chen:17}%
  \BibitemOpen
  \bibfield  {author} {\bibinfo {author} {\bibfnamefont {D.}~\bibnamefont {Chen}}, \bibinfo {author} {\bibfnamefont {Q.}~\bibnamefont {Liu}}, \ and\ \bibinfo {author} {\bibfnamefont {Z.}~\bibnamefont {He}},\ }\bibfield  {title} {\enquote {\bibinfo {title} {{Phase-detection distributed fiber-optic vibration sensor without fading-noise based on time-gated digital OFDR}},}\ }\href {\doibase 10.1364/OE.25.008315} {\bibfield  {journal} {\bibinfo  {journal} {Opt. Express}\ }\textbf {\bibinfo {volume} {25}},\ \bibinfo {pages} {8315--8325} (\bibinfo {year} {2017})}\BibitemShut {NoStop}%
\bibitem [{\citenamefont {Guerrier}\ \emph {et~al.}(2020)\citenamefont {Guerrier}, \citenamefont {Dorize}, \citenamefont {Awwad},\ and\ \citenamefont {Renaudier}}]{Guerrier:20}%
  \BibitemOpen
  \bibfield  {author} {\bibinfo {author} {\bibfnamefont {S.}~\bibnamefont {Guerrier}}, \bibinfo {author} {\bibfnamefont {C.}~\bibnamefont {Dorize}}, \bibinfo {author} {\bibfnamefont {E.}~\bibnamefont {Awwad}}, \ and\ \bibinfo {author} {\bibfnamefont {J.}~\bibnamefont {Renaudier}},\ }\bibfield  {title} {\enquote {\bibinfo {title} {{Introducing coherent MIMO sensing, a fading-resilient, polarization-independent approach to} $\varphi${-OTDR}},}\ }\href {\doibase 10.1364/OE.396460} {\bibfield  {journal} {\bibinfo  {journal} {Opt. Express}\ }\textbf {\bibinfo {volume} {28}},\ \bibinfo {pages} {21081--21094} (\bibinfo {year} {2020})}\BibitemShut {NoStop}%
\bibitem [{\citenamefont {Westbrook}(2020)}]{Westbrook:20}%
  \BibitemOpen
  \bibfield  {author} {\bibinfo {author} {\bibfnamefont {P.}~\bibnamefont {Westbrook}},\ }\bibfield  {title} {\enquote {\bibinfo {title} {{Big data on the horizon from a new generation of distributed optical fiber sensors}},}\ }\href {\doibase 10.1063/1.5144123} {\bibfield  {journal} {\bibinfo  {journal} {APL Photonics}\ }\textbf {\bibinfo {volume} {5}},\ \bibinfo {pages} {020401} (\bibinfo {year} {2020})},\ \Eprint {http://arxiv.org/abs/https://pubs.aip.org/aip/app/article-pdf/doi/10.1063/1.5144123/13499281/020401\_1\_online.pdf} {https://pubs.aip.org/aip/app/article-pdf/doi/10.1063/1.5144123/13499281/020401\_1\_online.pdf} \BibitemShut {NoStop}%
\bibitem [{\citenamefont {Lu}\ \emph {et~al.}(2019)\citenamefont {Lu}, \citenamefont {Lalam}, \citenamefont {Badar}, \citenamefont {Liu}, \citenamefont {Chorpening}, \citenamefont {Buric},\ and\ \citenamefont {Ohodnicki}}]{Ping_Lu:19}%
  \BibitemOpen
  \bibfield  {author} {\bibinfo {author} {\bibfnamefont {P.}~\bibnamefont {Lu}}, \bibinfo {author} {\bibfnamefont {N.}~\bibnamefont {Lalam}}, \bibinfo {author} {\bibfnamefont {M.}~\bibnamefont {Badar}}, \bibinfo {author} {\bibfnamefont {B.}~\bibnamefont {Liu}}, \bibinfo {author} {\bibfnamefont {B.~T.}\ \bibnamefont {Chorpening}}, \bibinfo {author} {\bibfnamefont {M.~P.}\ \bibnamefont {Buric}}, \ and\ \bibinfo {author} {\bibfnamefont {P.~R.}\ \bibnamefont {Ohodnicki}},\ }\bibfield  {title} {\enquote {\bibinfo {title} {{Distributed optical fiber sensing: Review and perspective}},}\ }\href {\doibase 10.1063/1.5113955} {\bibfield  {journal} {\bibinfo  {journal} {Applied Physics Reviews}\ }\textbf {\bibinfo {volume} {6}},\ \bibinfo {pages} {041302} (\bibinfo {year} {2019})},\ \Eprint {http://arxiv.org/abs/https://pubs.aip.org/aip/apr/article-pdf/doi/10.1063/1.5113955/14575982/041302\_1\_online.pdf} {https://pubs.aip.org/aip/apr/article-pdf/doi/10.1063/1.5113955/14575982/041302\_1\_online.pdf} \BibitemShut
  {NoStop}%
\bibitem [{\citenamefont {Donadello}\ \emph {et~al.}(2024)\citenamefont {Donadello}, \citenamefont {Clivati}, \citenamefont {Govoni}, \citenamefont {Margheriti}, \citenamefont {Vassallo}, \citenamefont {Brenda}, \citenamefont {Hovsepyan}, \citenamefont {Bertacco}, \citenamefont {Concas}, \citenamefont {Levi}, \citenamefont {Mura}, \citenamefont {Herrero}, \citenamefont {Carpentieri},\ and\ \citenamefont {Calonico}}]{Donadello:24}%
  \BibitemOpen
  \bibfield  {author} {\bibinfo {author} {\bibfnamefont {S.}~\bibnamefont {Donadello}}, \bibinfo {author} {\bibfnamefont {C.}~\bibnamefont {Clivati}}, \bibinfo {author} {\bibfnamefont {A.}~\bibnamefont {Govoni}}, \bibinfo {author} {\bibfnamefont {L.}~\bibnamefont {Margheriti}}, \bibinfo {author} {\bibfnamefont {M.}~\bibnamefont {Vassallo}}, \bibinfo {author} {\bibfnamefont {D.}~\bibnamefont {Brenda}}, \bibinfo {author} {\bibfnamefont {M.}~\bibnamefont {Hovsepyan}}, \bibinfo {author} {\bibfnamefont {E.~K.}\ \bibnamefont {Bertacco}}, \bibinfo {author} {\bibfnamefont {R.}~\bibnamefont {Concas}}, \bibinfo {author} {\bibfnamefont {F.}~\bibnamefont {Levi}}, \bibinfo {author} {\bibfnamefont {A.}~\bibnamefont {Mura}}, \bibinfo {author} {\bibfnamefont {A.}~\bibnamefont {Herrero}}, \bibinfo {author} {\bibfnamefont {F.}~\bibnamefont {Carpentieri}}, \ and\ \bibinfo {author} {\bibfnamefont {D.}~\bibnamefont {Calonico}},\ }\bibfield  {title} {\enquote {\bibinfo {title} {Seismic monitoring using the telecom fiber
  network},}\ }\href {\doibase 10.1038/s43247-024-01338-2} {\bibfield  {journal} {\bibinfo  {journal} {Communications Earth {\&} Environment}\ }\textbf {\bibinfo {volume} {5}},\ \bibinfo {pages} {178} (\bibinfo {year} {2024})}\BibitemShut {NoStop}%
\bibitem [{\citenamefont {Sakurai}\ and\ \citenamefont {Napolitano}(2017)}]{sakurai:17}%
  \BibitemOpen
  \bibfield  {author} {\bibinfo {author} {\bibfnamefont {J.}~\bibnamefont {Sakurai}}\ and\ \bibinfo {author} {\bibfnamefont {J.}~\bibnamefont {Napolitano}},\ }\href {https://books.google.it/books?id=010yDwAAQBAJ} {\emph {\bibinfo {title} {Modern Quantum Mechanics}}}\ (\bibinfo  {publisher} {Cambridge University Press},\ \bibinfo {year} {2017})\BibitemShut {NoStop}%
\bibitem [{Note1()}]{Note1}%
  \BibitemOpen
  \bibinfo {note} {This property is an immediate consequence of the isomorphism between $\protect \mathrm {SU}(2)$ and $\protect \mathrm {SO}(3)$ (the group of three dimensional rotations around the origin of Stokes space) and of the fact that the concatenation of rotations is still a rotation.}\BibitemShut {Stop}%
\bibitem [{\citenamefont {Gordon}\ and\ \citenamefont {Kogelnik}(2000)}]{Gordon:00}%
  \BibitemOpen
  \bibfield  {author} {\bibinfo {author} {\bibfnamefont {J.~P.}\ \bibnamefont {Gordon}}\ and\ \bibinfo {author} {\bibfnamefont {H.}~\bibnamefont {Kogelnik}},\ }\bibfield  {title} {\enquote {\bibinfo {title} {{PMD} fundamentals: Polarization mode dispersion in optical fibers},}\ }\href {\doibase 10.1073/pnas.97.9.4541} {\bibfield  {journal} {\bibinfo  {journal} {Proceedings of the National Academy of Sciences}\ }\textbf {\bibinfo {volume} {97}},\ \bibinfo {pages} {4541--4550} (\bibinfo {year} {2000})},\ \Eprint {http://arxiv.org/abs/https://www.pnas.org/content/97/9/4541.full.pdf} {https://www.pnas.org/content/97/9/4541.full.pdf} \BibitemShut {NoStop}%
\bibitem [{Note2()}]{Note2}%
  \BibitemOpen
  \bibinfo {note} {The property that the root mean square of the frequency noise is independent of the number of averaged samples $n$ is a direct consequence of the white noise character of the frequency noise of the laser.}\BibitemShut {Stop}%
\bibitem [{\citenamefont {Mazur}\ \emph {et~al.}(2024)\citenamefont {Mazur}, \citenamefont {Fontaine}, \citenamefont {Kelleher}, \citenamefont {Kamalov}, \citenamefont {Ryf}, \citenamefont {Dallachiesa}, \citenamefont {Chen}, \citenamefont {Neilson},\ and\ \citenamefont {Quinlan}}]{Mazur:24}%
  \BibitemOpen
  \bibfield  {author} {\bibinfo {author} {\bibfnamefont {M.}~\bibnamefont {Mazur}}, \bibinfo {author} {\bibfnamefont {N.~K.}\ \bibnamefont {Fontaine}}, \bibinfo {author} {\bibfnamefont {M.}~\bibnamefont {Kelleher}}, \bibinfo {author} {\bibfnamefont {V.}~\bibnamefont {Kamalov}}, \bibinfo {author} {\bibfnamefont {R.}~\bibnamefont {Ryf}}, \bibinfo {author} {\bibfnamefont {L.}~\bibnamefont {Dallachiesa}}, \bibinfo {author} {\bibfnamefont {H.}~\bibnamefont {Chen}}, \bibinfo {author} {\bibfnamefont {D.~T.}\ \bibnamefont {Neilson}}, \ and\ \bibinfo {author} {\bibfnamefont {F.}~\bibnamefont {Quinlan}},\ }\bibfield  {title} {\enquote {\bibinfo {title} {Continuous distributed phase and polarization monitoring of trans-atlantic submarine fiber optic cable},}\ }in\ \href {\doibase 10.1364/OFC.2024.Tu3J.1} {\emph {\bibinfo {booktitle} {Optical Fiber Communication Conference (OFC) 2024}}}\ (\bibinfo  {publisher} {Optica Publishing Group},\ \bibinfo {year} {2024})\ p.\ \bibinfo {pages} {Tu3J.1}\BibitemShut {NoStop}%
\bibitem [{\citenamefont {Oppenheim}, \citenamefont {Schafer},\ and\ \citenamefont {Buck}(1999)}]{Oppenheim:99}%
  \BibitemOpen
  \bibfield  {author} {\bibinfo {author} {\bibfnamefont {A.~V.}\ \bibnamefont {Oppenheim}}, \bibinfo {author} {\bibfnamefont {R.~W.}\ \bibnamefont {Schafer}}, \ and\ \bibinfo {author} {\bibfnamefont {J.~R.}\ \bibnamefont {Buck}},\ }\href {https://books.google.it/books?id=geTn5W47KEsC} {\emph {\bibinfo {title} {Discrete-Time Signal Processing}}},\ \bibinfo {edition} {2nd}\ ed.,\ Pearson education signal processing series\ (\bibinfo  {publisher} {Upper Saddle River, N.J.: Prentice Hall},\ \bibinfo {year} {1999})\BibitemShut {NoStop}%
\bibitem [{\citenamefont {Gordon}(2005)}]{Gordon:05}%
  \BibitemOpen
  \bibfield  {author} {\bibinfo {author} {\bibfnamefont {J.~P.}\ \bibnamefont {Gordon}},\ }\enquote {\bibinfo {title} {Statistical properties of polarization mode dispersion},}\ in\ \href {\doibase 10.1007/0-387-26307-1_3} {\emph {\bibinfo {booktitle} {Polarization Mode Dispersion}}}\ (\bibinfo  {publisher} {Springer New York},\ \bibinfo {address} {New York, NY},\ \bibinfo {year} {2005})\ pp.\ \bibinfo {pages} {52--59}\BibitemShut {NoStop}%
\bibitem [{\citenamefont {Galtarossa}\ \emph {et~al.}(2001)\citenamefont {Galtarossa}, \citenamefont {Palmieri}, \citenamefont {Schiano},\ and\ \citenamefont {Tambosso}}]{Galtarossa:01}%
  \BibitemOpen
  \bibfield  {author} {\bibinfo {author} {\bibfnamefont {A.}~\bibnamefont {Galtarossa}}, \bibinfo {author} {\bibfnamefont {L.}~\bibnamefont {Palmieri}}, \bibinfo {author} {\bibfnamefont {M.}~\bibnamefont {Schiano}}, \ and\ \bibinfo {author} {\bibfnamefont {T.}~\bibnamefont {Tambosso}},\ }\bibfield  {title} {\enquote {\bibinfo {title} {Measurement of birefringence correlation length in long, single-mode fibers},}\ }\href {\doibase 10.1364/OL.26.000962} {\bibfield  {journal} {\bibinfo  {journal} {Opt. Lett.}\ }\textbf {\bibinfo {volume} {26}},\ \bibinfo {pages} {962--964} (\bibinfo {year} {2001})}\BibitemShut {NoStop}%
\bibitem [{\citenamefont {Wuttke}, \citenamefont {Krummrich},\ and\ \citenamefont {Rosch}(2003)}]{Wuttke:03}%
  \BibitemOpen
  \bibfield  {author} {\bibinfo {author} {\bibfnamefont {J.}~\bibnamefont {Wuttke}}, \bibinfo {author} {\bibfnamefont {P.}~\bibnamefont {Krummrich}}, \ and\ \bibinfo {author} {\bibfnamefont {J.}~\bibnamefont {Rosch}},\ }\bibfield  {title} {\enquote {\bibinfo {title} {Polarization oscillations in aerial fiber caused by wind and power-line current},}\ }\href {\doibase 10.1109/LPT.2003.811143} {\bibfield  {journal} {\bibinfo  {journal} {IEEE Photonics Technology Letters}\ }\textbf {\bibinfo {volume} {15}},\ \bibinfo {pages} {882--884} (\bibinfo {year} {2003})}\BibitemShut {NoStop}%
\bibitem [{\citenamefont {Shtaif}\ and\ \citenamefont {Mecozzi}(2000)}]{Shtaif:00}%
  \BibitemOpen
  \bibfield  {author} {\bibinfo {author} {\bibfnamefont {M.}~\bibnamefont {Shtaif}}\ and\ \bibinfo {author} {\bibfnamefont {A.}~\bibnamefont {Mecozzi}},\ }\bibfield  {title} {\enquote {\bibinfo {title} {Study of the frequency autocorrelation of the differential group delay in fibers with polarization mode dispersion},}\ }\href {\doibase 10.1364/OL.25.000707} {\bibfield  {journal} {\bibinfo  {journal} {Opt. Lett.}\ }\textbf {\bibinfo {volume} {25}},\ \bibinfo {pages} {707--709} (\bibinfo {year} {2000})}\BibitemShut {NoStop}%
\bibitem [{\citenamefont {Mecozzi}\ \emph {et~al.}(2023)\citenamefont {Mecozzi}, \citenamefont {Antonelli}, \citenamefont {Mazur}, \citenamefont {Fontaine}, \citenamefont {Chen}, \citenamefont {Dallachiesa},\ and\ \citenamefont {Ryf}}]{Mecozzi:23}%
  \BibitemOpen
  \bibfield  {author} {\bibinfo {author} {\bibfnamefont {A.}~\bibnamefont {Mecozzi}}, \bibinfo {author} {\bibfnamefont {C.}~\bibnamefont {Antonelli}}, \bibinfo {author} {\bibfnamefont {M.}~\bibnamefont {Mazur}}, \bibinfo {author} {\bibfnamefont {N.}~\bibnamefont {Fontaine}}, \bibinfo {author} {\bibfnamefont {H.}~\bibnamefont {Chen}}, \bibinfo {author} {\bibfnamefont {L.}~\bibnamefont {Dallachiesa}}, \ and\ \bibinfo {author} {\bibfnamefont {R.}~\bibnamefont {Ryf}},\ }\bibfield  {title} {\enquote {\bibinfo {title} {Use of optical coherent detection for environmental sensing},}\ }\href {https://opg.optica.org/jlt/abstract.cfm?URI=jlt-41-11-3350} {\bibfield  {journal} {\bibinfo  {journal} {J. Lightwave Technol.}\ }\textbf {\bibinfo {volume} {41}},\ \bibinfo {pages} {3350--3357} (\bibinfo {year} {2023})}\BibitemShut {NoStop}%
\bibitem [{\citenamefont {Mazur}\ \emph {et~al.}(2022)\citenamefont {Mazur}, \citenamefont {Castellanos}, \citenamefont {Ryf}, \citenamefont {B\"{o}rjeson}, \citenamefont {Chodkiewicz}, \citenamefont {Kamalov}, \citenamefont {Yin}, \citenamefont {Fontaine}, \citenamefont {Chen}, \citenamefont {Dallachiesa}, \citenamefont {Corteselli}, \citenamefont {Copping}, \citenamefont {Gripp}, \citenamefont {Mortelette}, \citenamefont {Kowalski}, \citenamefont {Dellinger}, \citenamefont {Neilson},\ and\ \citenamefont {Larsson-Edefors}}]{Mazur:22a}%
  \BibitemOpen
  \bibfield  {author} {\bibinfo {author} {\bibfnamefont {M.}~\bibnamefont {Mazur}}, \bibinfo {author} {\bibfnamefont {J.~C.}\ \bibnamefont {Castellanos}}, \bibinfo {author} {\bibfnamefont {R.}~\bibnamefont {Ryf}}, \bibinfo {author} {\bibfnamefont {E.}~\bibnamefont {B\"{o}rjeson}}, \bibinfo {author} {\bibfnamefont {T.}~\bibnamefont {Chodkiewicz}}, \bibinfo {author} {\bibfnamefont {V.}~\bibnamefont {Kamalov}}, \bibinfo {author} {\bibfnamefont {S.}~\bibnamefont {Yin}}, \bibinfo {author} {\bibfnamefont {N.~K.}\ \bibnamefont {Fontaine}}, \bibinfo {author} {\bibfnamefont {H.}~\bibnamefont {Chen}}, \bibinfo {author} {\bibfnamefont {L.}~\bibnamefont {Dallachiesa}}, \bibinfo {author} {\bibfnamefont {S.}~\bibnamefont {Corteselli}}, \bibinfo {author} {\bibfnamefont {P.}~\bibnamefont {Copping}}, \bibinfo {author} {\bibfnamefont {J.}~\bibnamefont {Gripp}}, \bibinfo {author} {\bibfnamefont {A.}~\bibnamefont {Mortelette}}, \bibinfo {author} {\bibfnamefont {B.}~\bibnamefont {Kowalski}}, \bibinfo {author} {\bibfnamefont
  {R.}~\bibnamefont {Dellinger}}, \bibinfo {author} {\bibfnamefont {D.~T.}\ \bibnamefont {Neilson}}, \ and\ \bibinfo {author} {\bibfnamefont {P.}~\bibnamefont {Larsson-Edefors}},\ }\bibfield  {title} {\enquote {\bibinfo {title} {Transoceanic phase and polarization fiber sensing using real-time coherent transceiver},}\ }in\ \href {\doibase 10.1364/OFC.2022.M2F.2} {\emph {\bibinfo {booktitle} {Optical Fiber Communication Conference (OFC) 2022}}}\ (\bibinfo  {publisher} {Optica Publishing Group},\ \bibinfo {year} {2022})\ p.\ \bibinfo {pages} {M2F.2},\ \Eprint {http://arxiv.org/abs/https://opg.optica.org/abstract.cfm?URI=OFC-2022-M2F.2} {https://opg.optica.org/abstract.cfm?URI=OFC-2022-M2F.2} \BibitemShut {NoStop}%
\bibitem [{\citenamefont {Mazur}\ \emph {et~al.}(2023)\citenamefont {Mazur}, \citenamefont {Fontaine}, \citenamefont {Kelleher}, \citenamefont {Kamalov}, \citenamefont {Ryf}, \citenamefont {Dallachiesa}, \citenamefont {Chen}, \citenamefont {Neilson},\ and\ \citenamefont {Quinlan}}]{Mazur:23}%
  \BibitemOpen
  \bibfield  {author} {\bibinfo {author} {\bibfnamefont {M.}~\bibnamefont {Mazur}}, \bibinfo {author} {\bibfnamefont {N.~K.}\ \bibnamefont {Fontaine}}, \bibinfo {author} {\bibfnamefont {M.}~\bibnamefont {Kelleher}}, \bibinfo {author} {\bibfnamefont {V.}~\bibnamefont {Kamalov}}, \bibinfo {author} {\bibfnamefont {R.}~\bibnamefont {Ryf}}, \bibinfo {author} {\bibfnamefont {L.}~\bibnamefont {Dallachiesa}}, \bibinfo {author} {\bibfnamefont {H.}~\bibnamefont {Chen}}, \bibinfo {author} {\bibfnamefont {D.~T.}\ \bibnamefont {Neilson}}, \ and\ \bibinfo {author} {\bibfnamefont {F.}~\bibnamefont {Quinlan}},\ }\href@noop {} {\enquote {\bibinfo {title} {Advanced distributed submarine cable monitoring and environmental sensing using constant power probe signals and coherent detection},}\ } (\bibinfo {year} {2023}),\ \Eprint {http://arxiv.org/abs/2303.06528} {arXiv:2303.06528 [eess.SP]} \BibitemShut {NoStop}%
\bibitem [{\citenamefont {Yaman}\ \emph {et~al.}(2023)\citenamefont {Yaman}, \citenamefont {Li}, \citenamefont {Han}, \citenamefont {Inoue}, \citenamefont {Mateo},\ and\ \citenamefont {Inada}}]{Yaman:23}%
  \BibitemOpen
  \bibfield  {author} {\bibinfo {author} {\bibfnamefont {F.}~\bibnamefont {Yaman}}, \bibinfo {author} {\bibfnamefont {Y.}~\bibnamefont {Li}}, \bibinfo {author} {\bibfnamefont {S.}~\bibnamefont {Han}}, \bibinfo {author} {\bibfnamefont {T.}~\bibnamefont {Inoue}}, \bibinfo {author} {\bibfnamefont {E.}~\bibnamefont {Mateo}}, \ and\ \bibinfo {author} {\bibfnamefont {Y.}~\bibnamefont {Inada}},\ }\bibfield  {title} {\enquote {\bibinfo {title} {Polarization sensing using polarization rotation matrix eigenvalue method},}\ }in\ \href {\doibase 10.1364/OFC.2023.W1J.7} {\emph {\bibinfo {booktitle} {Optical Fiber Communication Conference (OFC) 2023}}}\ (\bibinfo  {publisher} {Optica Publishing Group},\ \bibinfo {year} {2023})\ p.\ \bibinfo {pages} {W1J.7}\BibitemShut {NoStop}%
\bibitem [{\citenamefont {Costa}\ \emph {et~al.}(2023)\citenamefont {Costa}, \citenamefont {Varughese}, \citenamefont {Mertz}, \citenamefont {Kamalov},\ and\ \citenamefont {Zhan}}]{Costa:23}%
  \BibitemOpen
  \bibfield  {author} {\bibinfo {author} {\bibfnamefont {L.}~\bibnamefont {Costa}}, \bibinfo {author} {\bibfnamefont {S.}~\bibnamefont {Varughese}}, \bibinfo {author} {\bibfnamefont {P.}~\bibnamefont {Mertz}}, \bibinfo {author} {\bibfnamefont {V.}~\bibnamefont {Kamalov}}, \ and\ \bibinfo {author} {\bibfnamefont {Z.}~\bibnamefont {Zhan}},\ }\bibfield  {title} {\enquote {\bibinfo {title} {Localization of seismic waves with submarine fiber optics using polarization-only measurements},}\ }\href {\doibase 10.1038/s44172-023-00138-4} {\bibfield  {journal} {\bibinfo  {journal} {Communications Engineering}\ }\textbf {\bibinfo {volume} {2}},\ \bibinfo {pages} {86} (\bibinfo {year} {2023})}\BibitemShut {NoStop}%
\bibitem [{\citenamefont {Galtarossa}\ \emph {et~al.}(2008)\citenamefont {Galtarossa}, \citenamefont {Grosso}, \citenamefont {Palmieri},\ and\ \citenamefont {Schenato}}]{Galtarossa:08}%
  \BibitemOpen
  \bibfield  {author} {\bibinfo {author} {\bibfnamefont {A.}~\bibnamefont {Galtarossa}}, \bibinfo {author} {\bibfnamefont {D.}~\bibnamefont {Grosso}}, \bibinfo {author} {\bibfnamefont {L.}~\bibnamefont {Palmieri}}, \ and\ \bibinfo {author} {\bibfnamefont {L.}~\bibnamefont {Schenato}},\ }\bibfield  {title} {\enquote {\bibinfo {title} {Reflectometric characterization of hinges in optical fiber links},}\ }\href {\doibase 10.1109/LPT.2008.921845} {\bibfield  {journal} {\bibinfo  {journal} {IEEE Photonics Technology Letters}\ }\textbf {\bibinfo {volume} {20}},\ \bibinfo {pages} {854--856} (\bibinfo {year} {2008})}\BibitemShut {NoStop}%
\bibitem [{Note3()}]{Note3}%
  \BibitemOpen
  \bibinfo {note} {A regularized version of a triangular window is $h(t) = (1/T_\protect \mathrm {win})[1-\protect \sqrt {\epsilon ^2+(t/T_\protect \mathrm {win})^2}]$ for $|T/T_\protect \mathrm {win}|\le \protect \sqrt {1-\epsilon ^2}$, and zero elsewhere, with $\epsilon \ll 1$}\BibitemShut {NoStop}%
\bibitem [{Note4()}]{Note4}%
  \BibitemOpen
  \bibinfo {note} {The inclusion of a window function modifies the Fourier transform of the time derivative of a tone of amplitude $\varphi (\protect \bar \Omega )$ and frequency $\protect \bar \Omega $ into $i \protect \bar \Omega \protect \tilde h(\Omega -\protect \bar \Omega ) \protect \tilde \varphi (\protect \bar \Omega )$, where $\protect \tilde h(\Omega )$ is the Fourier transform of the window function. Being $\protect \tilde h(0) = 1$ for the normalization condition (\ref {unitnor}), the amplitude of the spectrum of the time derivative at frequency $\protect \bar \Omega $ is not affected by the window function.}\BibitemShut {Stop}%
\end{thebibliography}%

\end{document}